\documentclass[acmsmall]{acmart}

\AtBeginDocument{%
  }

\setcopyright{acmlicensed}
\acmJournal{PACMMOD}
\acmYear{2023} \acmVolume{1} \acmNumber{1} \acmArticle{22} \acmMonth{5} \acmPrice{15.00}\acmDOI{10.1145/3588702}

\usepackage{algorithm}
\usepackage{algorithmic}
\usepackage{soul}
\usepackage[utf8]{inputenc}
\usepackage{amsmath}
\usepackage{amsthm}
\usepackage{booktabs}
\usepackage{multirow}
\usepackage{dsfont}
\usepackage{bm}
\usepackage{color}
\usepackage{xspace}
\usepackage{subfig}
\usepackage{xcolor}
\usepackage{makecell}

\newtheorem{definition}{Definition}
\newtheorem{theorem}{Theorem}
\newtheorem{example}{Example}
\newtheorem{lemma}{Lemma}

\newenvironment{repeatresult}[2]
{\vskip0.5em\par\textsc{#1} #2.\em}
{\vskip1em}

\newenvironment{reptheorem}[1]{\begin{repeatresult}{Theorem}{#1}}{\end{repeatresult}}
\newenvironment{replemma}[1]{\begin{repeatresult}{Lemma}{#1}}{\end{repeatresult}}

\newcommand*{\e}[1]{\emph{#1}}
\newcommand*{\txt}[1]{\texttt{#1}}

\newcommand*{\set}[1]{\{ #1 \}}
\newcommand*{\angs}[1]{\langle #1 \rangle}
\newcommand*{\enum}[2][m] {#2_1, \ldots, #2_{#1}}

\newcommand*{\defeq}{\ \mathrel{\mathop:}= \ }
\newcommand*{\pluseq}{{\ \mathrel{{+}{=}}} \ }

\newcommand*{\ranking}{\bm{\tau}}                 
\newcommand*{\partialOrder}{\bm{\nu}}             
\newcommand*{\fullpar}{\partialOrder^{\text{FP}}} 
\newcommand*{\parpar}{\partialOrder^{\text{PP}}}  
\newcommand*{\pchain}{\partialOrder^{\text{PC}}}  

\newcommand*{\numPW}{z}             
\newcommand*{\bsigma}{\bm{\sigma}}  

\newcommand*{\mew}{\text{MEW}\xspace}
\newcommand*{\mpw}{\text{MPW}\xspace}
\newcommand*{\esc}{\text{ESC}\xspace}
\newcommand*{\ESC}{Expected Score Computation\xspace}
\newcommand*{\mEw}{Most Expected Winner\xspace}
\newcommand*{\mPw}{Most Probable Winner\xspace}
\newcommand*{\mEws}{Most Expected Winners\xspace}

\newcommand*{\model}{\mathsf{M}}
\newcommand*{\RIM}{\mathsf{RIM}}
\newcommand*{\mallows}{\mathsf{MAL}}
\newcommand*{\RSM}{\mathsf{rRSM}}

\newcommand*{\vr}{\bm{r}}           
\newcommand*{\score}{s}             
\newcommand*{\winners}{\text{W}}    

\newcommand*{\VP}{\bm{P}}                               
\newcommand*{\completeVP}{\VP}                          
\newcommand*{\trunVP}{\VP^{\text{TR}}}                  
\newcommand*{\parparVP}{\VP^{\text{PP}}}                
\newcommand*{\fullparVP}{\VP^{\text{FP}}}               
\newcommand*{\pchainVP}{\VP^{\text{PC}}}                
\newcommand*{\partialVP}{\VP^{\text{PO}}}               
\newcommand*{\probaVP}{\VP^\model}                      
\newcommand*{\uniVP}{\VP^{\text{U}}}                    
\newcommand*{\rimVP}{\VP^{\RIM}}                        
\newcommand*{\rimTrunVP}{\VP^{\RIM \text{+TR}}}         
\newcommand*{\rimPartitionVP}{\VP^{\RIM \text{+FP}}}    
\newcommand*{\rimPChainVP}{\VP^{\RIM \text{+PC}}}       
\newcommand*{\rimPartialVP}{\VP^{\RIM \text{+PO}}}      
\newcommand*{\rsmVP}{\VP^{\RSM}}                        
\newcommand*{\mallowsVP}{\VP^{\mallows}}                
\newcommand*{\mallowsPartitionVP}{\VP^{\mallows \text{+FP}}}    

\newcommand*{\ifff}{if and only if\xspace}
\newcommand*{\eg}{e.g.,\xspace}
\newcommand*{\ie}{i.e.,\xspace}

\newcommand*{\etal}{\emph{et~al.}\xspace}

\newcommand{\haoyue}[1]{}

\newcommand{\rev}[1]{{#1}}
\newcommand{\revv}[1]{{#1}}

\begin{document}
  
\title{\mEw: An Interpretation of Winners over Uncertain Voter Preferences}

\author{Haoyue Ping}
\affiliation{%
  \institution{New York University}
 \country{New York, NY USA}
 }
 \email{hp1326@nyu.edu}
 \orcid{0000-0002-2694-3301}

\author{Julia Stoyanovich}\authornote{This research was supported in part by NSF awards No. 1916647, 1934464, and 1916505.}
 \affiliation{%
   \institution{New York University}
  \country{New York, NY USA}
 }
 \email{stoyanovich@nyu.edu}
 \orcid{0000-0002-1587-0450}


\begin{abstract}
It remains an open question how to determine the winner of an election when voter preferences are incomplete or uncertain. One option is to assume some probability space over the voting profile and select the \e{\mPw} (\mpw) --- the candidate or candidates with the best chance of winning.  In this paper, we propose an alternative winner interpretation, selecting the \e{\mEw} (\mew) according to the expected performance of the candidates.

We separate the uncertainty in voter preferences into the generation step and the observation step, which gives rise to a unified voting profile combining both incomplete and probabilistic voting profiles. We use this framework to establish the theoretical hardness of \mew over incomplete voter preferences, and then identify a collection of tractable cases for a variety of voting profiles, including those based on the popular Repeated Insertion Model (RIM) and its special case, the Mallows model. We develop solvers customized for various voter preference types to quantify the candidate performance for the individual voters, and propose a pruning strategy that optimizes computation.  The performance of the proposed solvers and pruning strategy is evaluated extensively on real and synthetic benchmarks, showing that our methods are practical. 
\end{abstract}



\maketitle

\section{Introduction}
\label{sec:intro}

Voting is a mechanism to determine winners among the candidates in an election by aggregating voter preferences. In classical voting theory, each voter gives a complete preference (most frequently, a ranking) of all candidates.   How voter preferences are aggregated is determined by a voting rule~\cite{DBLP:journals/scw/PritchardW07,DBLP:conf/aaai/GoldsmithLMP14,DBLP:journals/jasss/Seidl18}.
A prominent class of voting rules, which assign scores to candidates based on their positions in the rankings and then sum up the scores for each candidate, are positional scoring rules, on which we focus in this paper.

In practice, voter preferences may well be incomplete and represented by partial orders.
Since voting rules are defined over (complete) rankings, the solution is to replace each partial order with all of its linear extensions, each of which is a \e{completion} or a \e{possible world}\footnote{We use \e{completion} and \e{possible world} interchangeably.} of the partial order.
The preferences of all voters in an election are referred to as a \e{voting profile}.
A voting profile of complete rankings is a \e{complete voting profile}, while that of incomplete preferences is an \e{incomplete voting profile}.
Voters are assumed to cast their preferences independently.
A \e{completion} of an incomplete voting profile is a complete voting profile obtained by replacing each voter's partial order with one of its completions.

There have been various interpretations of winners proposed for this setting.
For example, Konczak \etal \cite{konczak2005voting} proposed the necessary winners (NWs) and possible winners (PWs): the
NWs are the candidates winning in all possible worlds, while the PWs are the ones winning in at least one possible world. Bachrach \etal \cite{DBLP:conf/aaai/BachrachBF10} assumed that the partial orders correspond to the uniform distribution over their completions, and evaluated the candidates by the number of possible worlds in which they win. We refer to this winner semantics as the \e{\mPw} (\mpw).

In this paper, we propose the \e{\mEw} (\mew) as an alternative winner interpretation for incomplete voter preferences under positional scoring rules. Like \mpw, it adopts the  possible world semantics of incomplete voting profiles.  However, in contrast to \mpw that determines a winner by a (weighted) count of the possible worlds in which she wins, \mew follows the principle of score-based rules that high-scoring candidates should be favored.  Specifically, \e{an \mew is the candidate who has the highest expected score in a random possible world}.

\mew and \mpw are similar in that they both aggregate election results over all possible worlds and give a balanced evaluation of the candidates.  Their difference lies in the aggregation methods, which will be discussed in detail in Section~\ref{sec:mew_vs_mpw}.
In practice, \mew and \mpw  may yield the same result, but this is not always the case.

\begin{figure}[tb!]
	\centering
	\small 
	\subfloat[A probabilistic voting profile of voters $x$ and $y$.]{
		\label{tab:profile}
		\begin{tabular}{@{}c|cccc@{}}
			\toprule
			Voter & $\ranking_1 {=} \angs{a, b, c}$ & $\ranking_2 {=} \angs{b, a, c}$ & $\ranking_3 {=} \angs{b, c, a}$ & $\ranking_4 {=} \angs{c, b, a}$ \\ \midrule
			$x$  &              $0.7$              &              $0.3$              &               $0$               &               $0$               \\
			$y$  &               $0$               &               $0$               &              $0.5$              &              $0.5$              \\ \bottomrule
		\end{tabular}
	}\\
	\subfloat[Each voter casts her vote independently of the other, leading to 4 possible worlds and 4 possible outcomes for (co-)winners under the plurality rule. The probability of a possible world is $\Pr(\ranking_x, \ranking_y)$. Under the plurality rule, candidate $a$ is the \mpw with a winning probability of $\Pr (\ranking_1, \ranking_3) {+} \Pr (\ranking_1, \ranking_4) {=} 0.7$, while candidate $b$ is the \mew with expected score $\Pr(\ranking_2 \mid x){+}\Pr(\ranking_3 \mid y) {=} 0.3{+}0.5{=}0.8$.]{
		\label{tab:possible_worlds}
		\begin{tabular}{@{}c|c|c@{}}
			\toprule
			Possible world                 & (Co-)winners by Plurality & $\Pr(\ranking_x, \ranking_y)$ \\ \midrule
			$x$ casts $\ranking_1$, $y$ casts $\ranking_3$ &        $\{a, b\}$         &    $0.7 \times 0.5 = 0.35$    \\
			$x$ casts $\ranking_1$, $y$ casts $\ranking_4$ &        $\{a, c\}$         &    $0.7 \times 0.5 = 0.35$    \\
			$x$ casts $\ranking_2$, $y$ casts $\ranking_3$ &          $\{b\}$          &    $0.3 \times 0.5 = 0.15$    \\
			$x$ casts $\ranking_2$, $y$ casts $\ranking_4$ &        $\{b, c\}$         &    $0.3 \times 0.5 = 0.15$    \\ \bottomrule
		\end{tabular}
	}~
	\caption{A probabilistic voting profile of 2 voters.}
	\label{fig:example_profile}
\end{figure}

\begin{example}\label{ex:1}
Figure~\ref{fig:example_profile} gives an example where \mew and \mpw select different winners in an election with two voters and three candidates, under the plurality rule. 
In this election, each voter produces a full ranking drawn from a probability distribution over  $\ranking_1 = \langle a, b, c \rangle$, $\ranking_2 = \langle b, a, c \rangle$,  $\ranking_3 = \langle b, c, a \rangle$, $\ranking_4 = \langle c, b, a \rangle$, with probabilities of each ranking for each voter given in Figure~\ref{tab:profile}.
Since voter preferences are probability distributions over rankings, the corresponding voting profile is named a \e{probabilistic voting profile}.

Let $\Pr(\ranking_x, \ranking_y)$ denote the probability that voters $x$ and $y$ cast rankings $\ranking_x$ and $\ranking_y$, respectively.  If we assume that $x$ and $y$ cast their votes independently, then $\Pr(\ranking_x, \ranking_y) = \Pr(\ranking_x \mid x) \cdot \Pr(\ranking_y \mid y)$. For example, $\Pr (\ranking_1, \ranking_3) = 0.7 \cdot 0.5 = 0.35$.
The voting profile in Figure~\ref{tab:profile} generates 4 possible worlds in Figure~\ref{tab:possible_worlds}.
The plurality rule rewards a candidate with 1 point every time she is ranked at the top of a ranking in the profile.
So, in the possible world of $\Pr (\ranking_1, \ranking_3)$, candidate $a$ obtains 1 point from $\ranking_1$, candidate $b$ obtains 1 point from $\ranking_3$, and both of them become (co-)winners in this possible world.
After enumerating all 4 possible worlds, we find that candidate $a$ is the \mpw with probability 0.7 to be a (co-)winner, while candidate $b$ is the \mew with an expected score of 0.8 points.

If the Borda rule, which rewards a candidate with the number of candidates ranked below her, is used, then, in the possible world of $\Pr (\ranking_1, \ranking_3)$, candidate $a$ would obtain 2 points from $\ranking_1$, while candidate $b$ would obtain 1 point from $\ranking_1$ and 2 points from $\ranking_3$. Candidate $b$ would then be the only NW of this profile, and she would also be the \mew with an expected score of 2.8 points.
\end{example}

{\bf Contributions and Roadmap.} We present related work in Section~\ref{sec:related}, and provide the necessary background on preferences and voting in Section~\ref{sec:preliminaries}.  

Then, in Section~\ref{sec:voting_profiles}, we present our \emph{first contribution: a unified framework for representing uncertainty in voter preferences}, where we distinguish between the uncertainty from voters themselves (\eg their preferences are actually probability distributions~\cite{marden1995analyzing}) and that due to the voting mechanism (\eg approval ballots do not allow the voters to fully reveal their preferences). We classify voting profiles into probabilistic profiles, where uncertainty is due to the voters themselves, incomplete profiles, where uncertainty is due to the voting mechanism, and combined  profiles, with both kinds of uncertainty. This classification gives us a framework within which to study the complexity of identifying \mew, by computing expected scores of the candidates.

In Section~\ref{sec:mew}, we discuss several alternative interpretations of \mew and formally state the \mew problem.  Then, we present our \emph{second contribution: an investigation of the computational complexity of \mew, and its related \ESC (\esc) problem}, in Section~\ref{sec:complexity}. The \mew problem turns out to be FP$^{\#P}$-complete under plurality, veto, and $k$-approval rules for the general case of uncertain profiles.\footnote{Recall that FP$^{\#P}$ is a class of functions efficiently solvable with an oracle to some \#P problem. A function $f$ is FP$^{\#P}$-hard if there is a polynomial-time Turing reduction from any FP$^{\#P}$ function to $f$.}

In Sections~\ref{sec:algorithms} and~\ref{sec:optimize}, we present our \emph{third contribution:  exact solvers for \mew computation over different voting profiles}. We first give a solver with exponential complexity for the partial voting profiles studied in Section~\ref{sec:complexity}. We then identify many interesting cases, such as the RIM~\cite{Doignon2004} voting profiles and the combined profiles of Mallows~\cite{Mallows1957} and partitioned preferences, where it is tractable to compute the expected scores and determine the \mew (Section~\ref{sec:algorithms}). In Section~\ref{sec:optimize}, we present \rev{performance optimizations to speed up the computation of MEW for large voting profiles.} 

In Section~\ref{sec:exp} we present our \emph{fourth contribution: an extensive experimental evaluation of the proposed solvers}, demonstrating that our proposed methods are practical. 

\rev{In Section~\ref{sec:mew_vs_mpw} we present a thorough comparison between \mew and \mpw, describing cases where \mew treats candidates differently from \mpw, and highlighting the computational advantages of \mew that can lead to more efficient computation in practice.}

We conclude and propose future directions in Section~\ref{sec:conc}.
\section{Related Work}
\label{sec:related}

\e{Winner semantics under incompleteness.} Among the winner interpretations for incomplete preferences, the most thoroughly studied are the necessary and possible winners~\cite{konczak2005voting}. A candidate is a necessary winner (NW) if she wins in every possible world;  she is a possible winner (PW) if she wins in at least one possible world.
Chakraborty \etal~\cite{DBLP:journals/tdasci/ChakrabortyDKKR21} recently developed practical techniques for NW and PW computation. NW and PW have substantial shortcomings: The requirement for NW is so strict that there are often no  winners available in a voting profile under this interpretation, while the requirement for PW does not differentiate between a candidate who only wins in one possible world and another candidate who only loses in one possible world. To address these limitations, alternative winner semantics for uncertain preferences have been proposed in the literature, discussed next.

Bachrach \etal \cite{DBLP:conf/aaai/BachrachBF10} assume that an incomplete voting profile of partial orders represents a uniform distribution over its completions, and prefer the candidates who enjoy victory in more possible worlds. This winner semantics is named the \e{\mPw} (\mpw).
While this semantics is well defined under any voting rule, and while it can be extended in a straight-forward way to incorporate the probability of a completion of a voting profile, computing a winner under \mpw is known to be intractable already under plurality~\cite{DBLP:conf/aaai/BachrachBF10}. Kenig and Kimelfeld~\cite{DBLP:conf/aaai/KenigK19} study the probability of the complement event, namely, losing an election, and devise an approach based on the Karp-Luby-Madras algorithm~\cite{DBLP:journals/jal/KarpLM89} for multiplicative polynomial-time approximations.

Hazon \etal \cite{DBLP:journals/ai/HazonAKW12} also investigate \mpw but in a different setting, where voter preferences are specified explicitly by rankings and their associated probabilities. They prove that it is \#P-hard to compute the winning probabilities under plurality, k-approval, Borda, Copeland, and Bucklin, and provide an approximation algorithm.

Imber and Kimelfeld~\cite{DBLP:conf/atal/ImberK21a} investigate the minimal and maximal possible ranks of candidates after rank aggregation of partial voting profiles, and prove intractability for every positional scoring rule.

\e{Preference models.} In this paper we model the uncertainty in voter preferences using a distance-based model known as the Mallows~\cite{Mallows1957}, and its generalizations, the Repeated Insertion Model~\cite{Doignon2004} and the Repeated Selection Model~\cite{DBLP:journals/tdasci/ChakrabortyDKKR21}.
Others have considered uncertainty in voter preferences under the Random Utility Models (RUMs) such as the Thurstone-Mosteller (TM)~\cite{Thurstone1927-THUALO-2,RePEc:spr:psycho:v:16:y:1951:i:1:p:3-9} and the Plackett-Luce (PL)~\cite{plackett1975analysis,luce1959individual}.
RUMs quantify the preferences over each item with a modal utility randomized with noise (e.g., Gaussian noise for TM and Gumbel distributions for PL), and the modal utilities are regarded as the ground truth.

Two recent papers proposed preference aggregation semantics for PL and TM models that are similar to \mew, in that they evaluate the candidates in an election based on their expected utility. 
Noothigattu \etal \cite{DBLP:conf/aaai/NoothigattuGADR18} first aggregate  preferences over the TM or PL models that correspond to each voter into a summary model to represent the entire voter base (without explicit use of a scoring rule at this step), and then select the highest modal utility candidate.  The authors show that their aggregation method is equivalent to Borda and Copeland.
Zhao \etal \cite{DBLP:conf/uai/ZhaoLWKMSX18}  apply randomized voting rules that sample a winner from the candidates with a probability proportional to their expected scores.  They demonstrate that the expected utilities of the candidates  can be determined efficiently under plurality and Borda, for PL models.  These papers share motivation with our work, but they do not consider distance-based preference models such as the Mallows, or their popular generalization like RIM, and do not study the complexity of winner determination for specific kinds of uncertain voting profiles.  Understanding the complexity of evaluation of \mew for TM and PL models for different kinds of incomplete voting profiles and voting rules is an interesting direction for future work.

\e{Querying probabilistic preferences.}
Voter preferences are a special case of preference data that has been studied in the database community~\cite{DBLP:journals/pvldb/JacobKS14}. Kenig \etal \cite{DBLP:conf/pods/KenigKPS17} propose RIMPPD, a database framework to incorporate probabilistic preferences represented by RIM models, and identify a class of tractable queries. Then Kenig \etal \cite{DBLP:conf/aaai/KenigIPKS18} further optimize the query engine with lifted inference. For a more general class of queries that are intractable in RIMPPD, Ping \etal \cite{DBLP:journals/pvldb/PingSK20} develop a number of exact solvers, as well as approximate techniques based on Multiple Importance Sampling.  In this work, we build on some of the technical insights of these papers, and develop solvers for \mew.
\section{Preliminaries}
\label{sec:preliminaries}

\subsection{Preferences}

We denote by $C = \set{\enum{c}}$ a set of \e{items} or \e{candidates} (used interchangeably.) For any $a, b, c \in C$, preference is a binary relation $\succ$ that is transitive ($a \succ b$ and $b \succ c$ imply $a \succ c$), irreflexive ($a \not\succ a$), and asymmetric ($a \succ b$ implies $b \not\succ a$).
A \e{preference pair} is an instance of this relation.

A (strict) \e{partial order} or \e{poset} is a set of preference pairs that corresponds to a directed acyclic graph with edges being the preference pairs.
For example, $\partialOrder_0 = \set{c_1 \succ c_2, c_1 \succ c_3}$ states that candidate $c_1$ is preferred to both $c_2$ and $c_3$.

A \e{ranking} or \e{permutation} is a list $\ranking = \angs{\enum{\tau}}$ such that $\forall i < j, \tau_i \succ \tau_j$.
It defines a bijection between the candidates and the ranks: $\ranking(i) = \tau_i$ and $\ranking^{-1}(\tau_i) = i$.

The \e{linear extensions} of a partial order $\partialOrder$, denoted by $\Omega(\partialOrder)$, is a set of rankings consistent with $\partialOrder$.
For example, $C_0 = \set{c_1, c_2, c_3}$ and $\partialOrder_0 = \set{c_1 \succ c_2, c_1 \succ c_3}$, then $\Omega(\partialOrder_0) = \set{\angs{c_1, c_2, c_3}, \angs{c_1, c_3, c_2}}$.

\subsection{Voting and Winners}
\label{sec:preliminaries:vote}

We denote by $V = \set{\enum[n]{v}}$ a set of voters and by $\completeVP = (\enum[n]{\ranking})$ a \e{(complete) voting profile} where $\ranking_i$ is a ranking over $C$ by voter $v_i$.

Among various voting rules, the \e{positional scoring rules} are arguably the most thoroughly studied.
Let $\vr_m = (\vr_m(1), ..., \vr_m(m))$ denote a positional scoring rule where $\forall 1 \leq i < j \leq m, \vr_m(i) \geq \vr_m(j)$ and $\vr_m(1) > \vr_m(m)$.
It assigns a score $\vr_m(i)$ to the candidate at rank $i$.  The performance of a candidate $c$ is the sum of her scores across the voting profile $\completeVP$:
$\score(c, \completeVP) = \sum_{\ranking \in \completeVP}{\score(c, \ranking)}$ where $\score(c, \ranking) = \vr_m(\ranking^{-1}(c))$.
Candidate $w$ is a (co-)winner if her score is not less than that of any other candidate.
We use $\winners(\vr_m, \completeVP)$ to denote the set of co-winners.

\begin{table}[tb!]
  \caption{Examples of positional scoring rules.}
  \begin{tabular}{@{}ll@{}}
    \toprule
    Voting rule          & $\vr_m$                    \\ \midrule
    \textit{plurality}   & $(1, 0, \ldots, 0, 0)$     \\
    \textit{veto}        & $(1, 1, \ldots, 1, 0)$     \\
    \textit{2-approval } & $(1, 1, 0, \ldots, 0, 0)$  \\
    \textit{Borda}       & $(m-1, m-2, \ldots, 1, 0)$ \\ \bottomrule
  \end{tabular}
  \centering
  \label{tab:voting_rules}
\end{table}

\begin{example}
  \rev{Given the voting profile in Table~\ref{tab:complete_profile}, under the Borda rule, \txt{Biden} obtains 3 points from \txt{Ann} and \txt{Dave}, and 0 point from \txt{Bob}, which makes him the winner with a total score of 6.}
\end{example}

\begin{table}[b!]
  \centering
  \caption{\rev{A voting profile of 3 voters over 4 candidates.}}
  \begin{tabular}{@{}ll@{}}
    \toprule
    voter      & ranking                                       \\ \midrule
    \txt{Ann}  & $\angs{\txt{Biden, Sanders, Weld, Trump}}$ \\
    \txt{Bob}  & $\angs{\txt{Trump, Weld, Sanders, Biden}}$ \\
    \txt{Dave} & $\angs{\txt{Biden, Sanders, Weld, Trump}}$ \\ \bottomrule
  \end{tabular}
  \label{tab:complete_profile}
\end{table}

\paragraph{\mPw.}

Voters may only partially express their preferences.
Let $\partialVP = (\enum[n]{\partialOrder})$ denote a \e{partial voting profile} of $n$ partial orders.
Recall that $\Omega(\partialOrder)$ is the set of \e{linear extensions} of $\partialOrder$, which are also called the \e{completions} of $\partialOrder$.
A complete voting profile $\completeVP = (\enum[n]{\ranking})$ is a \e{completion} or a \e{possible world} of a partial voting profile $\partialVP = (\enum[n]{\partialOrder})$ if $\forall \ranking_i \in \completeVP, \ranking_i \in \Omega(\partialOrder_i)$.
Let $\Omega(\partialVP) = \set{\enum[\numPW]{\completeVP}}$ denote the set of completions of $\partialVP$.
Assume that $\partialVP$ represents a uniform distribution of its possible worlds.
The \mPw (\mpw)~\cite{DBLP:conf/aaai/BachrachBF10,DBLP:journals/ai/HazonAKW12} is defined as follows.

\begin{definition} [\mpw] \label{def:mpw}
  Given a partial voting profile $\partialVP$ and a positional scoring rule $\vr_m$, candidate $w \in C$ is the \mPw (\mpw), \ifff, $\Pr(w \mid \partialVP) = \max_{c \in C} \Pr(c \mid \partialVP)$ where $\Pr(c \mid \partialVP) = \sum_{\completeVP \in \Omega(\partialVP)} \mathds{1}(c \in \winners(\vr_m, \completeVP)) \cdot \Pr(\completeVP \mid \partialVP)$.
\end{definition}

A candidate is the \mpw if her chance of winning in a random possible world is no less than any other candidate.

\begin{example}\label{eg:nw_pw}
  \rev{Given the partial voting profile in Table~\ref{tab:partial_profile}, we first construct its completions listed in Table~\ref{tab:completions_of_partial_profile}. Under the plurality rule, \txt{Biden} is the only winner in completion No.1, while completion No.2 has winners \txt{Biden}, \txt{Sanders} and \txt{Trump}. Both completions have the probability 0.5. Thus, Biden is the \mpw with winning probability 1, while the winning probability for \txt{Sanders} and \txt{Trump} is 0.5.}
\end{example}

\begin{table}[t!]
  \centering
  \caption{\rev{A partial voting profile of 3 voters over 4 candidates. \txt{Ann} gives a partial order, while \txt{Bob} and \txt{Dave} give rankings.}}
  \begin{tabular}{@{}ll@{}}
    \toprule
    voter      & partial order                                 \\ \midrule
    \txt{Ann}  & $\set{\txt{Biden} \succ \txt{Weld}, \txt{Sanders} \succ \txt{Weld}, \txt{Weld} \succ \txt{Trump}}$  \\
    \txt{Bob}  & $\angs{\txt{Trump, Weld, Sanders, Biden}}$ \\
    \txt{Dave} & $\angs{\txt{Biden, Sanders, Weld, Trump}}$ \\ \bottomrule
  \end{tabular}
  \label{tab:partial_profile}
\end{table}

\begin{table}[t!]
  \centering
  \caption{\rev{The two completions of the voting profile in Table~\ref{tab:partial_profile}.}}
  \begin{tabular}{@{}lll}
    \toprule
    completion & voter      & ranking                                       \\ \midrule
    & \txt{Ann}  & $\angs{\txt{Biden, Sanders, Weld, Trump}}$ \\
    No. 1      & \txt{Bob}  & $\angs{\txt{Trump, Weld, Sanders, Biden}}$ \\
    & \txt{Dave} & $\angs{\txt{Biden, Sanders, Weld, Trump}}$ \\ \midrule
    & \txt{Ann}  & $\angs{\txt{Sanders, Biden, Weld, Trump}}$ \\
    No. 2      & \txt{Bob}  & $\angs{\txt{Trump, Weld, Sanders, Biden}}$ \\
    & \txt{Dave} & $\angs{\txt{Biden, Sanders, Weld, Trump}}$ \\ \bottomrule
    &            &
  \end{tabular}
  \label{tab:completions_of_partial_profile}
\end{table}

\subsection{Preference Models}
\label{sec:preliminaries:pref}

\paragraph{Repeated Insertion Model (RIM)}

Doignon \etal \cite{Doignon2004} proposed a generative model $\RIM(\bsigma, \Pi)$ that defines a probability distribution over permutations.  It is parameterized by a reference ranking $\bsigma=\angs{\enum{\sigma}}$ and a probability function $\Pi$ where $\Pi(i, j)$ is the probability of inserting $\sigma_i$ at position $j$.
Algorithm~\ref{alg:rim} presents the RIM generative procedure.
It starts with an empty ranking, inserts items in the order of $\bsigma$, and places $\sigma_i$ at the $j^{th}$ position of the incomplete ranking $\ranking$ with probability $\Pi(i, j)$.

\begin{algorithm}[tb!]
\small 
  \raggedright
  \caption{Repeated Insertion Model}
  \label{alg:rim}
  \textbf{Input}: $\RIM(\bsigma, \Pi)$ where $|\bsigma| = m$ \\
  \textbf{Output}:  Ranking $\ranking$
  \begin{algorithmic}[1]
    \STATE Initialize an empty ranking $\ranking = \angs{}$.
    \FOR {$i = 1, \ldots, m$}
      \STATE Insert $\bsigma(i)$ into $\ranking$ at $j \in [1, i]$ with probability $\Pi(i, j)$.
    \ENDFOR
    \RETURN  $\ranking$
  \end{algorithmic}
\end{algorithm}

\begin{example}
    $\RIM(\angs{a, b, c}, \Pi)$ generates $\ranking' {=} \angs{c, a, b}$ as follows.
    Initialize $\ranking_0 {=} \angs{}$. 
    When $i=1$, $\ranking_1 {=} \angs{a}$ by inserting $a$ into $\ranking_0$ with probability $\Pi(1,1)$. 
    When $i=2$, $\ranking_2 {=} \angs{a, b}$ by inserting $b$ into $\ranking_1$ at position 2 with probability $\Pi(2,2)$.
    When $i=3$, $\ranking' {=} \angs{c, a, b}$ by inserting $c$ into $\ranking_2$ at position 1 with probability $\Pi(3,1)$. 
    Overall, $\Pr(\ranking' \mid \angs{a,b,c}, \Pi) {=} \Pi(1,1) \cdot \Pi(2,2) \cdot \Pi(3,1)$.
\end{example}

\paragraph{RIM Inference and Cover Width.}

Given $\RIM(\bsigma, \Pi)$ and a partial order $\partialOrder$, RIM inference is used to calculate the marginal probability of $\partialOrder$ over $\RIM(\bsigma, \Pi)$, or, equivalently, the probability that a sample from $\RIM(\bsigma, \Pi)$ is a linear extension of $\partialOrder$.

RIMDP~\cite{DBLP:conf/aaai/KenigIPKS18} is a RIM inference solver that uses Dynamic Programming (DP).
It runs the RIM insertion procedure (Algorithm~\ref{alg:rim}) to generate all possible rankings that satisfy the partial order. Given $m$ items, there are potentially $m!$ permutations to generate.
RIMDP prunes the space based on an insight that it is only necessary to track the positions of items directly related to some not-yet-inserted item in the partial order.
Two items $\set{a, b}$ are \e{directly related} and $a$ is a \e{cover} of $b$, \ifff $a \succ b$ and $\nexists c \in C, a \succ c \succ b$.
As a DP approach, the states of RIMDP are mappings from tracked items to their positions.
The maximum number of tracked items during RIM procedure is called the \e{cover width}, denoted $\text{cw}(\bsigma, \partialOrder)$.
The complexity of RIMDP is $O(m^{\text{cw}(\bsigma, \partialOrder)+2})$. 
RIMDP runs in polynomial time for partial orders whose cover widths are bounded.

\begin{example}
    Let $\bsigma=\angs{\enum[9]{\sigma}}$ and $\partialOrder=\set{\sigma_3 \succ \sigma_5, \sigma_5 \succ \sigma_8}$.
    RIMDP does not track the positions of $\sigma_1$ and $\sigma_2$ since they are not directly related to any item in $\partialOrder$.
    The insertion of $\sigma_3$ generates 3 states: $\set{\sigma_3 \rightarrow 1}$, $\set{\sigma_3 \rightarrow 2}$, and $\set{\sigma_3 \rightarrow 3}$.
    The position of $\sigma_3$ is the boundary for $\sigma_5$.
    After inserting $\sigma_5$, RIMDP will stop tracking $\sigma_3$ by merging all states sharing the same position of $\sigma_5$, \eg merging $\set{\sigma_3 \rightarrow 1, \sigma_5 \rightarrow 3}$ and $\set{\sigma_3 \rightarrow 2, \sigma_5 \rightarrow 3}$ into $\set{\sigma_5 \rightarrow 3}$.
    RIMDP will keep tracking $\sigma_5$ until $\sigma_8$ is inserted.
    The cover width in this example is 1, since RIMDP tracks at most 1 item for each insertion.
\end{example}

\begin{algorithm}[b!]
	\small 
	\raggedright
	\caption{Repeated Selection Model}
	\label{alg:rsm}
	\textbf{Input}: $\mathsf{RSM}(\bsigma, \Pi, p)$ where $|\bsigma|=m$ \\
	\textbf{Output}:  Partial order $\partialOrder$
	\begin{algorithmic}[1]
		\STATE Initialize an empty partial order $\partialOrder = \set{}$.
		\FOR {$i = 1, \ldots, m-1$}
		\STATE Sample $j \in [1, m-i+1]$ with probability $\Pi(i, j)$.
		\STATE Obtain item $\sigma = \bsigma(j)$ and remove it from $\bsigma$.
		\FOR {$k = 1, \ldots, m-i$}
		\STATE Add $\sigma \succ \bsigma(k)$ into $\partialOrder$ with probability $p(i)$.
		\ENDFOR
		\ENDFOR
		\RETURN  $\partialOrder$
	\end{algorithmic}
\end{algorithm}

\paragraph{Repeated Selection Model (RSM)}

Algorithm~\ref{alg:rsm} presents a generative model for partial orders called $\mathsf{RSM}(\bsigma, \Pi, p)$~\cite{DBLP:journals/tdasci/ChakrabortyDKKR21}.
The model is parameterized by a reference ranking $\bsigma$, a probability function $\Pi$ where $\Pi(i, j)$ is the probability of the $j^{th}$ item selected at step $i$, and a probability function $p:\set{1,...,m-1} \rightarrow [0,1]$ determining whether the selected item is preferred to each remaining item.
In contrast to RIM that randomizes insertion positions, RSM randomizes the insertion order.
In this paper, we will use RSM with $p \equiv 1$, such that it only outputs rankings, and will denote it  $\RSM(\bsigma, \Pi)$.

\paragraph{Mallows Model.}

Denoted by $\mallows(\bsigma, \phi)$, where $0 \leq \phi \leq 1$, the Mallows model~\cite{Mallows1957} defines a probability distribution over rankings: reference ranking $\bsigma$ at the center and other rankings closer to $\bsigma$ having higher probabilities.
For a given ranking $\ranking$, $\Pr(\ranking|\bsigma, \phi) \propto \phi^{D(\bsigma, \ranking)}$, where $D(\bsigma, \ranking) = |{(a, a') | a \succ_{\bsigma} a', a' \succ_{\ranking} a}|$ is the Kendall-tau distance counting the number of disagreed preference pairs.
If $\phi=1$, the Mallows becomes a uniform distribution.
The $\mallows(\bsigma, \phi)$ is a special case for both $\RIM(\bsigma, \Pi)$ by $\Pi(i, j) = \frac{\phi^{i-j}}{1+\phi+...+\phi^{i-1}}$ and $\RSM(\bsigma, \Pi)$ by $\Pi(i, j) = \frac{\phi^{j-1}}{1+\phi+...+\phi^{m-i}}$.

\section{Uncertain Voting Profiles}
\label{sec:voting_profiles}

In classical voting theory, voters give complete rankings over candidates.
However, in practice only partial preferences may be observed, due to the voting mechanism (\eg when approval ballots or a ranking of at most $k<m$ candidates are elicited), the uncertainty in preferences themselves~\cite{marden1995analyzing}, or both.  Figure~\ref{fig:venn_diagrams_of_voting_profiles}  represents uncertainty as two distinct steps: preference generation (Figure~\ref{fig:venn_generation}) and preference observation (Figure~\ref{fig:venn_observation}). Important special cases of voting profiles are discussed next.

\begin{figure}[b!]
	\centering
	\subfloat[Generation step]{
		\label{fig:venn_generation}
		\includegraphics[width=0.22\linewidth]{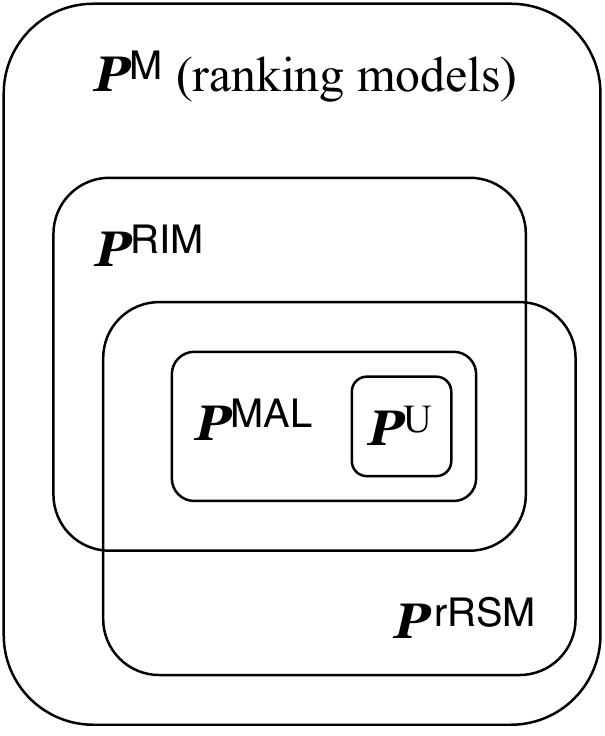}
	}
	\subfloat[Observation step]{
		\label{fig:venn_observation}
		\includegraphics[width=0.34\linewidth]{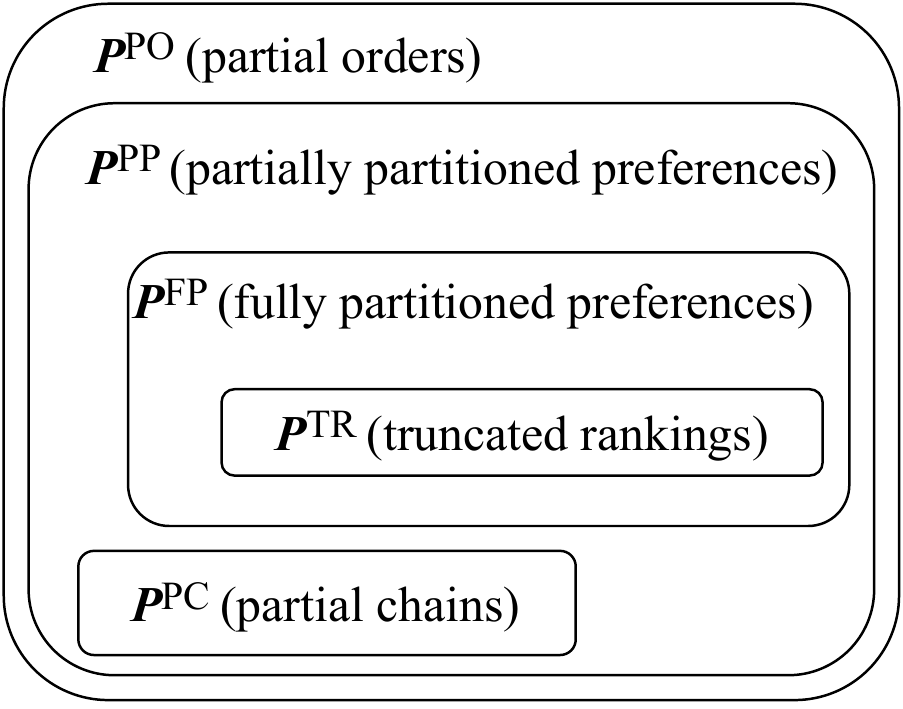}
	}~
	\caption{Uncertain voting profiles. (a) Uncertainty in the preferences by the voters themselves is represented by the ranking models in the generation step. (b) Uncertainty introduced by the preference elicitation mechanism is represented by the incomplete voting profiles in the observation step.}
	\label{fig:venn_diagrams_of_voting_profiles}
\end{figure}

\subsection{Uncertainty in profile generation}
The most general form of voter preferences over rankings is a non-parametric probability distribution such as that given in Figure~\ref{tab:profile}. 
Let $\probaVP=(\enum[n]{\model})$ denote a \e{probabilistic voting profile} where $\model_i$ is the ranking model of voter $v_i$.
A \e{possible world} of $\probaVP$ is a complete voting profile $\completeVP = (\enum[n]{\ranking})$ where each $\ranking_i$ is sampled from $\model_i$.
It is assumed that the voters cast their ballots independently, \ie $\Pr(\completeVP \mid \probaVP) = \prod_{i=1}^{n} {\Pr(\ranking_i \mid \model_i)}$.
So $\probaVP$ represents a probability distribution of its possible worlds $\Omega(\probaVP) = \set{\enum[\numPW]{\completeVP}}$.

Let $\score(c, \model)$ and $\score(c, \probaVP)$ denote the scores assigned to candidate $c$ by model $\model$ and  probabilistic voting profile $\probaVP$, respectively.
(Note that both are random variables.)
Partial voting profiles are a special case of probabilistic voting profiles, since they are based on the assumption that all completions are equally likely. 
Below are a few other cases.

In a \e{uniform voting profile}, denoted by $\uniVP$, no voter has any preference between any two candidates. Voter preferences are sampled from the uniform distribution over the rankings of all candidates.

There are specific important ranking models that we will describe in Section~\ref{sec:algorithms}.
The Mallows model~\cite{Mallows1957} is the best known, and it is generalized by the Repeated Insertion Model (RIM)~\cite{Doignon2004} and the ranking version of the Repeated Selection Model (RSM)~\cite{DBLP:journals/tdasci/ChakrabortyDKKR21}.
Voting profiles consisting of Mallows, RIMs, and RSMs (ranking version) are denoted by $\mallowsVP$, $\rimVP$, and $\rsmVP$, respectively.
We will see that, when the generation step is based on these models, we have a computational advantage for computing the \mew.

\subsection{Uncertainty in profile observation}

A \e{partial voting profile}, denoted by $\partialVP$, consists of partial orders and represents a uniform distribution of its completions.
What follows are important special cases.

A \e{(fully) partitioned voting profile}, denoted by $\fullparVP$, records preferences in the form of partitions: linear orders of item buckets, with no preference among the items within a bucket.

A \e{partially partitioned voting profile}, denoted by $\parparVP$, generalizes the fully partitioned preferences by only considering a subset of items, with no preference information for the missing ones.

A \e{partial chain voting profile}, denoted by $\pchainVP$, records preferences in the form of partial chains. A partial chain is a linear order of a subset of items, with no preference over the remaining items. The partial chains are a special case of the partially partitioned preferences, where each partition only consists of one item.

A \e{truncated voting profile}, denoted by $\trunVP$, records preferences in the form of truncated rankings.
Let $\ranking^{(t, b)}$ denote a truncated ranking with $t$ items at the top and $b$ items at the bottom, with no preference specified for the middle part of the ranking.
The $\ranking^{(t, b)}$ is a special case of the partitioned preferences, with $t+b+1$ partitions.

\paragraph{Combined voting profiles.}

Most research works have assumed that a partial voting profile represents a uniform distribution over its completions. However, the assumption that all completions are equally likely may not hold in practice.
We propose \e{combined voting profiles} $\VP^{\model+P}$, where each voter is associated with both her original incomplete preferences $P$ and a ranking model $\model$.  The ranking model $\model$ is her prior preference distribution, which may be obtained from historical data.
But after observing new evidence $P$, the probabilities of rankings that violate $P$ collapse to zero, while the remaining rankings that satisfy $P$ have the same relative probabilities among each other.
Formally speaking, the preferences of this voter become the posterior distribution of $\model$ conditioned on $P$. 

In another sense, the combined voting profiles are also the most general form of voting profiles (same as $\probaVP$) that unify all voting profiles so far.
For example, a partitioned voting profile $\fullparVP$ is essentially $\VP^{\text{U+FP}}$, and a RIM voting profile $\rimVP$ is essentially $\VP^{\RIM+\emptyset}$ where $\emptyset$ means that the partial orders are the empty ones.

Among the tractable cases in Section~\ref{sec:algorithms}, there are also combined voting profiles, which means that applying combined voting profiles both brings the benefit of more customized ranking distributions, and can also be practical.

\section{\mEw}
\label{sec:mew}

For a general voting profile and a positional scoring rule, the performance of a candidate can be quantified by the expectation of her score in a random possible world.  We define this formally below.

\begin{definition} [\mew] \label{def:mew}
  Given a general voting profile $\probaVP$ and a positional scoring rule $\vr_m$, candidate $w$ is a \mEw, \ifff, $\mathds{E}(\score(w, \probaVP)) = \max_{c \in C} \mathds{E}(\score(c, \probaVP))$.
\end{definition}

We denote the set of \mEws by $\mew(\vr_m, \probaVP)$.

\subsection{Alternative Interpretations}
\label{sec:mew:alternative_interpretations}

To gain an intuition for \mEw~(\mew), we will now give two winner definitions that are equivalent to \mew. Proofs can be found in Appendix~\ref{sec:appendix:proofs}.

\paragraph{Least Expected Regret Winner.}

The MEW can also be regarded as the candidate who minimizes the expected regret in a random possible world.
Let $\txt{Regret}(w, \completeVP)$ denote the regret value of choosing $w \in C$ as the winner given a complete voting profile $\completeVP$.
\[
\txt{Regret}(w, \completeVP) = \max_{c \in C} \score(c, \completeVP) - \score(w, \completeVP)
\]
Accordingly, the regret value $\txt{Regret}(w, \probaVP)$ over a probabilistic voting profile becomes a random variable.
\[
\mathds{E}(\txt{Regret}(w, \probaVP)) = \sum_{\completeVP \in \Omega(\probaVP)} \txt{Regret}(c, \completeVP) \cdot \Pr(\completeVP \mid \probaVP)
\]

Candidate $w$ is a \e{Least Expected Regret Winner}, \ifff, she minimizes the expected regret $\mathds{E}(\txt{Regret}(w, \probaVP))$.

\def\theoremLeastExpectedRegretWinner{
  Least Expected Regret Winner is equivalent to \mew.
}

\begin{theorem}
  \label{theorem:least_expected_regret_winner}
  \theoremLeastExpectedRegretWinner
\end{theorem}

\revv{Our use of expected regret as an alternative interpretation of MEW is inspired by Lou and Boutillier~\cite{DBLP:conf/ijcai/LuB11a}, who were the first to use regret in winner determination.  They proposed  MMR, the winner that minimized regret in the worst-case completion.  In contrast, MEW minimizes regret in expectation, across all completions.}

\paragraph{Meta-Election Winner.}
Recall that a voting profile $\probaVP$ represents a probability distribution of possible worlds $\Omega(\probaVP) = \set{\completeVP_1, \ldots, \completeVP_\numPW}$.
The Meta-Election Winner is defined as the candidate who wins a meta election with a large meta profile $\completeVP_M = (\completeVP_1, \ldots, \completeVP_\numPW)$ where rankings in $\completeVP_i$ are weighted by $\Pr(\completeVP_i \mid \probaVP)$.

\def\theoremGiganticElectionWinner{
  Meta-Election Winner is equivalent to \mew.
}

\begin{theorem}
  \label{theorem:gigantic_election_winner}
  \theoremGiganticElectionWinner
\end{theorem}

\subsection{Problem Statement}
\label{sec:problem}

The \mew is determined based on the expected performance of the candidates. \rev{Thus, the winner determination problem of \mew can be reduced to the problem of \ESC (\esc), stated below and addressed in the remainder of the paper.}

\begin{definition} [\esc] \label{def:esc}
  Given a general voting profile $\probaVP$, a positional scoring rule $\vr_m$, and a candidate $c \in C$, compute $\mathds{E}(\score(c, \probaVP))$ the expected score of the candidate $c$.
\end{definition}

\section{Hardness of \esc}
\label{sec:complexity}

This section investigates the complexity of the \esc problem. We first prove the hardness of two closely related problems, the Fixed-rank Counting Problem and the Rank Estimation Problem, and then demonstrate that the \esc problem is hard as well. Proofs can be found in Appendix~\ref{sec:appendix:proofs}.

\subsection{Fixed-rank Counting Problem}

Counting the number of linear extensions of a partial order is well-known to be \#P-complete~\cite{DBLP:conf/stoc/BrightwellW91}.
The Fixed-rank Counting Problem (FCP) is interested in the number of linear extensions where an item is placed at a specific rank.
Let $\Omega(\partialOrder)$ denote the set of linear extensions of $\partialOrder$, and $N(c {\rightarrow} j \mid \partialOrder)$ denote the number of linear extensions in $\Omega(\partialOrder)$ where item $c$ is placed at rank $j$.

\begin{definition}[FCP]
    Given a partial order $\partialOrder$ over $m$ items, an item $c$, and an integer $j \in [1, m]$, calculate $N(c {\rightarrow} j \mid \partialOrder)$, the number of linear extensions of $\partialOrder$ where item $c$ is placed at rank $j$.
\end{definition}

De Loof's doctoral dissertation~\cite{Loof2009EfficientCO} discusses this problem (Section 4.2.1), proposing exact and approximate algorithms, but does not provide proof of complexity.

\def\theoremFCPhardnessOverPartialOrders{
    The FCP is \#P-complete.
}

\begin{theorem} \label{theorem:FCP_shaPcomplete_over_partialOrders}
    \theoremFCPhardnessOverPartialOrders
\end{theorem}

The hardness of FCP facilitates the hardness proofs for the Rank Estimation Problem.

\subsection{Rank Estimation Problem}

Now we move on to the Rank Estimation Problem (REP).
This problem can be regarded as the probabilistic version of the FCP.
It calculates the probability that a given item is placed at a specific rank.
But the REP is generalized from partial orders to arbitrary ranking distributions.

\begin{definition}[REP]
    Given a ranking model $\model$ over $m$ items, an item $c$, and an integer $j \in [1, m]$, calculate $\Pr(c {\rightarrow} j \mid \model)$, the probability that item $c$ is placed at rank $j$, by $\model$.
\end{definition}

For the convenience of discussions in the rest of this Section, we also define two special cases of the REP where items are placed at the top and bottom of the linear extensions.

\begin{definition}[REP-$t$]
    Given a ranking model $\model$ of $m$ items and an item $c$, calculate $\Pr(c {\rightarrow} 1 \mid \model)$, the probability that item $c$ is placed at the top of a linear extension, by $\model$.
\end{definition}

\begin{definition}[REP-$b$]
	Given a ranking model $\model$ of $m$ items and an item $c$, calculate $\Pr(c {\rightarrow} m \mid \model)$, the probability that item $c$ is placed at the bottom of a linear extension, by $\model$.
\end{definition}

Lerche and S{\o}rensen~\cite{lerche2003evaluation} proposed an approximation for the REP over partial orders, but did not provide a formal complexity proof.
Bruggemann and Annoni~\cite{bruggemann2014average} and De Loof \etal \cite{de2011approximation} considered a related problem, calculating the expected rank of an item in the linear extensions of a partial order.
These works focused on approximation, lacking complexity proofs as well.

With the help of Theorem~\ref{theorem:FCP_shaPcomplete_over_partialOrders}, it turns out that the REP and its two variants are all FP$^{\#P}$-complete over partial orders.

\def\lemmaREPtHardnessOverPartialOrders{
    If ranking model $\model$ is a partial order $\partialOrder$ of $m$ items representing a uniform distribution of $\Omega(\partialOrder)$, the REP-t is FP$^{\#P}$-complete.
}

\begin{lemma} \label{lemma:REPt_shaPcomplete_over_partialOrders}
    \lemmaREPtHardnessOverPartialOrders
\end{lemma}

\def\lemmaREPbHardnessOverPartialOrders{
    If ranking model $\model$ is a partial order $\partialOrder$ of $m$ items representing a uniform distribution of $\Omega(\partialOrder)$, the REP-b is FP$^{\#P}$-complete.
}

\begin{lemma} \label{lemma:REPb_shaPcomplete_over_partialOrders}
	\lemmaREPbHardnessOverPartialOrders
\end{lemma}

\def\theoremRepHardnessOverPartialOrders{
    If ranking model $\model$ is a partial order $\partialOrder$ of $m$ items representing a uniform distribution of $\Omega(\partialOrder)$, the REP is FP$^{\#P}$-complete.
}

\begin{theorem}
    \label{theorem:REP_shaPcomplete_over_partialOrders}
    \theoremRepHardnessOverPartialOrders
\end{theorem}

\subsection{Complexity of \esc}

\esc is closely related to REP.  Firstly,  \esc  is no harder than REP over general voting profiles (Theorem~\ref{theorem:reduction}), which lays the foundation for the identification of tractable cases in Section~\ref{sec:algorithms}.

\def\theoremReductionEscToRep{
    Given a general voting profile $\probaVP$ and a positional scoring rule $\vr_m$, the \esc problem can be reduced to the REP.
}

\begin{theorem} \label{theorem:reduction}
    \theoremReductionEscToRep
\end{theorem}

\def\theoremRepMewEquivalentUnderKapproval{
    The REP for rank $k$ is equivalent to the \esc problem over either one or both of the $(k-1)$-approval and $k$-approval rules.
}

\begin{theorem} \label{theorem:REP_MEW_equivalent_under_k_approval}
    \theoremRepMewEquivalentUnderKapproval
\end{theorem}

Theorem~\ref{theorem:REP_MEW_equivalent_under_k_approval} provides an insight into the relation between REP and \esc in terms of computational complexity.
If a solver is available for the \esc problem over a collection of $k$-approval votes, this solver is computationally equivalent to the REP solver.
Note that Theorem~\ref{theorem:REP_MEW_equivalent_under_k_approval} is not limited to partial voting profiles.

\def\theoremHardnessForEscGivenPartialOrderAndPlurality{
    Given a partial voting profile $\partialVP$, a distinguished candidate $c$, and plurality rule $\vr_m$, the \esc problem of calculating $\mathds{E}(\score(c \mid \partialVP, \vr_m))$ is FP$^{\#P}$-complete.
}

\begin{theorem} \label{theorem:shaPcomplete_for_expected_score_given_partialOrder_and_plurality}
    \theoremHardnessForEscGivenPartialOrderAndPlurality
\end{theorem}

\def\theoremHardnessForEscGivenPartialOrderAndVeto{
    Given a partial voting profile $\partialVP$, a distinguished candidate $c$, and veto rule $\vr_m$, the \esc problem of calculating $\mathds{E}(\score(c \mid \partialVP, \vr_m))$ is FP$^{\#P}$-complete.
}

\begin{theorem} \label{theorem:shaPcomplete_for_expected_score_given_partialOrder_and_veto}
	\theoremHardnessForEscGivenPartialOrderAndVeto
\end{theorem}

\def\theoremHardnessForExpectedScoreGivenPartialOrderAndKApproval{
    Given a partial voting profile $\partialVP$, a distinguished candidate $c$, and $k$-approval rule $\vr_m$, the \esc problem of calculating $\mathds{E}(\score(c \mid \partialVP, \vr_m))$ is FP$^{\#P}$-complete.
}

\begin{theorem} \label{theorem:shaPcomplete_for_expected_score_given_partialOrder_and_k_approval}
    \theoremHardnessForExpectedScoreGivenPartialOrderAndKApproval
\end{theorem}

The three theorems above demonstrate the hardness of the \esc problem over partial voting profiles, under plurality, veto, and $k$-approval, respectively.
In particular, \esc is FP$^{\#P}$-complete even under plurality (Theorem~\ref{theorem:shaPcomplete_for_expected_score_given_partialOrder_and_plurality}).

\section{Most Expected Winner Solvers}
\label{sec:algorithms}

The problem of determining \mew can be reduced to the \esc problem (Definition~\ref{def:esc}), then further reduced to the REP by Theorem~\ref{theorem:reduction}.
This section will solve the \mew problem by solving the REP.
Proofs can be found in Appendix~\ref{sec:appendix:proofs}.

\subsection{Solver for RIM-based partial orders}
\label{sec:algorithms:exact}

We adopt the RIMDP algorithm~\cite{DBLP:conf/aaai/KenigIPKS18}, discussed in Section~\ref{sec:preliminaries:pref}, to develop an exact solver for the REP over posets, and more generally, for RIM combined with posets. The exact solver only modifies the RIMDP by always tracking the target candidate, such that when RIM insertions finish, RIMDP obtains the probabilities of the target candidate being placed at each rank.

The RIMDP has exponential complexity $O(m^{\text{cw}(\bsigma, \partialOrder)+2})$, where $\text{cw}(\bsigma, \partialOrder)$ is the cover width, given the reference ranking $\bsigma$ and the partial order $\partialOrder$ ~\cite{DBLP:conf/aaai/KenigIPKS18}. It takes $O(nm^{\text{cw}(\bsigma, \partialOrder)+3})$ to calculate \mew by executing RIMDP for all $m$ candidates and $n$ voters. 

\subsection{Solvers for Partitioned Preferences, Truncated Rankings,  and Partial Chains} 

It turns out that the \mew problem can be solved efficiently for all special cases of partial voting profiles in Figure~\ref{fig:venn_observation}, \ie the partitioned voting profile, 

\def\theoremTractabilityOfFP{
  Given a positional scoring rule $\vr_m$, a fully partitioned voting profile $\fullparVP {=} (\enum[n]{\fullpar})$, and candidate $w$, determining $w \in \mew(\vr_m, \fullparVP)$ is in $O(nm^2)$.
}

\begin{theorem} \label{theorem:tractability_of_fullparVP}
     \theoremTractabilityOfFP
\end{theorem}

The truncated voting profiles can be solved by the same approach above, since they are a special case of the fully partitioned profile.

\def\theoremTractabilityOfPC{
  Given a positional scoring rule $\vr_m$, a partial chain voting profile $\pchainVP = (\enum[n]{\pchain})$, and candidate $w$, determining $w \in \mew(\vr_m, \pchainVP)$ is in $O(nm^2)$.
}

\begin{theorem} \label{theorem:tractability_of_pchainVP}
    \theoremTractabilityOfPC
\end{theorem}

The \mew over partially partitioned preferences can be solved by extending the above approach for partial chain voting profiles.

\def\theoremTractabilityOfPP{
  Given a positional scoring rule $\vr_m$, a partially partitioned voting profile $\parparVP = (\enum[n]{\parpar})$, and candidate $w$, determining $w \in \mew(\vr_m, \parparVP)$ is in $O(nm^2)$.
}

\begin{theorem} \label{theorem:tractability_of_parparVP}
  \theoremTractabilityOfPP
\end{theorem}

\begin{algorithm}[tb!]
\small 
  \raggedright
  \caption{REP solver for RIM}
  \label{alg:rim_rank_estimation}
  \textbf{Input}: Item $c$, $\RIM(\bsigma, \Pi)$ where $|\bsigma| = m$ \\
  \textbf{Output}: $\set{k \rightarrow \Pr(c {\rightarrow} k \mid \bsigma, \Pi) \mid k \in [1, m]}$
  
  \begin{algorithmic}[1] 
    \STATE $\delta_0 \defeq \emptyset$, $\mathcal{P}_0 \defeq \set{\delta_0}$ and $q_0(\delta_0) \defeq 1$
    \FOR {$i=1, \ldots, m$}
    \STATE $\mathcal{P}_i \defeq \set{}$
    \FOR {$\delta \in \mathcal{P}_{i-1}$}
    \FOR {$j=1, \ldots, i$} \label{alg:rim_rank_estimation:j}
    \IF {$\bsigma(i)$ is $c$}
    \STATE $\delta' \defeq \set{c \rightarrow j}$
    \ELSIF {$\delta = \set{c \rightarrow k}$ and $j \leq k$}
    \STATE $\delta' \defeq \set{c \rightarrow k + 1}$
    \ELSE
    \STATE $\delta' \defeq \delta$
    \ENDIF
    \IF {$\delta' \notin \mathcal{P}_i$}
    \STATE $\mathcal{P}_i.add(\delta')$
    \STATE $q_i(\delta') \defeq 0$
    \ENDIF
    \STATE $q_i(\delta') \pluseq q_{i-1}(\delta) \cdot \Pi(i, j)$
    \ENDFOR
    \ENDFOR
    \ENDFOR
    \STATE $\forall k \in [1, m], \Pr(c {\rightarrow} k \mid \bsigma, \Pi) = q_m(\set{c \rightarrow k})$.
    \RETURN $\set{k \rightarrow \Pr(c {\rightarrow} k \mid \bsigma, \Pi) \mid k \in [1, m]}$
  \end{algorithmic}
\end{algorithm}

\subsection{Solver for Probabilistic Voting Profiles}

While the problem of \mew determination is hard in general, it is tractable over specific ranking models such as the Mallows~\cite{Mallows1957} and its generalizations RIM~\cite{Doignon2004} and RSM~\cite{DBLP:journals/tdasci/ChakrabortyDKKR21}.

\def\theoremTractabilityOfRIM{
  Given positional scoring rule $\vr_m$, a RIM voting profile $\rimVP = (\enum[n]{\RIM})$, and candidate $w$, determining $w \in \mew(\vr_m, \rimVP)$ is in $O(nm^4)$.
}

\begin{theorem} \label{theorem:tractability_of_rimVP}
    \theoremTractabilityOfRIM
\end{theorem}

The complexity of \mew determination over RSM voting profiles is $O(nm^4)$ as well.
Proof can be found in Appendix~\ref{sec:appendix:proofs}.

\subsection{Solver for Combined Voting Profiles}

It is usually harder to compute the expected scores over combined voting profiles.
Below are some cases where this problem is tractable.
The first case is the RIMs combined with truncated rankings.

\def\theoremTractabilityOfRimTrun{
  Given a positional scoring rule $\vr_m$, a voting profile $\rimTrunVP {=} \big( (\RIM_1, \ranking_1^{(t_1, b_1)}), \ldots, \\(\RIM_n, \ranking_n^{(t_n, b_n)}) \big)$, and candidate $w$, determining $w \in \mew(\vr_m, \rimTrunVP)$ is in $O(nm^4)$.
}

\begin{theorem} \label{theorem:tractability_of_rimTrunVP}
    \theoremTractabilityOfRimTrun
\end{theorem}

Mallows combined with partitioned preferences is also tractable.

\def\theoremTractabilityOfMallowsFP{
  Given a positional scoring rule $\vr_m$, a voting profile $\mallowsPartitionVP=\big( (\mallows_1, \fullpar_1), \ldots, \\(\mallows_n, \fullpar_n) \big)$, and candidate $w$, determining $w \in \mew(\vr_m, \mallowsPartitionVP)$ is in $O(nm^4)$.
}

\begin{theorem} \label{theorem:tractability_of_mallowsPartitionVP}
    \theoremTractabilityOfMallowsFP
\end{theorem}

\begin{table}[tb!]
  \caption{
    Complexity of exact \mew solvers under general positional scoring rules for various voting profiles, including partial orders (PO), partially partitioned preferences (PP), fully partitioned preferences (FP), partial chains (PC), RIMs, rRSMs, Mallows, and combined profiles. \rev{A dagger ($\dagger$) refers to results presented in Appendix~\ref{sec:appendix:rsm_profile}.}
    An asterisk (*) means that the RIM and rRSM models are effectively the Mallows models in the experimental benchmarks.
  }
\small 
  \begin{tabular}{@{}llll}
    \toprule
    Profile                           & Complexity            & Solver                                                                                   & Experiments                               \\ \midrule
    $\partialVP$                      & $O(nm^{\text{cw}+3})$ & RIMDP                                                                                    & Figures~\ref{fig:posets} and ~\ref{fig:cw} \\
    $\parparVP$                       & $O(nm^2)$             & Theorem~\ref{theorem:tractability_of_parparVP}                                           & Figure~\ref{fig:pp}                       \\
    $\fullparVP$                      & $O(nm^2)$             & Theorem~\ref{theorem:tractability_of_fullparVP}                                          & Figure~\ref{fig:fp}                       \\
    $\pchainVP$                       & $O(nm^2)$             & Theorem~\ref{theorem:tractability_of_pchainVP}                                           & Figure~\ref{fig:pc}                       \\
    \rev{$\trunVP$}                   & \rev{$O(nm^2)$}       & \rev{Theorem~\ref{theorem:tractability_of_fullparVP}, as $\fullparVP$}                   & Figure~\ref{fig:tr}                       \\ \midrule
    $\rimVP$                          & $O(nm^4)$             & Theorem~\ref{theorem:tractability_of_rimVP}                                              &                                           \\
    \rev{$\mallowsVP$}                & \rev{$O(nm^4)$}       & \rev{Theorem~\ref{theorem:tractability_of_rimVP}, as $\rimVP$}                           & Figure~\ref{fig:mallows}                  \\
    $\rsmVP$                          & $O(nm^4)$             & Theorem~\ref{theorem:tractability_of_rsmVP}$^\dagger$                                    & Figure~\ref{fig:rsm}*                     \\ \midrule
    $\rimPartialVP$                   & $O(nm^{\text{cw}+3})$ & RIMDP                                                                                    & Figure~\ref{fig:mallows|po}               \\
    \rev{$\VP^{\RIM \text{+PP}}$}     & $O(nm^{\text{cw}+3})$ & \rev{RIMDP, as $\rimPartialVP$}                                                          &                                           \\
    \rev{$\rimPChainVP$}              & $O(nm^{\text{cw}+3})$ & \rev{RIMDP, as $\rimPartialVP$}                                                          &                                           \\
    \rev{$\rimPartitionVP$}           & $O(nm^{\text{cw}+3})$ & \rev{RIMDP, as $\rimPartialVP$}                                                          &                                           \\
    $\rimTrunVP$                      & $O(nm^4)$             & Theorem~\ref{theorem:tractability_of_rimTrunVP}                                          & Figure~\ref{fig:mallows|tr}*              \\
    $\mallowsPartitionVP$             & $O(nm^4)$             & Theorem~\ref{theorem:tractability_of_mallowsPartitionVP}                                 & Figure~\ref{fig:mallows|fp}               \\
    \rev{$\VP^{\mallows \text{+TR}}$} & $O(nm^4)$             & \rev{Theorem~\ref{theorem:tractability_of_mallowsPartitionVP}, as $\mallowsPartitionVP$} &                                           \\
    \rev{$\VP^{\mallows \text{+PP}}$} & $O(nm^{\text{cw}+3})$ & \rev{RIMDP, as $\VP^{\RIM \text{+PP}}$}                                                  &                                           \\
    \rev{$\VP^{\mallows \text{+PC}}$} & $O(nm^{\text{cw}+3})$ & \rev{RIMDP, as $\rimPChainVP$}                                                           &                                           \\
    \rev{$\VP^{\mallows \text{+PO}}$} & $O(nm^{\text{cw}+3})$ & \rev{RIMDP, as $\rimPartialVP$}                                                          &                                           \\ \bottomrule
  \end{tabular}
  \centering
  \label{tab:complexity}
\end{table}

\subsection{Summary}
\label{sec:algorithms:summary}

\rev{Table~\ref{tab:complexity} summarizes all solvers in this Section, including additional conclusions for a large number of specialized voting profiles.}
For example, the complexity over Mallows voting profiles is bounded by $O(nm^4)$, since they are a special case of $\rimVP$.
An interesting observation is that the \mew complexity over $\rimPartitionVP$ is not tractable, but its special case $\mallowsPartitionVP$ has polynomial complexity.
Although this table demonstrates that evaluating \mew over probabilistic voting profiles has higher complexity than that over incomplete voting profiles (\eg $O(nm^4)$ for $\mallowsPartitionVP$ but only $O(nm^2)$ for $\fullparVP$), the tractability results over a collection of probabilistic and combined voting profiles give the \mew a computation advantage in practice.
\rev{Note that the combined voting profiles with RSM models are not covered by Table~\ref{tab:complexity}, since the complexity of determining the \mew for them is still an open question.}

\section{Optimization strategies}
\label{sec:optimize}

The algorithms in Section~\ref{sec:algorithms} find the MEW by computing candidate scores for a single voter at a time, and then aggregating the scores across the voting profile.
We now present \rev{two optimization strategies} for a voting profile solver in Algorithm~\ref{alg:pruning}.

\begin{algorithm}[tb!]
  \small 
  \raggedright
  \caption{Solver for voting profiles}
  \label{alg:pruning}
  \textbf{Input}: general voting profile $\probaVP=(\enum[n]{\model})$, voting rule $\vr_m$ \\
  \textbf{Output}: $\mew(\vr_m, \probaVP)$
  
  \begin{algorithmic}[1]
    \STATE \rev{Group identical votes in $\probaVP$, sort them by the group size in descending order, to generate a weighted voting profile $\VP_w$ where each vote $\model_i$ is associated with a weight $w_i$}
    \STATE Obtain candidates $C = \set{\enum{c}}$ from $\VP_w$
    \STATE $\forall c \in C, UB(c) {\defeq} 0, LB(c) {\defeq} 0$ \label{alg:pruning:bounds:1}
    \FOR {\rev{$(\model, w) \in \VP_w$}}
    \FOR {$c \in C$}
    \STATE Compute the best possible rank $a$ of $c$ over $\model$
    \STATE Compute the worst possible rank $b$ of $c$ over $\model$
    \STATE \rev{$UB(c) \pluseq w \cdot \vr_m(a)$, $LB(c) \pluseq w \cdot \vr_m(b)$} \label{alg:pruning:sum}
    \ENDFOR
    \ENDFOR \label{alg:pruning:bounds:2}
    \STATE Compute $max(LB)$, prune any candidate $c$ if $UB(c) < max(LB)$ \label{alg:pruning:prune:1}
    \FOR {\rev{$(\model, w) \in \VP_w$}} \label{alg:pruning:for:1}
    \STATE Let $C_t$ denote the candidates currently tracked by $UB$ and $LB$
    \FOR {$c \in C_t$}
    \STATE Compute the exact score of $c$ over $\model$
    \STATE Refine $UB(c)$ and $LB(c)$ with its exact score \rev{and the weight $w$}
    \ENDFOR
    \STATE Compute $max(LB)$, prune any candidate $c$ if $UB(c) < max(LB)$ \label{alg:pruning:prune:2}
    \IF{Only one candidate remains in $UB$ and $LB$}
    \STATE Let $c'$ denote the only remaining candidate
    \RETURN $\mew(\vr_m, \probaVP) = c'$ \label{alg:pruning:ret:1}
    \ENDIF
    \ENDFOR \label{alg:pruning:for:2}
    \STATE Let $C'$ denote the set of remaining candidates in $UB$ and $LB$
    \RETURN $\mew(\vr_m, \probaVP) = C'$
  \end{algorithmic}
\end{algorithm}

\paragraph{Candidate pruning}
We can save computation by iteratively pruning candidates who cannot be the MEW.
The key idea is to quickly compute score upper- and lower-bounds for all candidates across the voting profile,  progressively refine these scores during per-voter computation, and prune candidates who cannot be the MEW based on the upper- and lower-bounds computed so far.

Algorithm~\ref{alg:pruning} presents the candidate pruning procedure. The first step is to compute $UB(c)$ and $LB(c)$, the upper- and lower-bounds of the score of each candidate $c$ by quickly computing their best possible and worst possible ranks assigned by each voter (lines~\ref{alg:pruning:bounds:1} - \ref{alg:pruning:bounds:2}).  For example, in a given partial order $M$ (voter) in a partial voting profile, the best possible rank of candidate $c$ is bounded by the number of its ancestors in $M$, while its worst possible rank is bounded by the number of its descendants.  Combing this information with the voting rule, we can efficiently obtain score upper- and lower-bounds of each candidate for each voter.

\begin{example}
  Let $\partialOrder=\set{\sigma_3 \succ \sigma_5, \sigma_3 \succ \sigma_8}$ be a partial order, over $m=10$ items, and assume the voting rule is 2-approval.
  Item $\sigma_5$ has 1 ancestor $\sigma_3$ and no descendant in $\partialOrder$.
  So its best possible rank is 2 and its worst possible rank is $m$, which correspond to scores 1 and 0, respectively, under 2-approval.  Thus, 1 is the score upper-bound of $\sigma_5$, and 0 the score lower-bound, for $\partialOrder$.
\end{example}

Per-voter score upper-bounds and lower-bounds can be aggregated (summed up) across all voters in the profile (line~\ref{alg:pruning:sum} of Algorithm~\ref{alg:pruning}).  The pruning strategy can then confidently determine the losing candidates as those whose score upper-bounds are lower than the best observed score lower-bound (line~\ref{alg:pruning:prune:1} of Algorithm~\ref{alg:pruning}).

Score upper- and lower-bounds are refined as the algorithm iterates over the voters (lines~\ref{alg:pruning:for:1}-~\ref{alg:pruning:for:2}), computing the exact scores of viable (tracked) candidates, refining $UB(c)$ and $LB(c)$, and pruning any candidates that are no longer viable (line~\ref{alg:pruning:prune:2}).

If, at any point during the iteration, only a single tracked candidate $w$ remains, then it is returned as the $\mew$ (line~\ref{alg:pruning:ret:1}).  Once all voters $M$ have been iterated over, return all remaining tracked candidates as the winners.  (Note that if after all voters are iterated, and there are still multiple tracked items, then their upper- and lower-bounds are precisely their exact scores, and they all share the same exact score, otherwise the candidates with lower exact scores would have been pruned during voter iteration.)

\begin{example}
  Assume $UB=\set{\sigma_1 \rightarrow 15, \sigma_2 \rightarrow 6}$ and $LB = \set{\sigma_1 \rightarrow 5, \sigma_2 \rightarrow 2}$ at a time point during refining the $UB$ and $LB$.
  Assume that after computing the exact scores of $\sigma_1$ and $\sigma_2$ over the latest voter, the refined $UB=\set{\sigma_1 \rightarrow 12, \sigma_2 \rightarrow 5}$ and the refined $LB = \set{\sigma_1 \rightarrow 7, \sigma_2 \rightarrow 3}$.
  Now, candidate $\sigma_2$ can be pruned, since $UB(\sigma_2) < LB(\sigma_1)$.
  After pruning $\sigma_2$, only $\sigma_1$ is tracked by $UB$ and $LB$, thus we can confidently declare $\sigma_1$ as the winner.
\end{example}

\paragraph{\rev{Voter grouping}}
\rev{As another optimization, we observe that when multiple voters cast the same vote, we only need to compute the expected candidate scores for this vote once.  When computing the \mew over a voting profile, the first step is to group identical votes and sort them by their frequencies in descending order, which generates a new voting profile that is weighted and sorted.}

\section{Experiments}
\label{sec:exp}

We systematically tested the proposed algorithms and optimization strategies with both synthetic and real datasets, described below. All experiments were conducted on a Linux machine with two Intel Xeon Platinum 8268 24-core 2.9GHz Processors and 384GB RAM.  All code is available at \url{https://github.com/DataResponsibly/mew-supplementals}. 

\paragraph{Synthetic data generators}

The synthetic partial orders are generated by RSMs.
Let $\mathsf{RSM}(\bsigma, \Pi, p)$ denote the preference model of a voter where $\bsigma = \angs{\enum{\sigma}}$ and $\Pi(i, j) = \frac{\phi^{j-1}}{1+\phi+...+\phi^{m-i}}$, which makes its item selection order equivalent to $\mallows(\bsigma, \phi)$.
Then uniformly draw probabilities $p: \set{1, \ldots, m-1} \rightarrow \text{Uniform}(0, p_{max})$ where $p_{max}$ is part of the parameter setup in our experiments.

Partially partitioned preferences are generated with specified numbers of partitions. First, create an ordered list of partitions, and randomly choose an item for each partition to guarantee that each partition has at least one item. Then, create an additional partition for the missing items. Finally, assign each of the remaining items to a random partition, including the ``missing items'' partition.  Fully partitioned preferences are generated in a similar fashion, except that there is no partition for the missing items. Partial chains also use this generator, except that it stops at the first step where each partition has been assigned a single random item.

Truncated rankings are generated by first drawing a ranking from the uniform ranking distribution, and keeping only the top and the bottom parts of the ranking.

\paragraph{Real datasets} We worked with three real datasets with different voting profile types.
\e{CrowdRank}~\cite{DBLP:conf/webdb/StoyanovichJG15} is a dataset of partial chains over movies collected from Amazon Mechanical Turk.
It consists of 50 human intelligence tasks (HITs), each of which asked 100 users to rank 20 movies that they were familiar with.
\e{MovieLens} is a dataset of movie ratings collected by GroupLens (\url{www.grouplens.org}).
We used the most frequently rated movies and converted movie ratings by the same user into partially partitioned preferences. We then obtained a partially partitioned voting profile with 200 movies and 6040 users.
\e{Travel}~\cite{8635080} is a dataset of ratings of European attractions that belong to 24 categories in Google Reviews. The ratings are average values given by each user for each attraction category.
We converted these average ratings into a fully partitioned voting profile with 24 attraction categories and 5456 users.

\subsection{Performance optimizations}
\label{sec:exp:pruning}

\rev{We tested the performance of optimization strategies over synthetic partial voting profiles, with 10 candidates under three voting rules.
The partial orders are generated by $\mathsf{RSM}(\bsigma, \Pi, p)$ and $p_{max}$ as described above.
In each voting profile, all voters share the same reference ranking $\bsigma$ and insertion probabilities $\Pi$, and their item selection orders are equivalent to $\mallows(\bsigma, \phi)$.
For the convenience of discussion, we use $\phi$ as the setting that describes $\Pi$.}

\rev{\paragraph{Candidate pruning}
We turned off voter grouping to test the performance of candidate pruning.
For each parameter setting, we generated 10 voting profiles, fixing 10,000 voters and $\phi = 0.5$, varying $p_{max}$, and computed \mew with and without candidate pruning.
We present speedup as a function of $p_{max}$ for $10,000$ voters in Figure~\ref{fig:strategies:pruning}, for plurality, 2-approval and Borda.  We observe that pruning never hurts performance, and that it is more effective for higher values of $p_{max}$, the parameter that controls the density (number of preference pairs) in a partial order.  Pruning is most effective for plurality, followed by 2-approval and then by Borda.}     
\rev{We thus conclude that these performance optimizations are effective at reducing the running time of \mew computation, and we use both optimizations in all remaining experiments.}

\begin{figure}[tb!]
	\centering
	\subfloat[candidate pruning]{
		\label{fig:strategies:pruning}
		\includegraphics[width=0.3\textwidth]{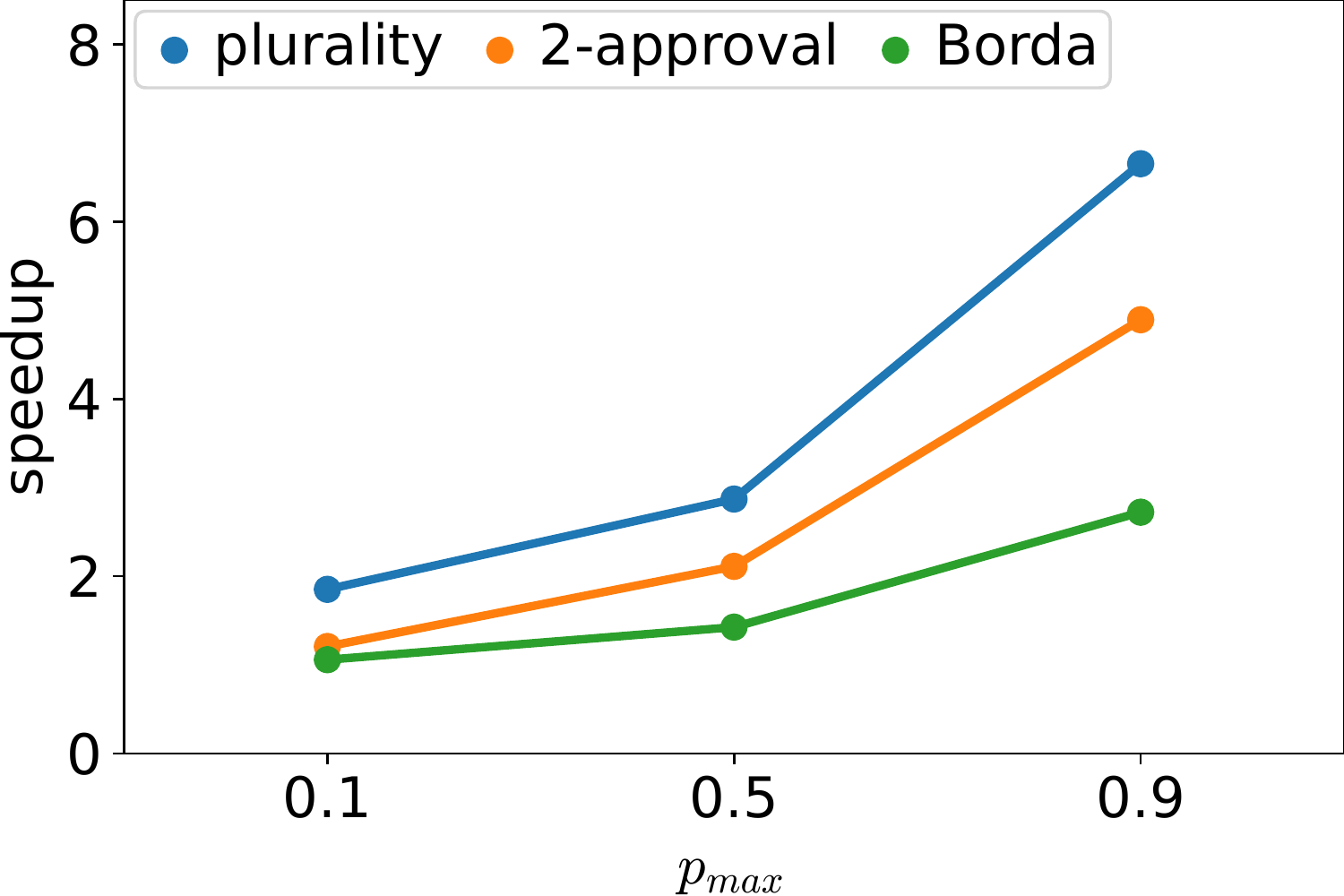}
	}\hspace{5em}
	\subfloat[voter grouping]{
		\label{fig:strategies:grouping}
		\includegraphics[width=0.3\textwidth]{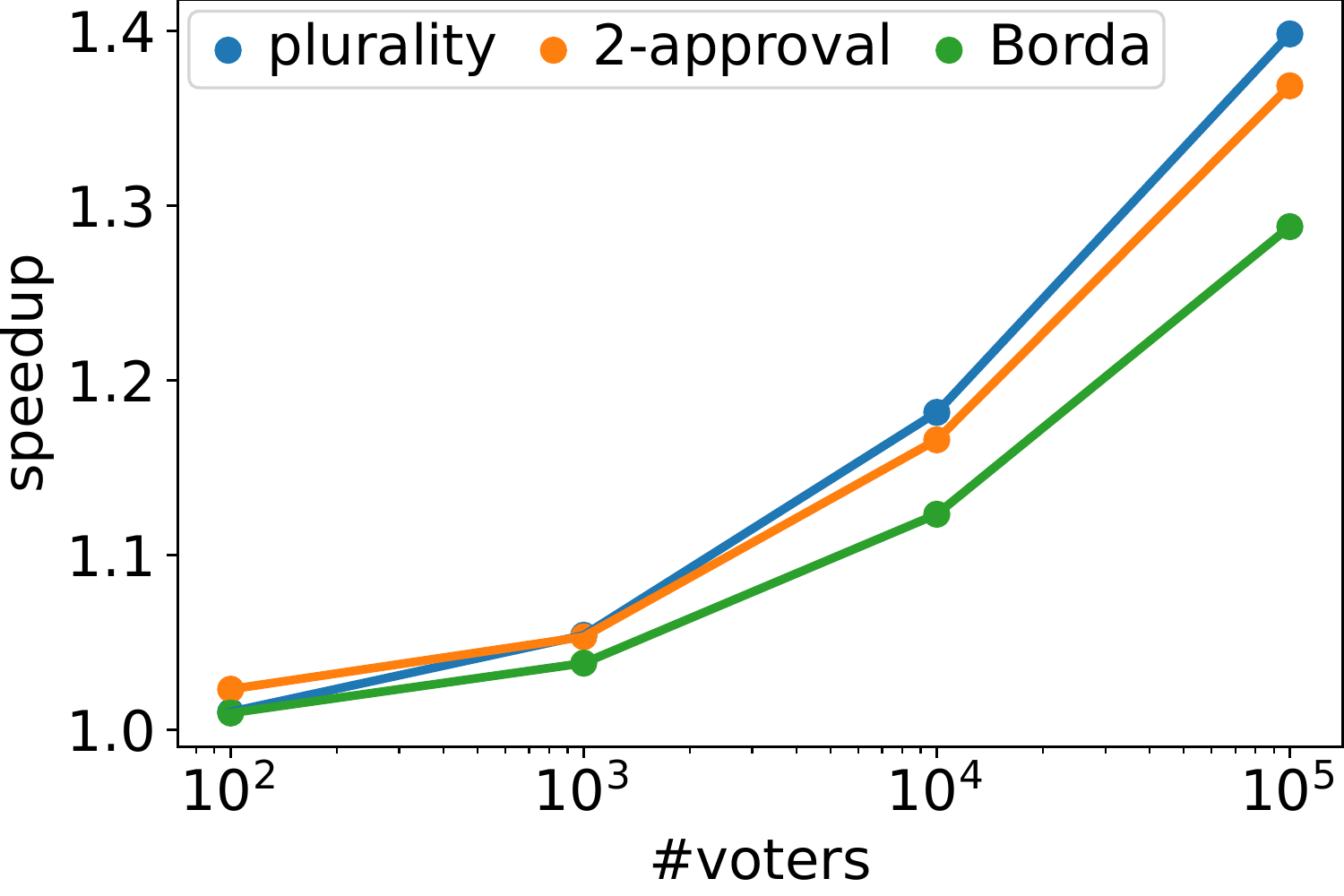}
	}
	\caption{\rev{Performance of optimization strategies over partial voting profiles, with 10 cand., $\phi=0.5$. Candidate pruning with 10,000 voters;  effectiveness improves with $p_{max}$. Voter grouping with $p_{max}=0.1$; effectiveness improves with \#voters.}}
	\label{fig:strategies}
\end{figure}

\rev{\paragraph{Voter grouping}
In this experiment, we turned off candidate pruning, and tested the voter grouping optimization.
For each parameter setting, we generated 10 voting profiles, fixing $\phi = 0.5$ and $p_{max} = 0.1$, varying the number of voters, and computed \mew with and without voter grouping. 
The speedup of voter grouping is the running time without voter grouping divided by the running time with voter grouping. Figure~\ref{fig:strategies:grouping} demonstrates that voter grouping never hurts performance, and that the speedup realized by this optimization increases with the number of voters.}

\begin{figure}[b!]
	\centering
	\subfloat[running time]{
		\label{fig:parallel:time}
		\includegraphics[width=0.3\textwidth]{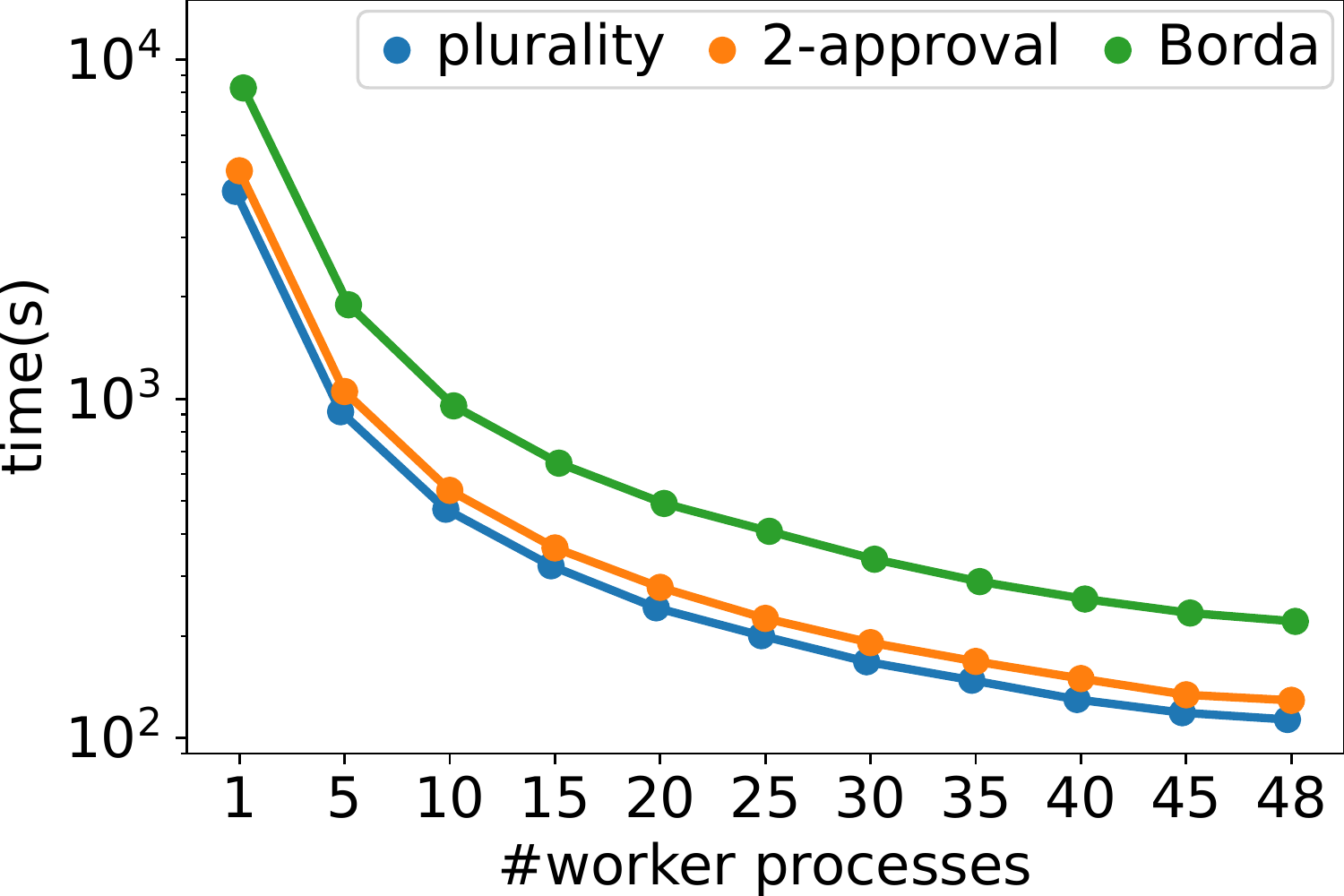}
	}\hspace{5em}
	\subfloat[speedup]{
		\label{fig:parallel:speedup}
		\includegraphics[width=0.3\textwidth]{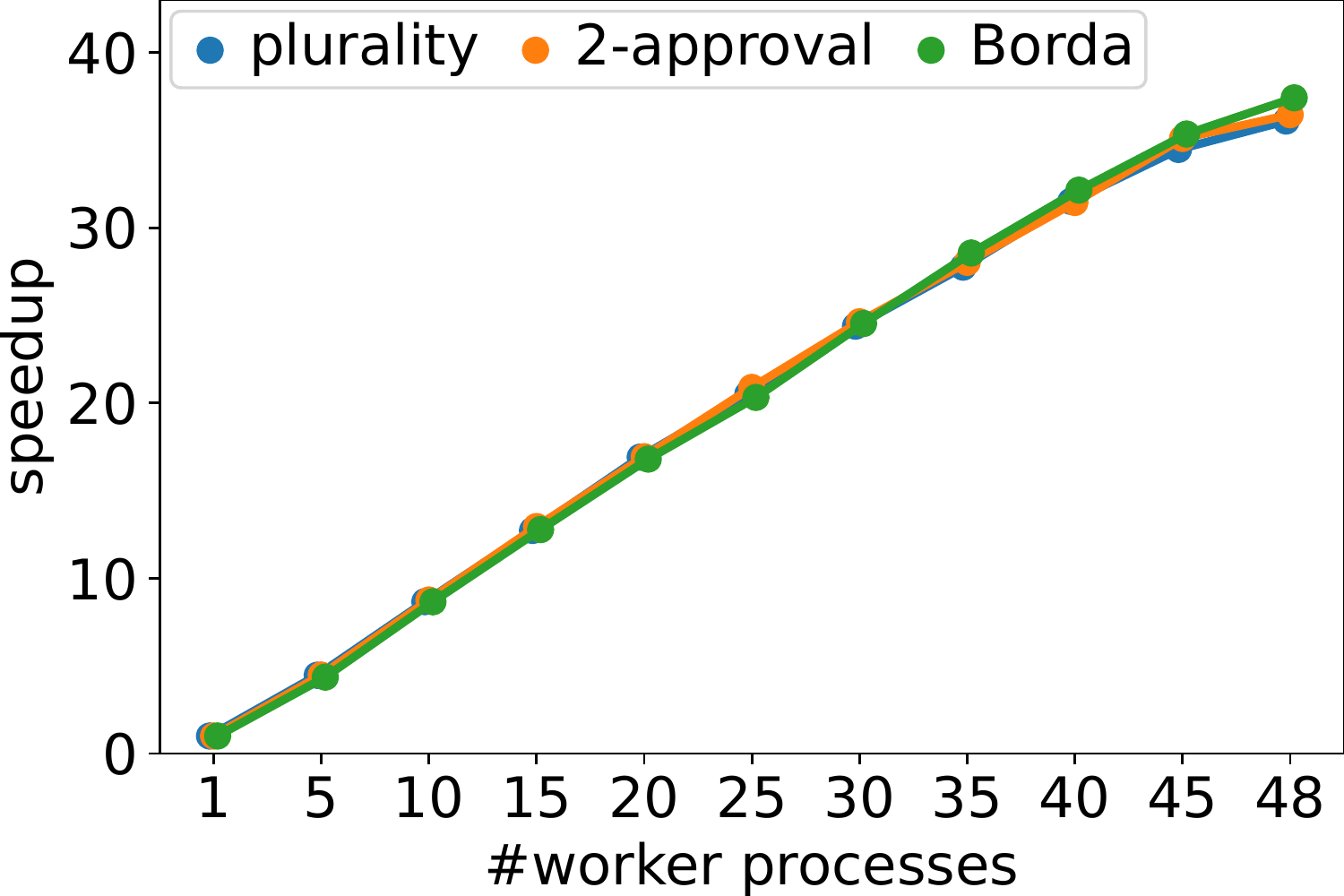}
	}~
	\caption{\rev{Parallel implementation for MEW over partial voting profiles of 10 candidates, 1 million voters, $\phi=0.5$, and $p_{max}=0.1$. The speedup due to parallelism is nearly linear in the number of worker processes.}}
	\label{fig:parallel}
\end{figure}

\paragraph{Parallel computation of \mew} \rev{Algorithm~\ref{alg:pruning} gives a sequential implementation of \mew, with linear complexity in the number of voters. To handle a large volume of voters in real-world settings, the computation of \mew can be easily parallelized by turning off the candidate pruning strategy and allocating all votes to different CPU cores.}
\rev{We tested the parallel implementation of \mew over 10 synthetic partial voting profiles of 10 candidates and 1 million voters.
The partial orders are generated with $\phi = 0.5$ and $p_{max} = 0.1$, using the same methodology as in Section~\ref{sec:exp:pruning}.
Figure~\ref{fig:parallel:time} demonstrates that the solver runs faster with more worker processes.
In Figure~\ref{fig:parallel:speedup}, the speedup of the parallel solver relative to the sequential solver increases linearly with the number of worker processes.}

\subsection{Incomplete voting profiles}

\paragraph{Partial voting profiles}
A collection of synthetic profiles of 10 candidates are generated to test the scalability of the solver in Section~\ref{sec:algorithms:exact}.
Synthetic partial orders are generated in the same way as in Section~\ref{sec:exp:pruning}.  We generated 10 voting profiles for each parameter setting, and calculated the \mew under three voting rules, with performance optimizations discussed in Section~\ref{sec:optimize}.

\begin{figure}[tb!]
	\centering
	\subfloat[1000 voters]{
		\label{fig:posets:phi}
		\includegraphics[width=0.3\textwidth]{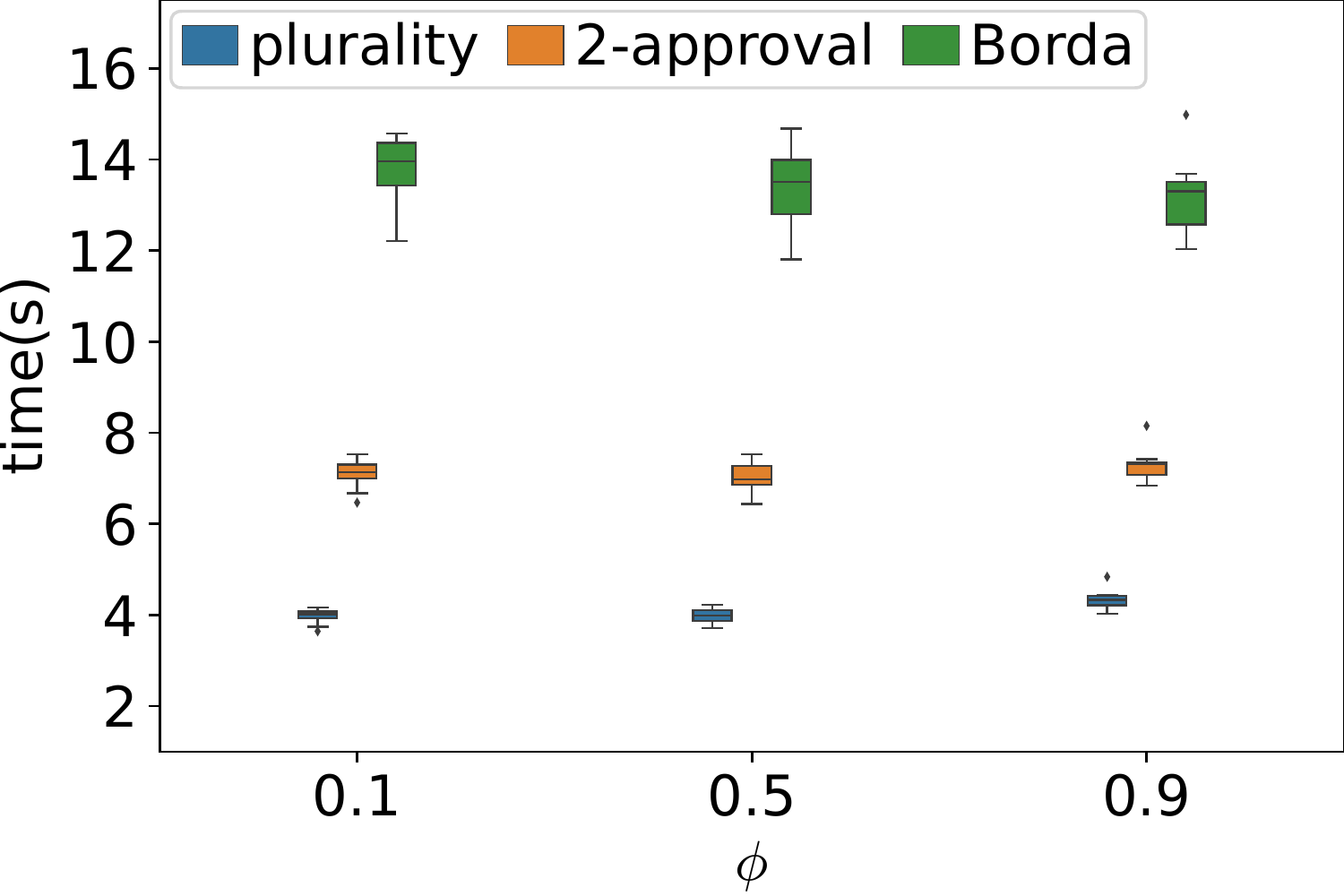}
	}\hspace{5em}
	\subfloat[$\phi = 0.5$]{
		\label{fig:posets:voters}
		\includegraphics[width=0.3\textwidth]{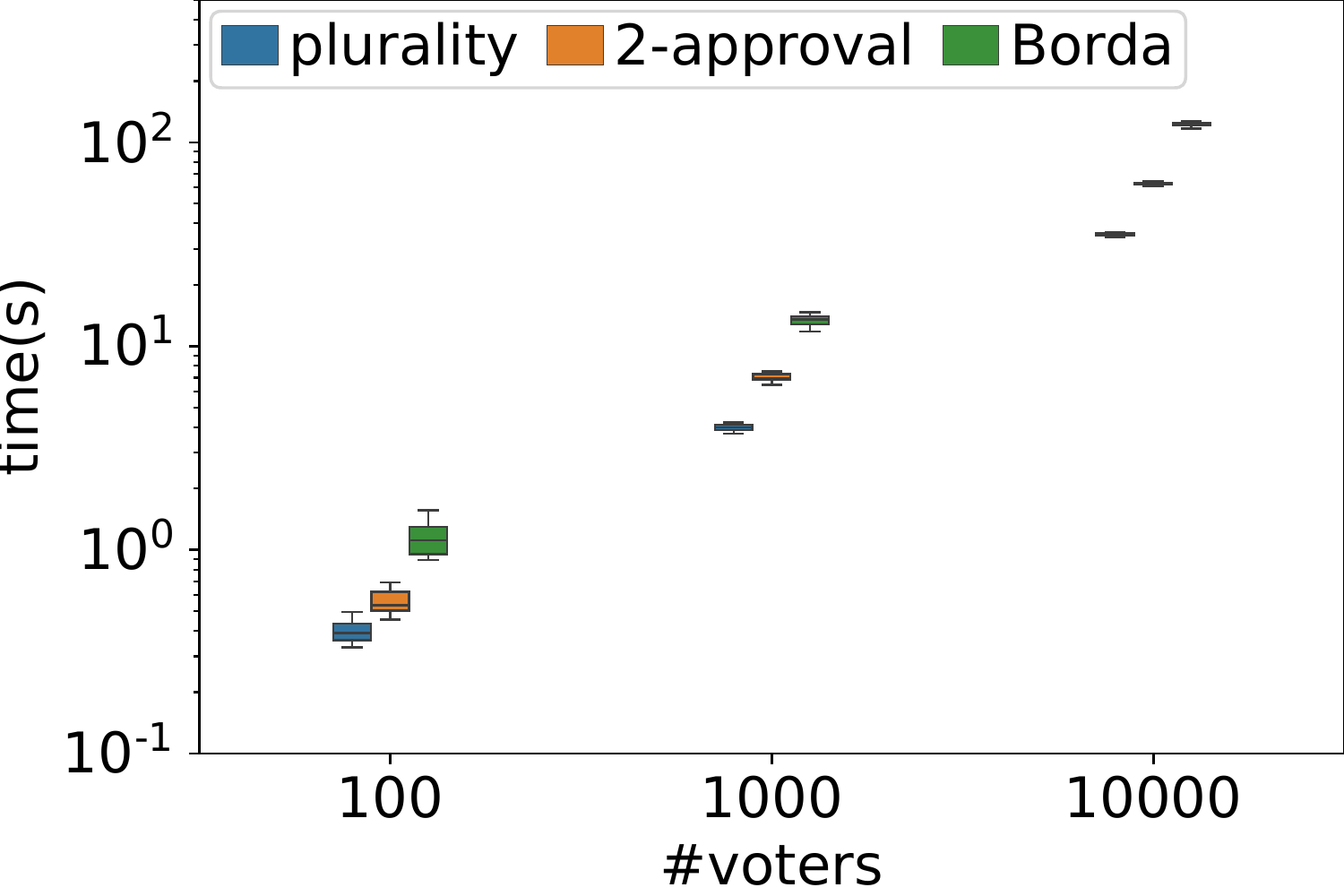}
	}~
	\caption{Running time over partial voting profiles, fixing 10 candidates and $p_{max}=0.1$. \mew is determined fastest under plurality, slowest under Borda. Running time is insensitive to changes in $\phi$ (voter preference diversity). Running time increases linearly with the number of voters.}
	\label{fig:posets}
\end{figure}

Figure~\ref{fig:posets} demonstrates the impact of voting rules on the running time.  \mew is determined fastest under plurality, followed by 2-approval, and finally, by Borda.  In Figure~\ref{fig:posets:phi}, the $\phi$ parameter of the Mallows model is varied to test the impact of voter consensus level on the running time, which turns out to be minor. Figure~\ref{fig:posets:voters} shows the linearly increasing running time with the number of voters.

\begin{figure}[b!]
	\centering
	\includegraphics[width=0.3\linewidth]{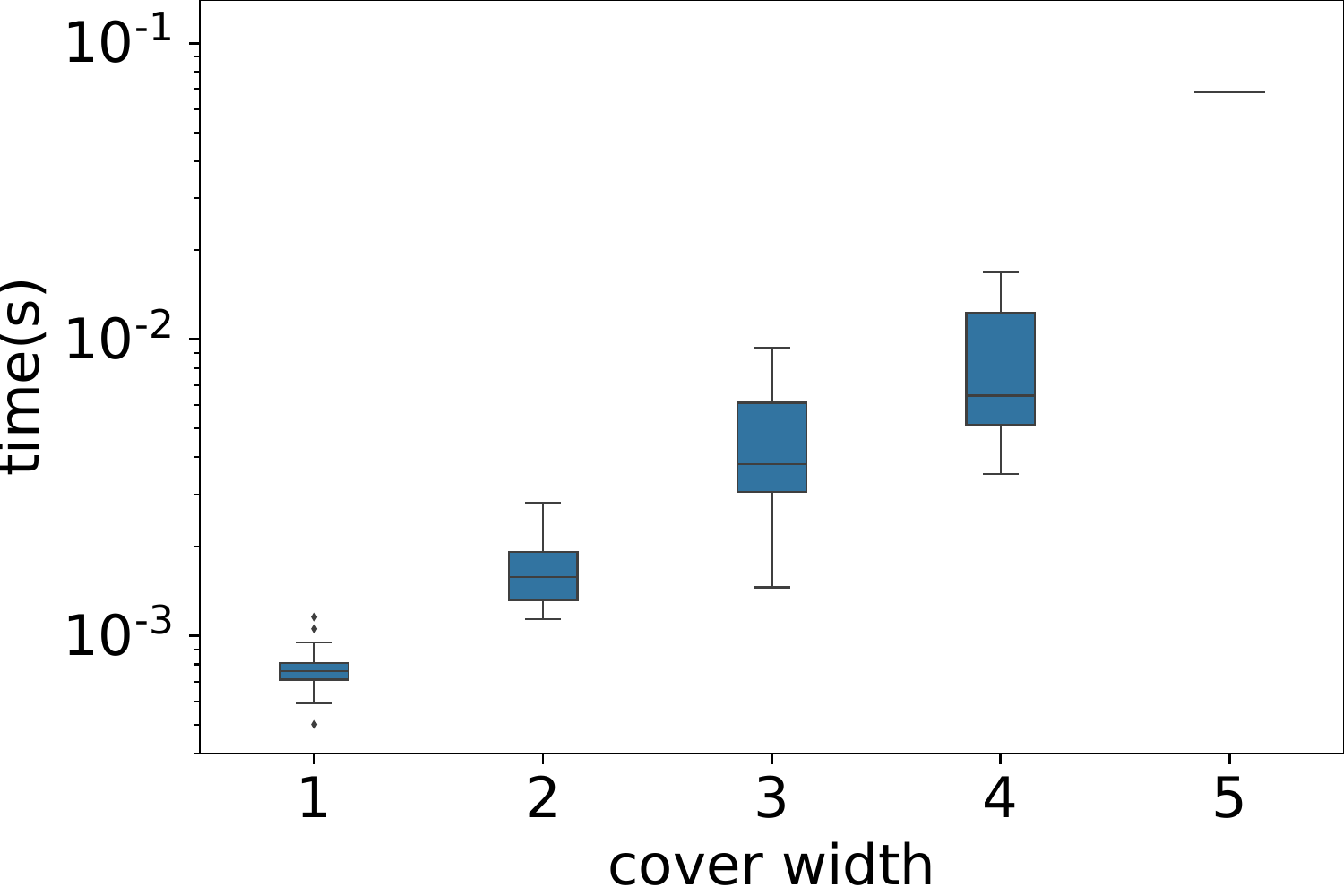}
	\caption{Running time over partial orders increases exponentially with cover widths.  100 partial orders generated with RSM over 10 candidates, $p_{max}=0.1$, and insertion probabilities equivalent to Mallows with $\phi=0.5$.}
	\label{fig:cw}
\end{figure}

\e{Partial orders and cover width.} A benchmark of partial orders is prepared to investigate the impact of the cover width parameter, discussed in Section~\ref{sec:preliminaries:pref}, on performance. We fixed the reference ranking of the RSM over 10 candidates, and made its insertion probabilities equivalent to $\phi = 0.5$ in Mallows.
For each partial order, we sample a new vector of edge construction probabilities with $p_{max} = 0.1$, generating 100 partial orders.
Figure~\ref{fig:cw} shows that the exact solver for partial orders has exponential complexity with the cover width, which is consistent with our complexity analysis.

\e{Partially partitioned voting profiles.} Synthetic profiles are generated with the parameter settings in Figure~\ref{fig:pp}.
For each  setup, we generated 10 profiles and calculated the \mew with the solver in Theorem~\ref{theorem:tractability_of_parparVP} and Algorithm~\ref{alg:pruning}.
Figure~\ref{fig:pp} shows their average running time to demonstrate the scalability w.r.t. the number of partitions, candidates, and voters.
In Figure~\ref{fig:pp:k}, the running time decreased with greater number of partitions.
Figures~\ref{fig:pp:m} and~\ref{fig:pp:n} show that the running time increases linearly with more candidates and voters.

\begin{figure}[tb!]
	\centering
	\subfloat[80 cand, 1000 voters, varying \#partitions]{
		\label{fig:pp:k}
		\includegraphics[width=0.3\linewidth]{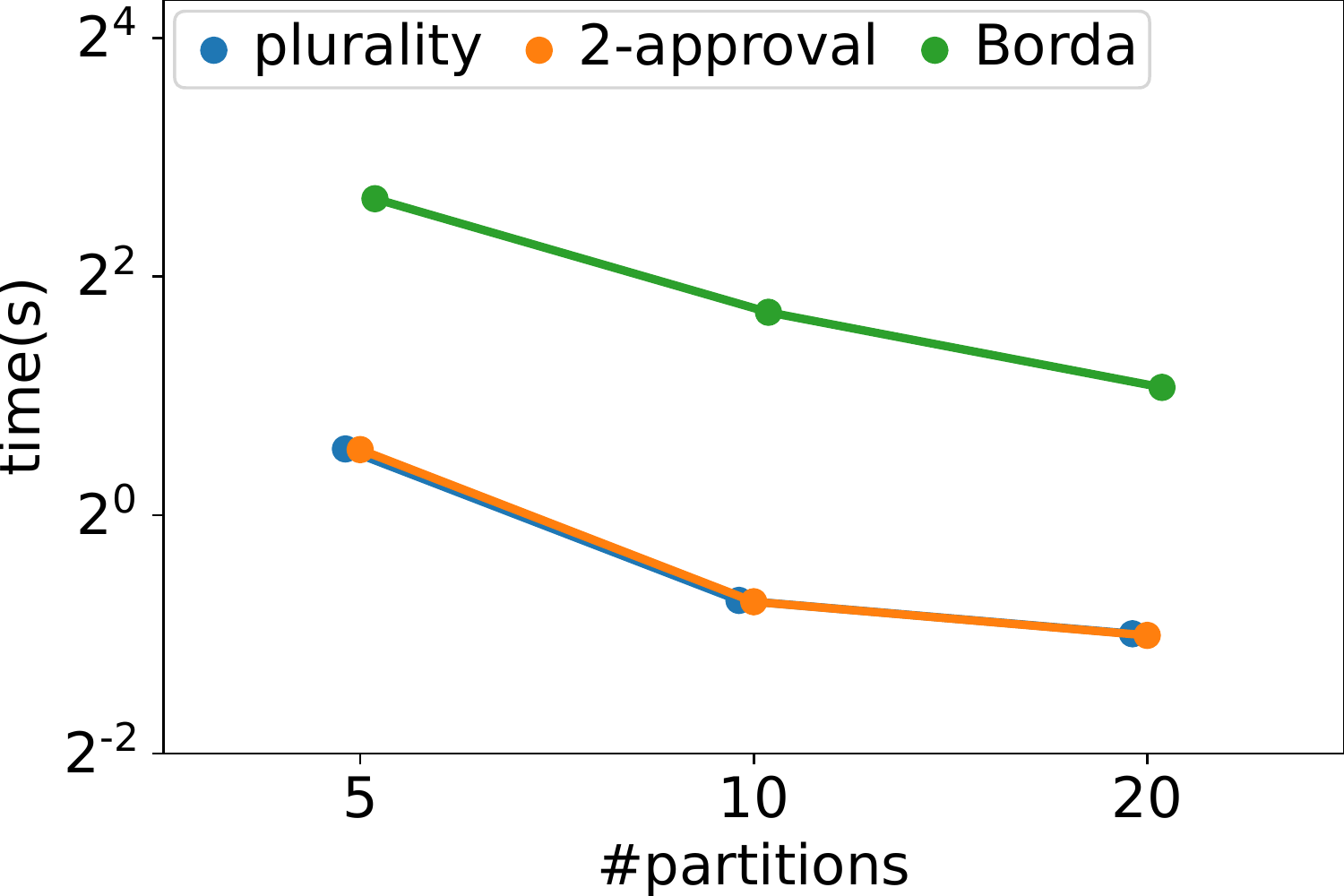}
	}\hfill
	\subfloat[1000 voters, 5 partitions, varying \#cand]{
		\label{fig:pp:m}
		\includegraphics[width=0.3\linewidth]{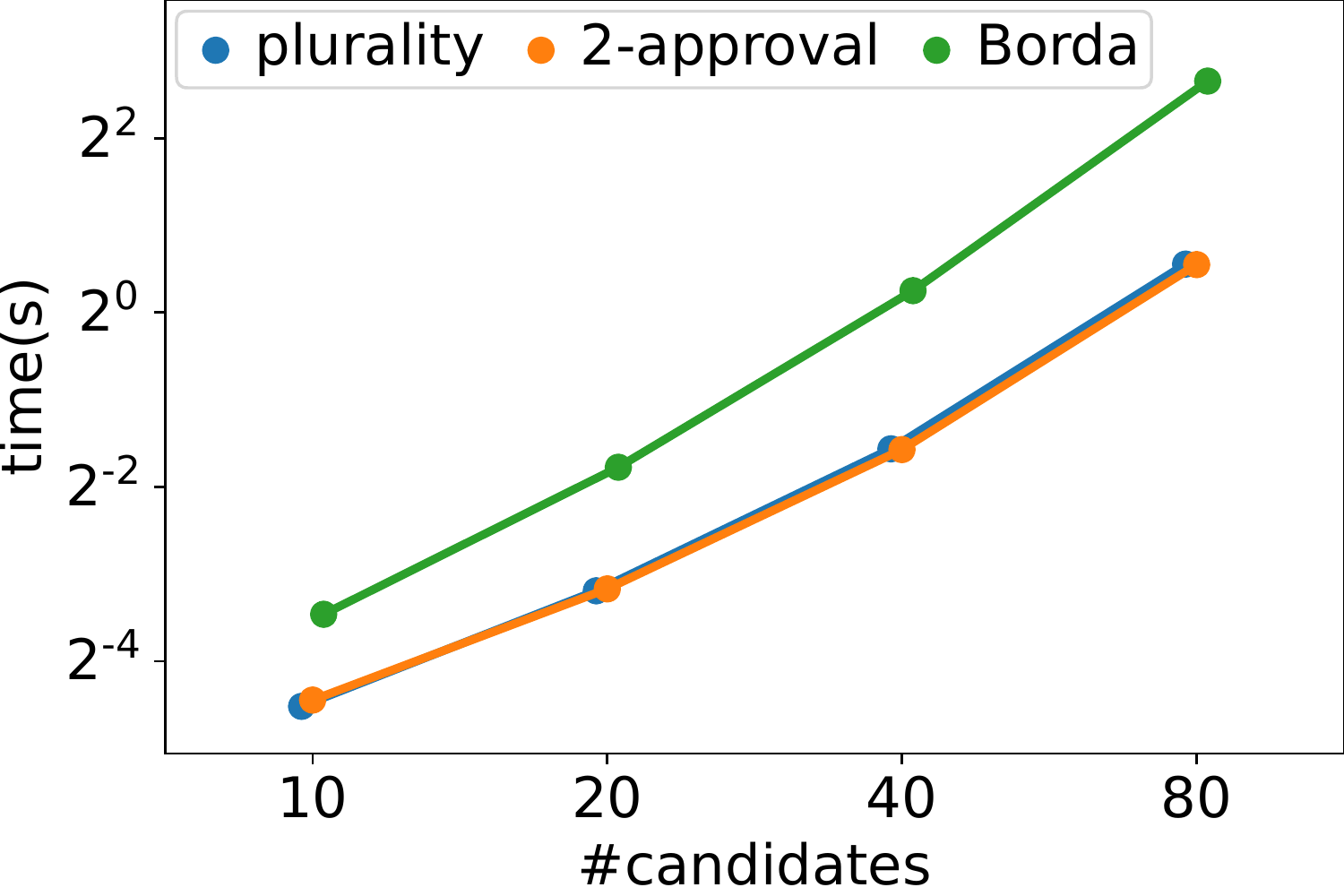}
	}\hfill
	\subfloat[\rev{80 cand, 5 partitions, varying \#voters}]{
		\label{fig:pp:n}
		\includegraphics[width=0.3\linewidth]{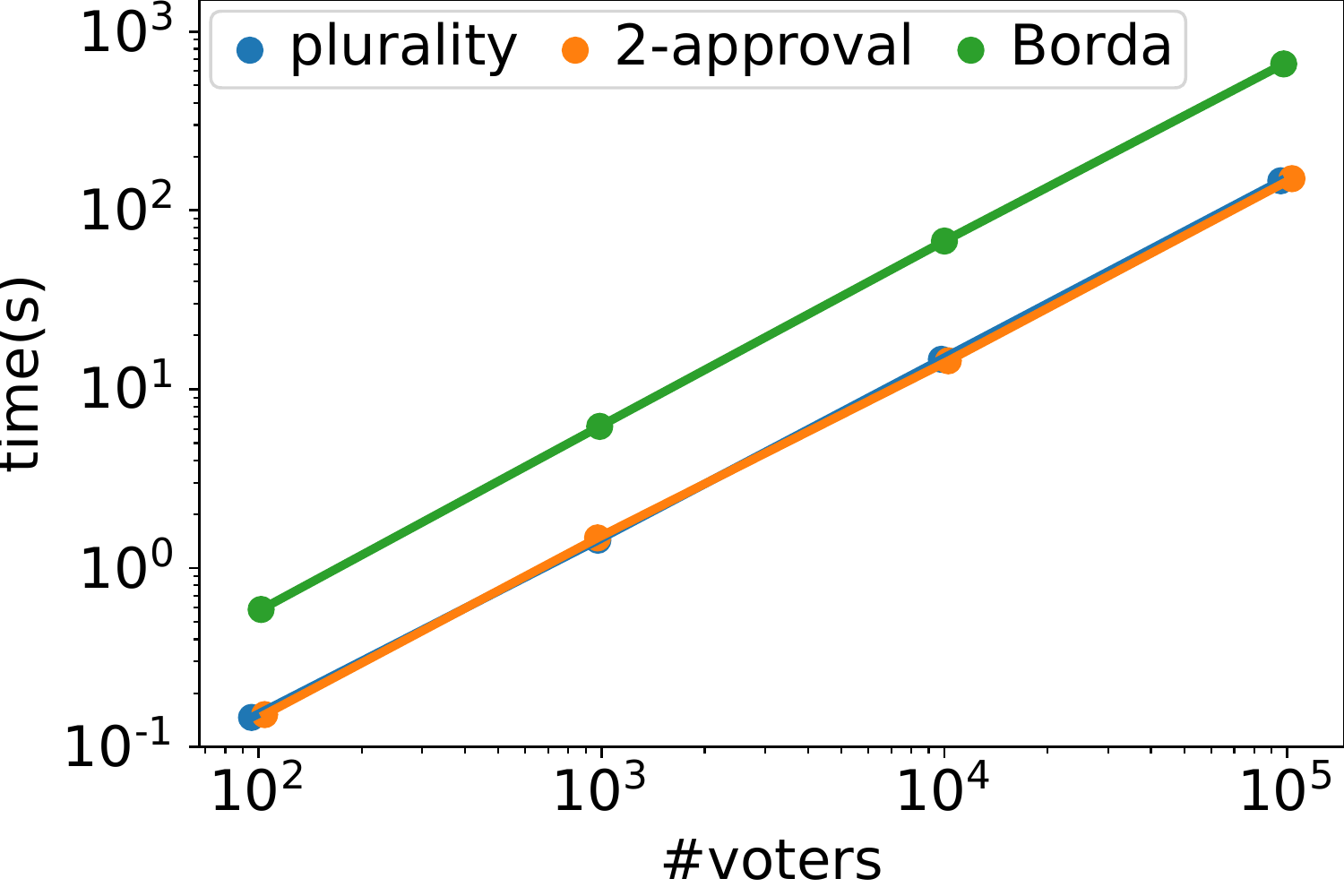}
	}\hfill
	\caption{Average time over partially partitioned profiles. It declines with more partitions in the voter preferences, increases with more candidates in the voting profile, and increases linearly with the number of voters.}
	\label{fig:pp}
\end{figure}

\e{Fully partitioned voting profiles.} The setup of this experiment is identical to that of the partially partitioned preferences above.
While Figure~\ref{fig:fp:m} also gives the linear  growth of running time with increasing number of candidates, Figure~\ref{fig:fp:k} shows opposite trends between Borda and the other two voting rules. It turns out that under Borda rule, the pruning strategy benefits from the increasing number of partitions.
In comparison, after turning off the pruning strategy, the running time of computing \mew under the Borda rule increased from 0.76 seconds to 0.79 seconds.

\begin{figure}[tb!]
	\centering
	\subfloat[80 cand, 1000 voters, varying \#partitions]{
		\label{fig:fp:k}
		\includegraphics[width=0.3\linewidth]{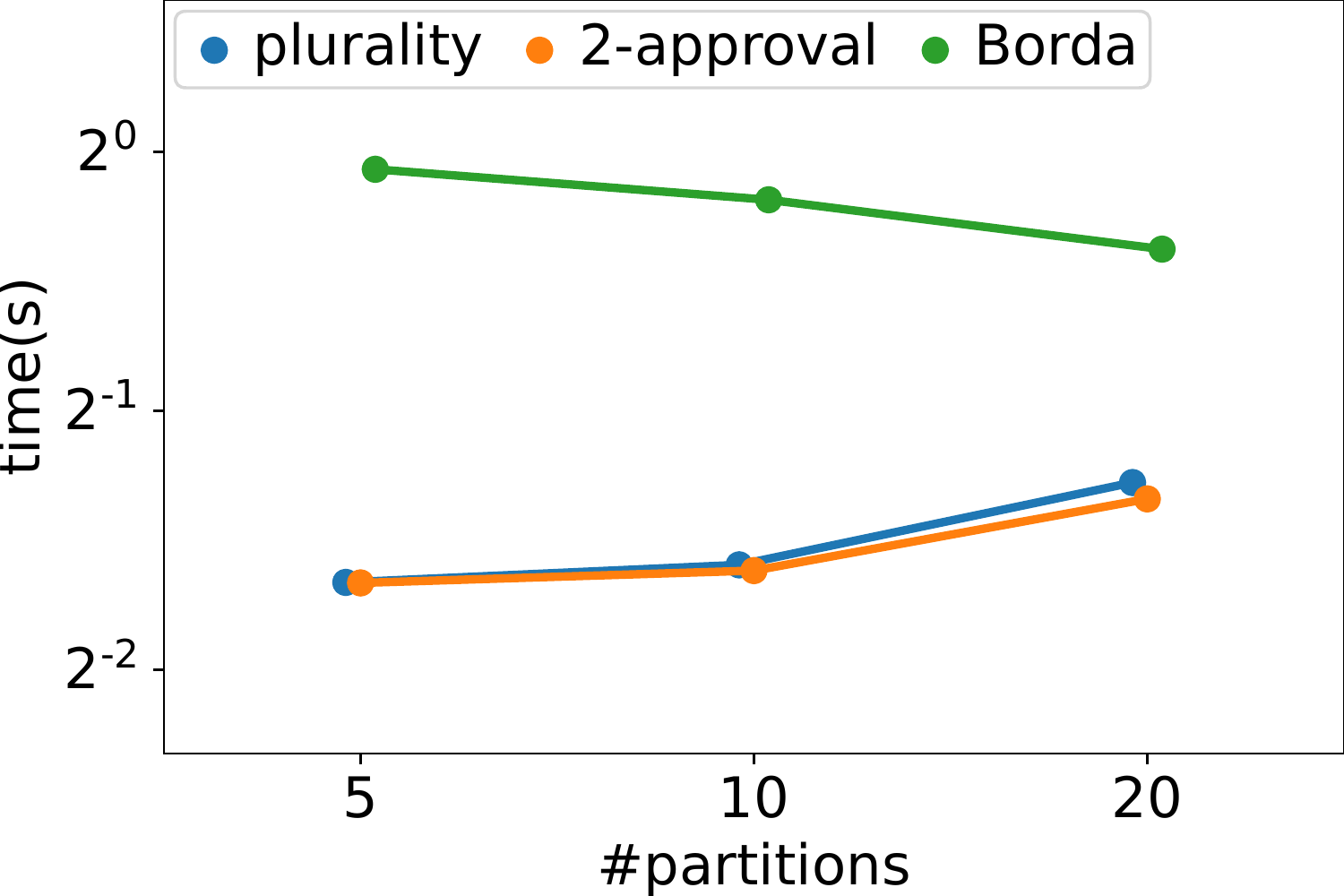}
	}\hfill
	\subfloat[1000 voters, 5 partitions, varying \#cand]{
		\label{fig:fp:m}
		\includegraphics[width=0.3\linewidth]{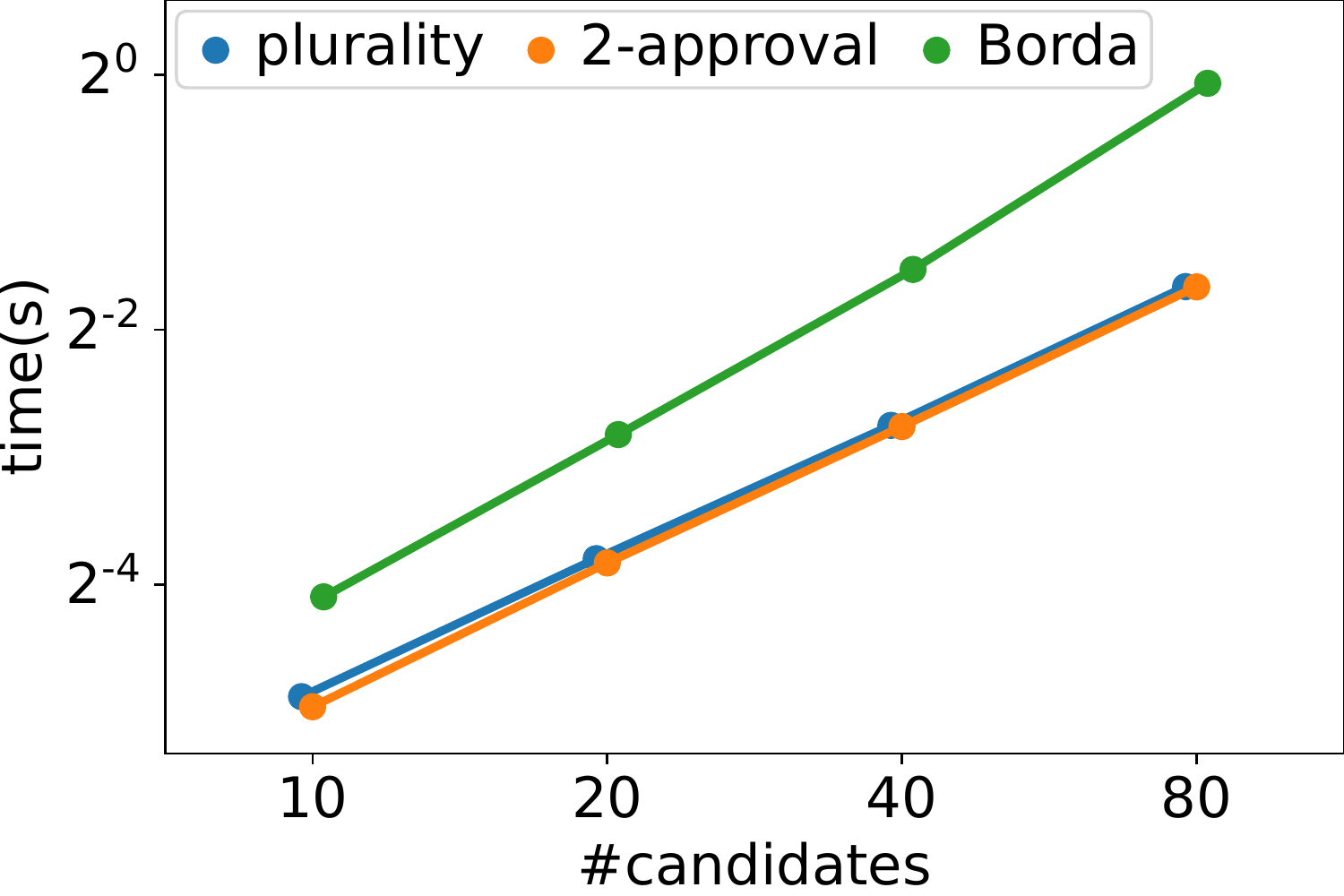}
	}\hfill
	\subfloat[\rev{80 cand, 5 partitions, varying \#voters}]{
		\label{fig:fp:n}
		\includegraphics[width=0.3\linewidth]{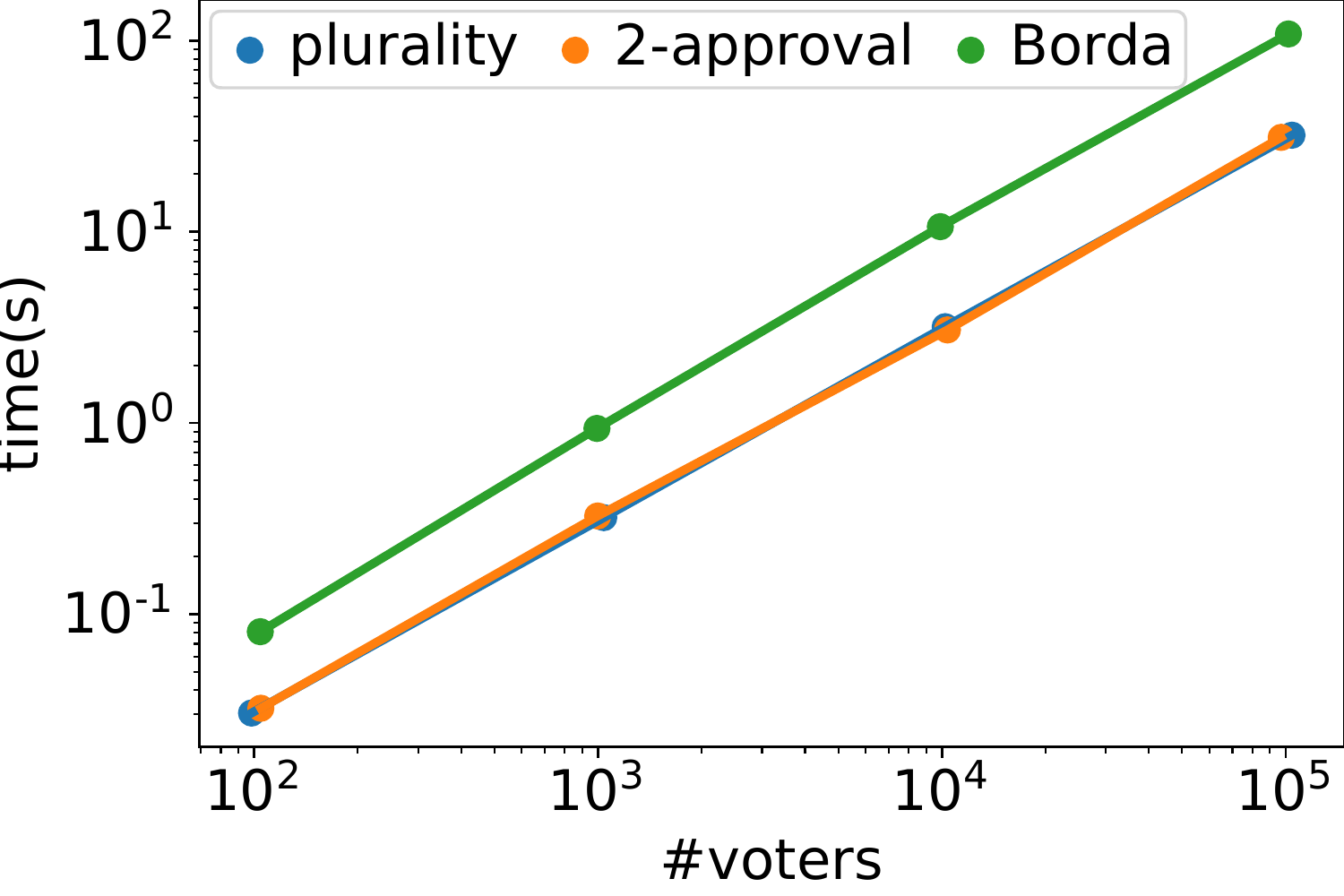}
	}\hfill
	\caption{Average time over fully partitioned profiles. It increases under plurality and 2-approval, but declines under Borda, with more partitions in the voter preferences. It increases linearly when adding more candidates or voters.}
	\label{fig:fp}
\end{figure}

\e{Partial chain voting profiles.} The setup of this experiment is almost identical to the experiments of partitioned voting profiles, except that the number of partitions is now the chain size. Figure~\ref{fig:pc} presents the average running time to compute the \mew with candidate pruning under various parameter settings.
In Figure~\ref{fig:pc:k}, running time increases mildly with chain size under Borda, while it even drops under plurality and 2-approval. \rev{In Figures~\ref{fig:pc:m} and~\ref{fig:pc:n}, the running time increases linearly with more candidates and voters.}

\begin{figure}[tb!]
	\centering
	\subfloat[80 cand, 1000 voters, varying chain size]{
		\label{fig:pc:k}
		\includegraphics[width=0.3\linewidth]{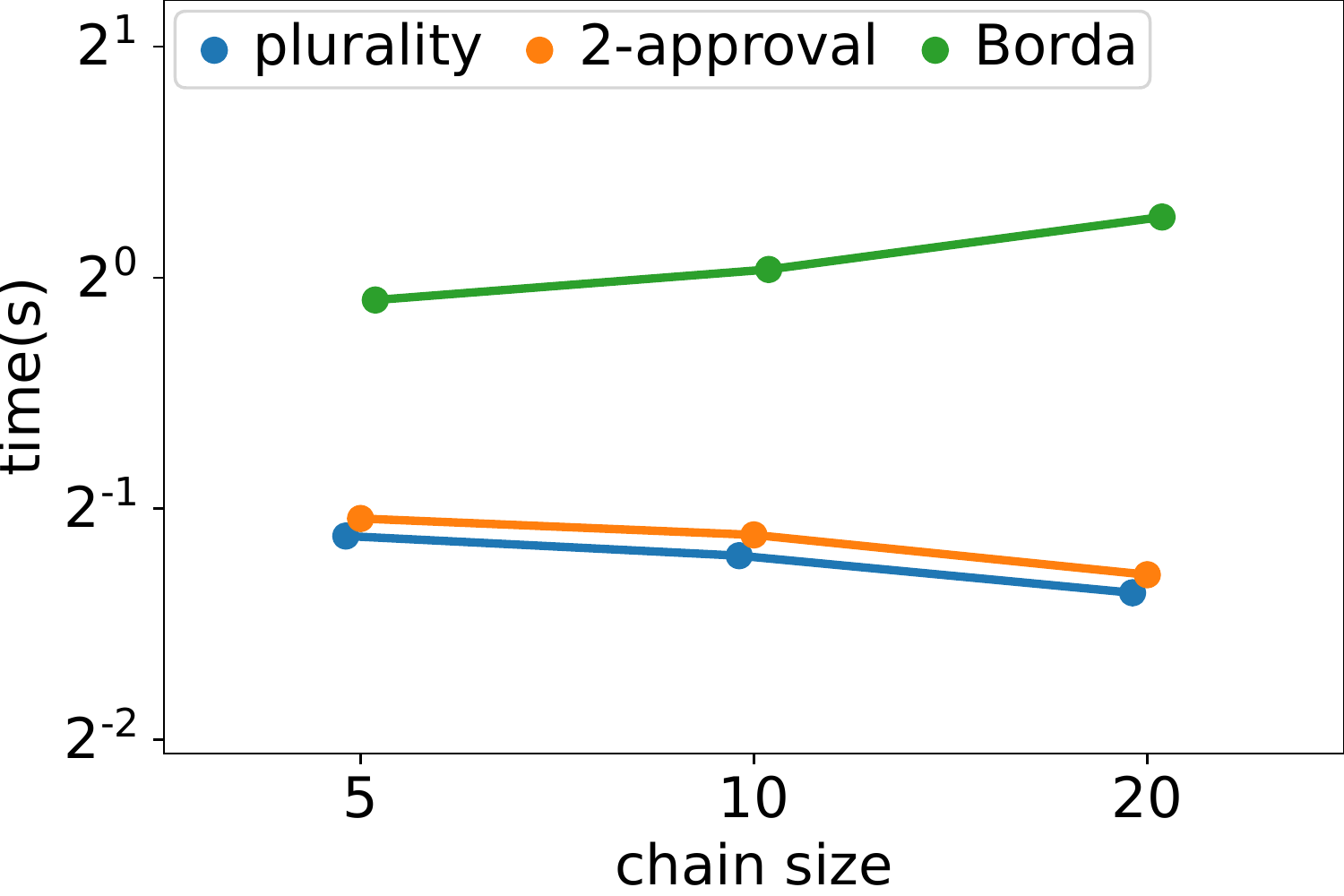}
	}\hfill
	\subfloat[1000 voters, chain size 5, varying \#cand]{
		\label{fig:pc:m}
		\includegraphics[width=0.3\linewidth]{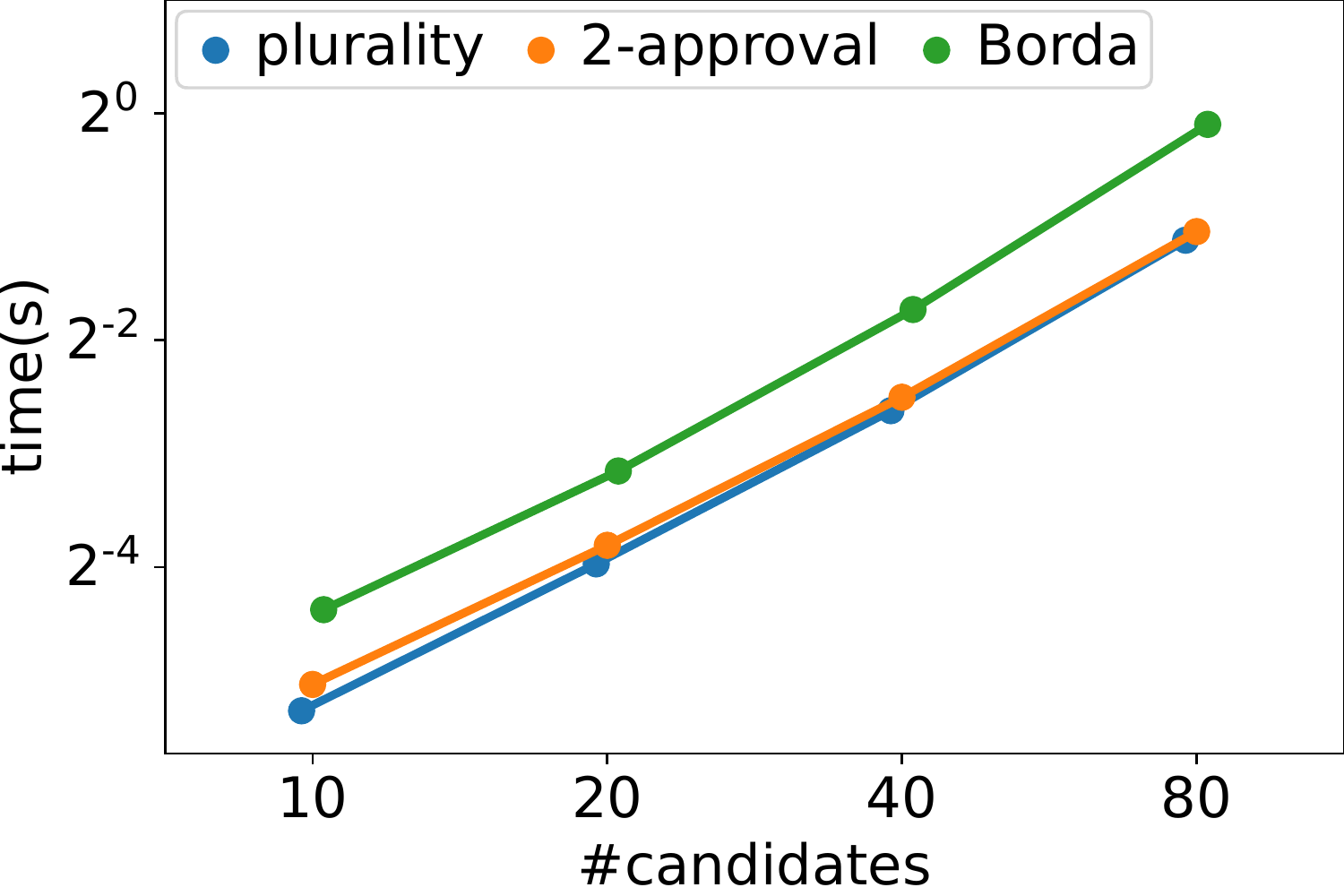}
	}\hfill
	\subfloat[\rev{80 cand, chain size 5, varying \#voters}]{
		\label{fig:pc:n}
		\includegraphics[width=0.3\linewidth]{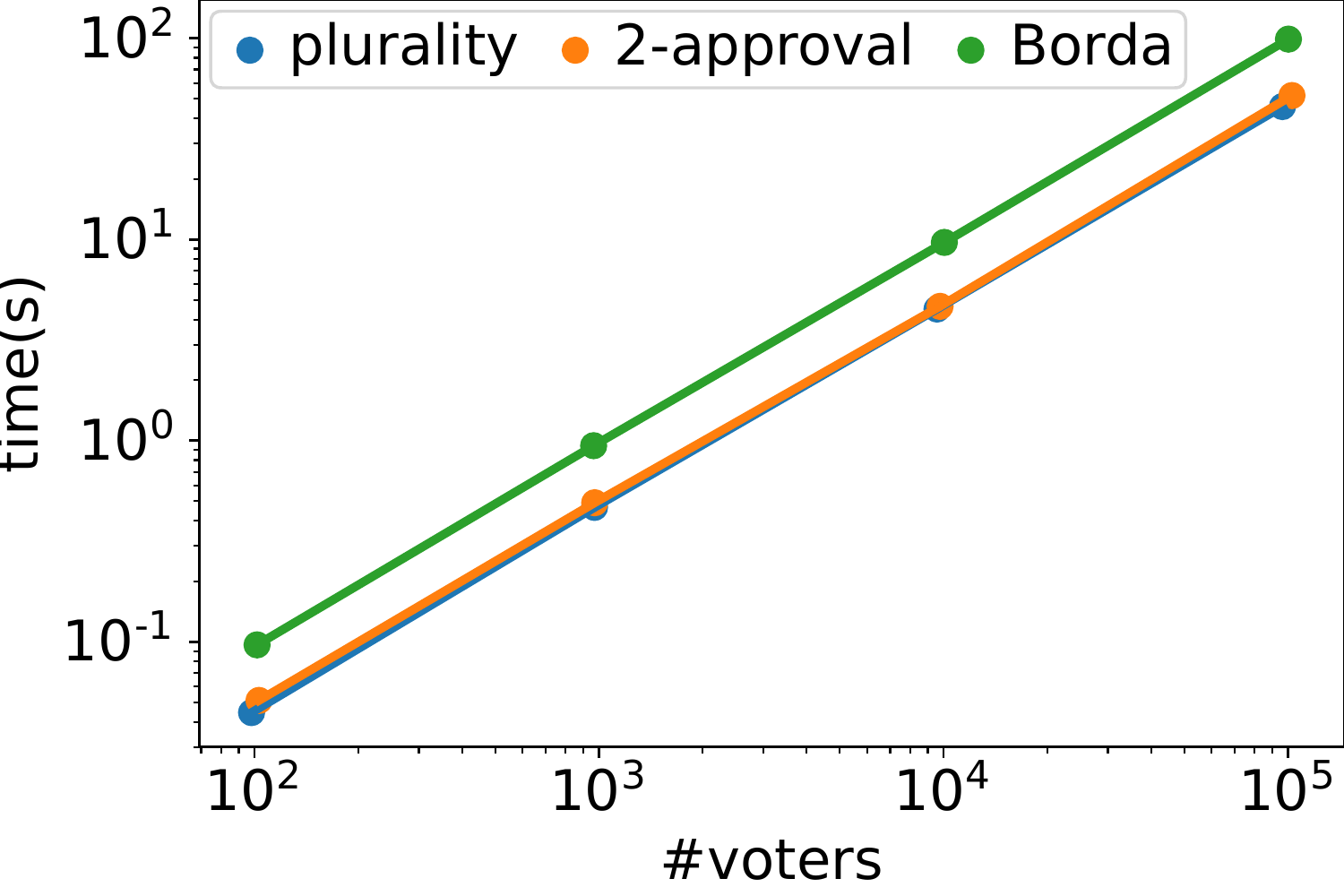}
	}\hfill
	\caption{Average time over partial chain profiles is more sensitive to the Borda rule when increasing the chain size, and increases linearly with the number of candidates or voters for all rules.}
	\label{fig:pc}
\end{figure}

\e{Truncated voting profiles.} In this experiment, we fixed the number of candidates to be 80, and varied the top and bottom sizes of the truncated rankings, and the number of voters.
For each parameter setting, we generated 10 voting profiles and computed the average running time of \mew with the candidate pruning strategy. 

Figure~\ref{fig:tr:tb} demonstrates that the top and bottom size has no impact on the running time over plurality and 2-approval, which is expected since these voting rules only need information about the top-1 or 2 ranked candidates. Under Borda, \mew computation is faster for higher top and bottom size, also as expected, since uncertainty in the truncated rankings decreases accordingly.  Finally, Figure~\ref{fig:tr:n} demonstrates that the running time over truncated voting profiles increases linearly with the number of voters.

\begin{figure}[tb!]
	\centering
	\subfloat[80 cand, 1000 voters]{
		\label{fig:tr:tb}
		\includegraphics[width=0.3\linewidth]{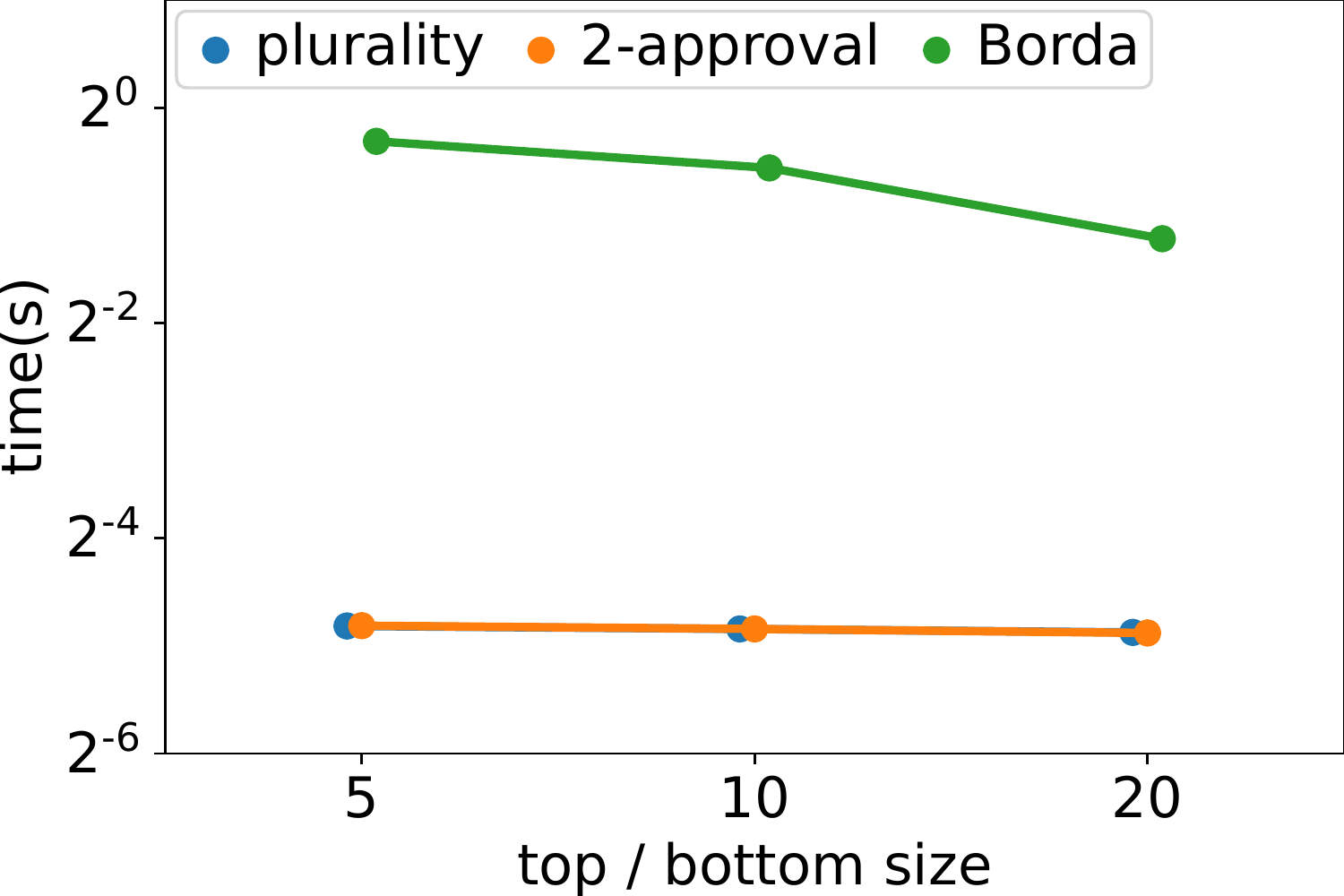}
	}\hspace{5em}
	\subfloat[\rev{80 cand, top / bottom size 5}]{
		\label{fig:tr:n}
		\includegraphics[width=0.3\linewidth]{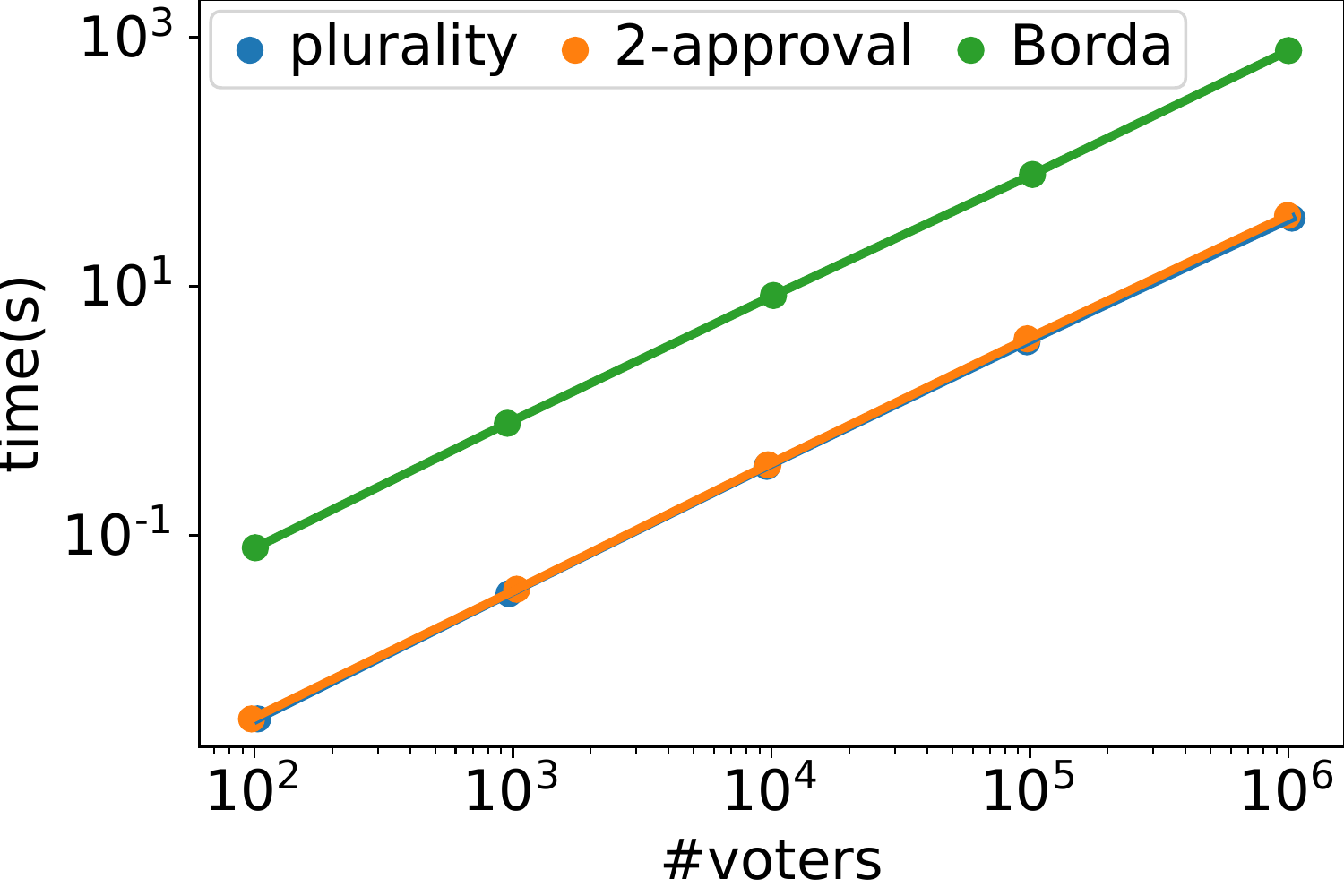}
	}~
	\caption{Average time over truncated voting profiles. It stays the same for plurality and 2-approval, but steadily decreases, when increasing the top and bottom sizes of the truncated rankings. It increases linearly with the number of voters.}
	\label{fig:tr}
\end{figure}

\subsection{Probabilistic voting profiles}

\e{Mallows voting profiles.}
We generate synthetic profiles to test the performance of the RIM solver over its special case, the Mallows model.  We vary the number of candidates from 10 to 80, and the dispersion parameter $\phi$ from 0.1 to 0.9, generate 10 profiles for each parameter setting, and report the average running time, with pruning, in Figure~\ref{fig:mallows}.
Figure~\ref{fig:mallows:m} shows that the solver scales well with the number of candidates, while Figure~\ref{fig:mallows:phi} shows that the time to compute \mew is insensitive to the Mallows parameter $\phi$.

\begin{figure}[tb!]
	\centering
	\subfloat[$\phi = 0.5$, varying \#candidates]{
		\label{fig:mallows:m}
		\includegraphics[width=0.3\linewidth]{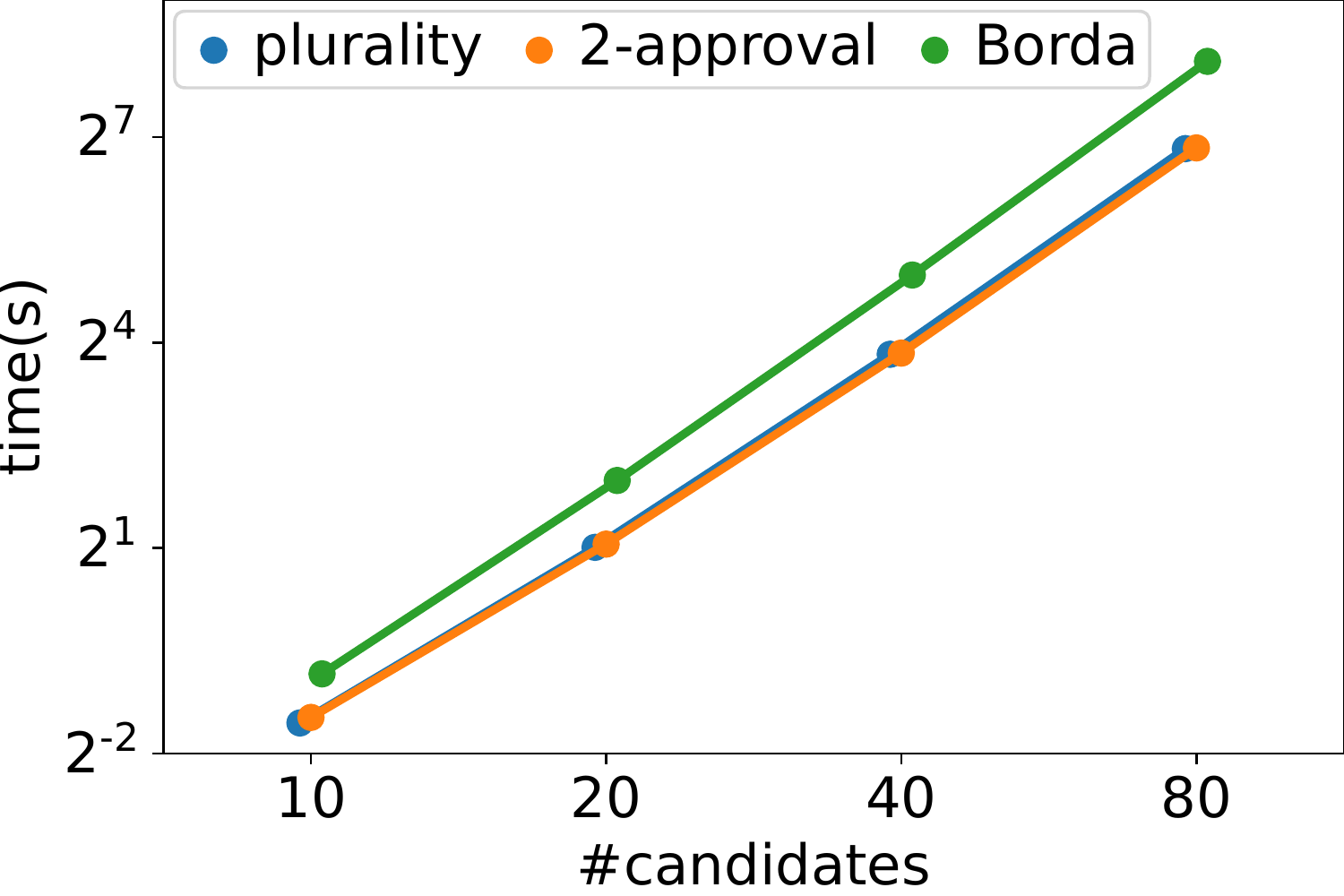}
	}\hspace{5em}
	\subfloat[10 candidates, varying $\phi$]{
		\label{fig:mallows:phi}
		\includegraphics[width=0.3\linewidth]{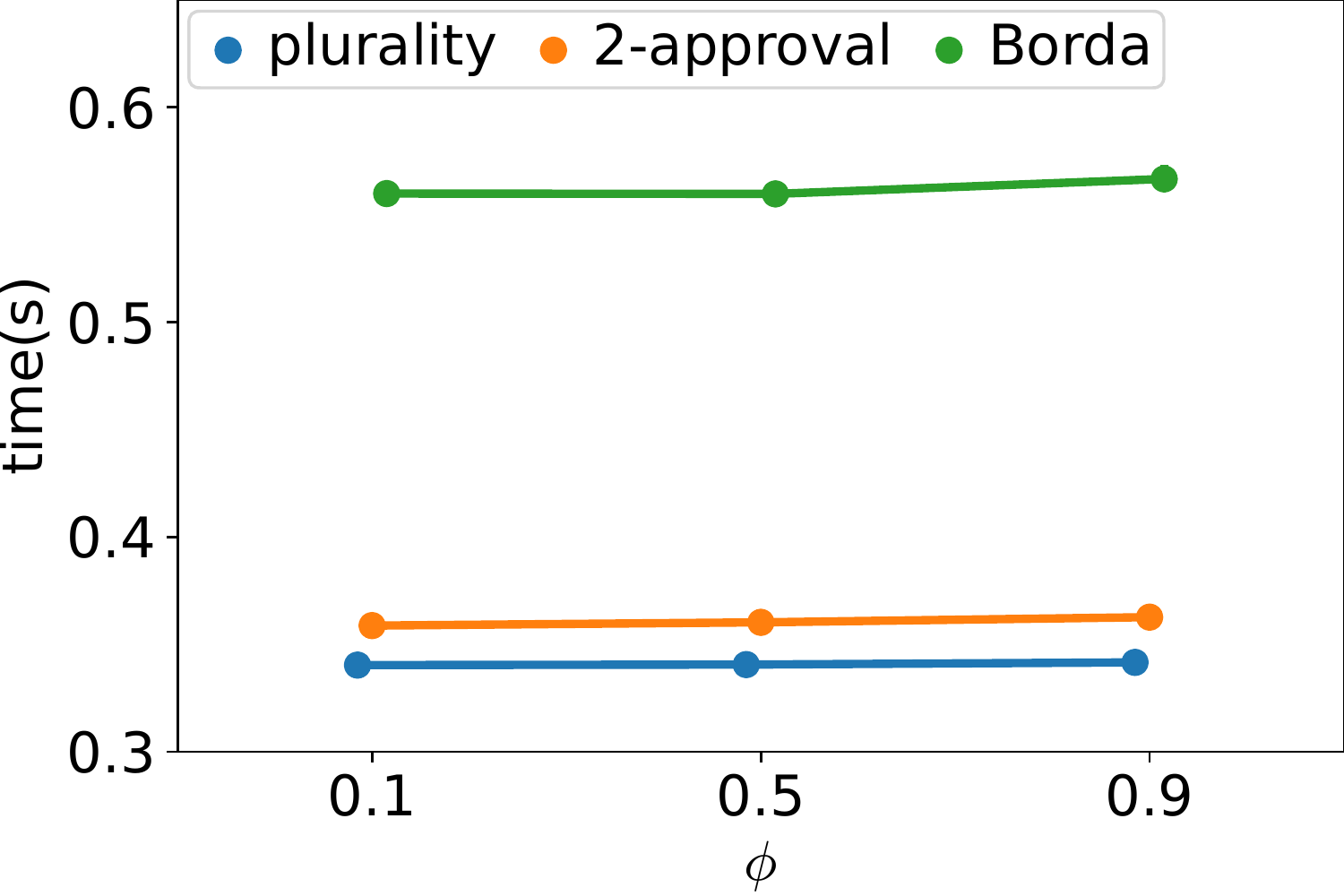}
	}~
	\caption{Average time over Mallows voting profiles, fixing 1000 voters. It increases linearly with the number of candidates, while the $\phi$ has little impact over the running time.}
	\label{fig:mallows}
\end{figure}

\e{RSM voting profiles.} This experiment uses exactly the same voting profiles as the experiment above for Mallows models. Since rRSM generalizes the Mallows model, we converted the Mallows models into rRSM instances and invoked the RSM solver for these converted voting profiles.
Compared to Figure~\ref{fig:mallows:m}, Figure~\ref{fig:rsm:m} shows better scalability with regards to increasing number of candidates, especially for the plurality and 2-approval rules.
Figure~\ref{fig:rsm:phi} gives a similar conclusion as Figure~\ref{fig:mallows:phi} that the dispersion of the preferences has little impact over the running time.

\begin{figure}[tb!]
	\centering
	\subfloat[$\phi = 0.5$, varying \#candidates]{
		\label{fig:rsm:m}
		\includegraphics[width=0.3\linewidth]{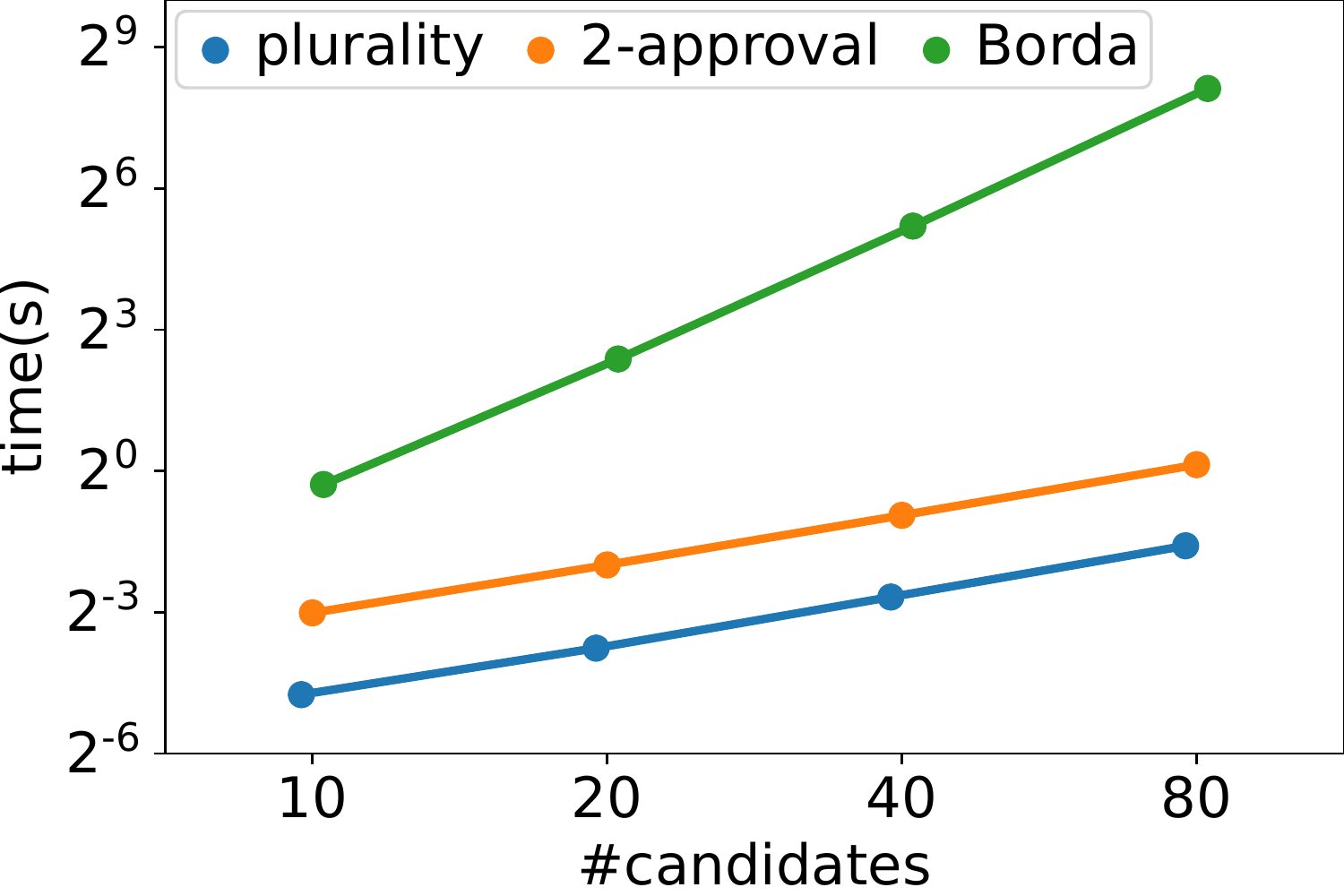}
	}\hspace{5em}
	\subfloat[10 candidates, varying $\phi$]{
		\label{fig:rsm:phi}
		\includegraphics[width=0.3\linewidth]{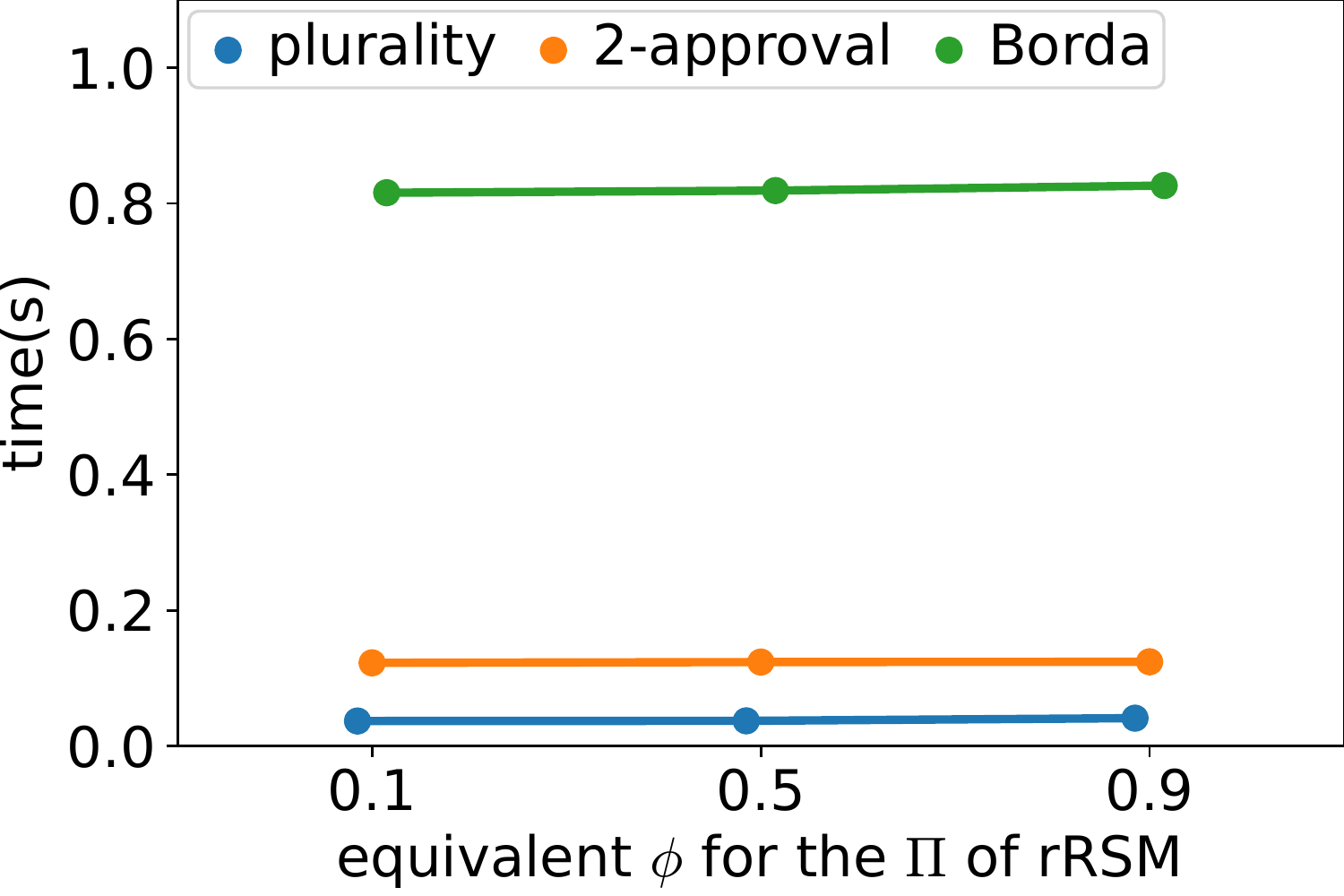}
	}~
	\caption{Average time over RSM voting profiles, fixing 1000 voters. It increases linearly with the number of candidates, while the equivalent $\phi$ has little impact over the running time.}
	\label{fig:rsm}
\end{figure}

\subsection{Combined voting profiles}

\e{Mallows with fully partitioned preferences.} In this experiment, we investigate the impact of the Mallows $\phi$ on the running time.
All voting profiles have 10 candidates, 1000 voters, and 2 partitions for the partitioned preferences.
Figure~\ref{fig:mallows|fp} shows that the Mallows $\phi$ has little impact on the running time, and that the running time under plurality and 2-approval rules are almost identical.  Similar trends were observed for more than 2 partitions, and we do not report these results here.

\e{Mallows with truncated rankings.} The setup of this experiment is similar to the above, for Mallows combined with fully partitioned preferences, except that the top and bottom sizes of the truncated rankings are fixed to be 3, instead of fixing the number of partitions.
Figure~\ref{fig:mallows|tr} demonstrates that the dispersion $\phi$ has little impact on the running time.  Similar trends were observed for other truncated ranking sizes, and we do not report these results here.

\e{Mallows with partial orders.} The setup of this experiment is similar to the previous two experiments for combined voting profiles. All voters in a profile have the same Mallows model. Their partial orders are generated by RSMs whose selection probabilities are derived from dispersion $\phi$ of their Mallows, and each voter has an independent edge construction probability vector sampled from $p_{max} = 0.9$. Figure~\ref{fig:mallows|po} shows that $\phi$ substantially impacts the running time: The computation of \mew is faster with a small $\phi$ value, especially for Borda.  This is expected, because the voters exhibit stronger consensus for lower values of $\phi$, and so the pruning strategy is more effective.

\begin{figure}[tb!]
	\centering
	\subfloat[$\mallowsPartitionVP$]{
		\label{fig:mallows|fp}
		\includegraphics[width=0.3\linewidth]{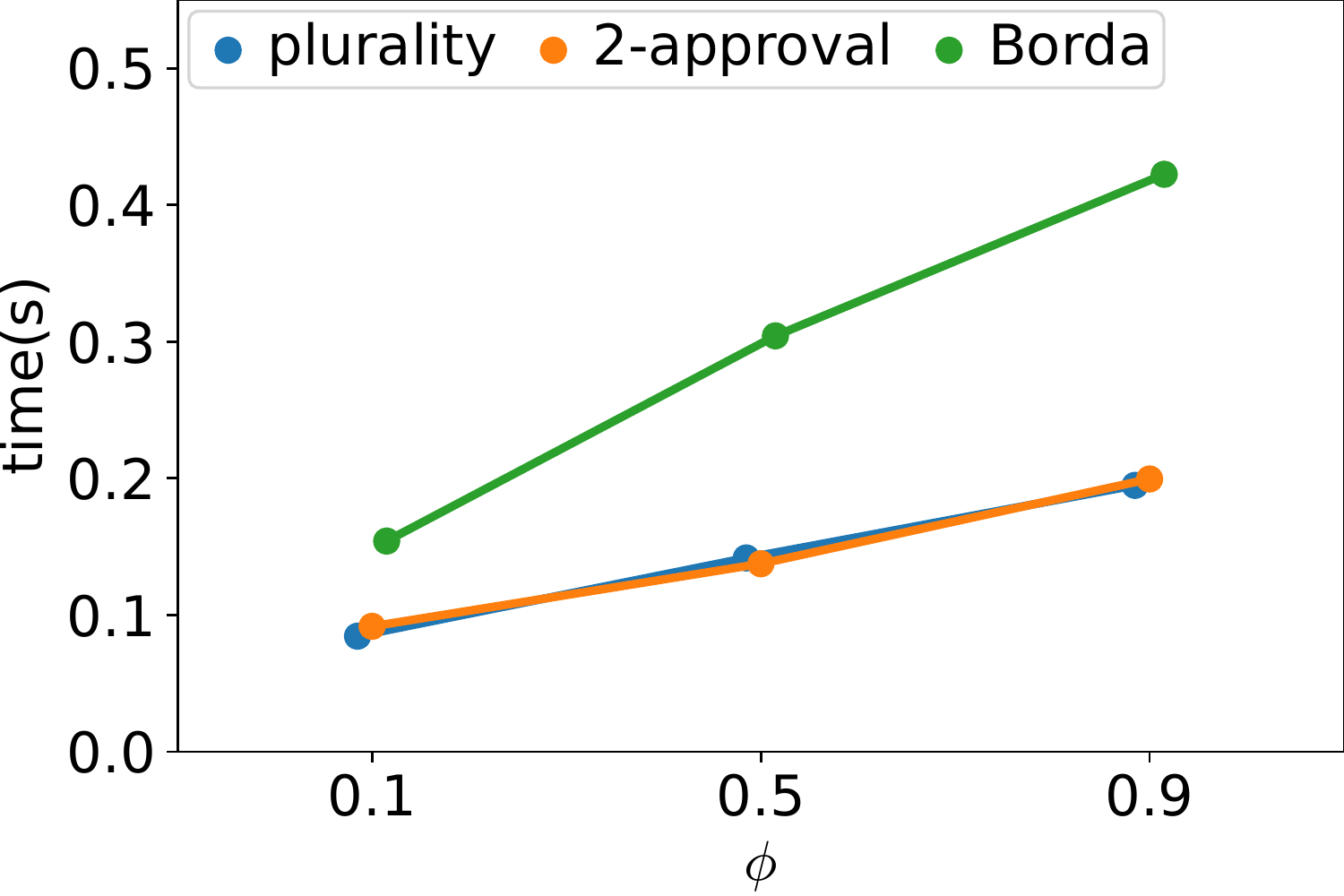}
	}\hfill
	\subfloat[$\VP^{\mallows \text{+TR}}$]{
		\label{fig:mallows|tr}
		\includegraphics[width=0.3\linewidth]{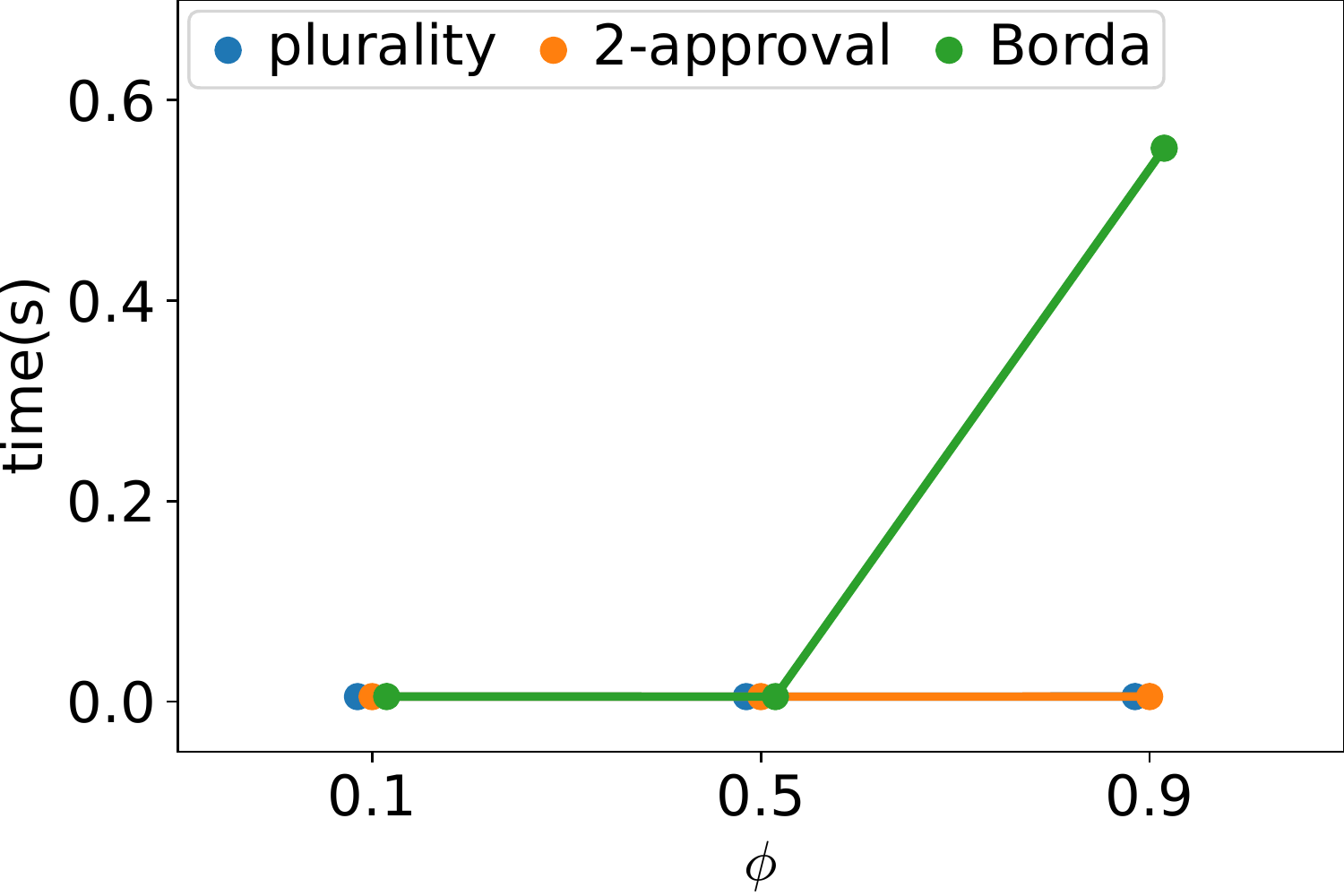}
	}\hfill
	\subfloat[$\VP^{\mallows \text{+PO}}$]{
		\label{fig:mallows|po}
		\includegraphics[width=0.3\linewidth]{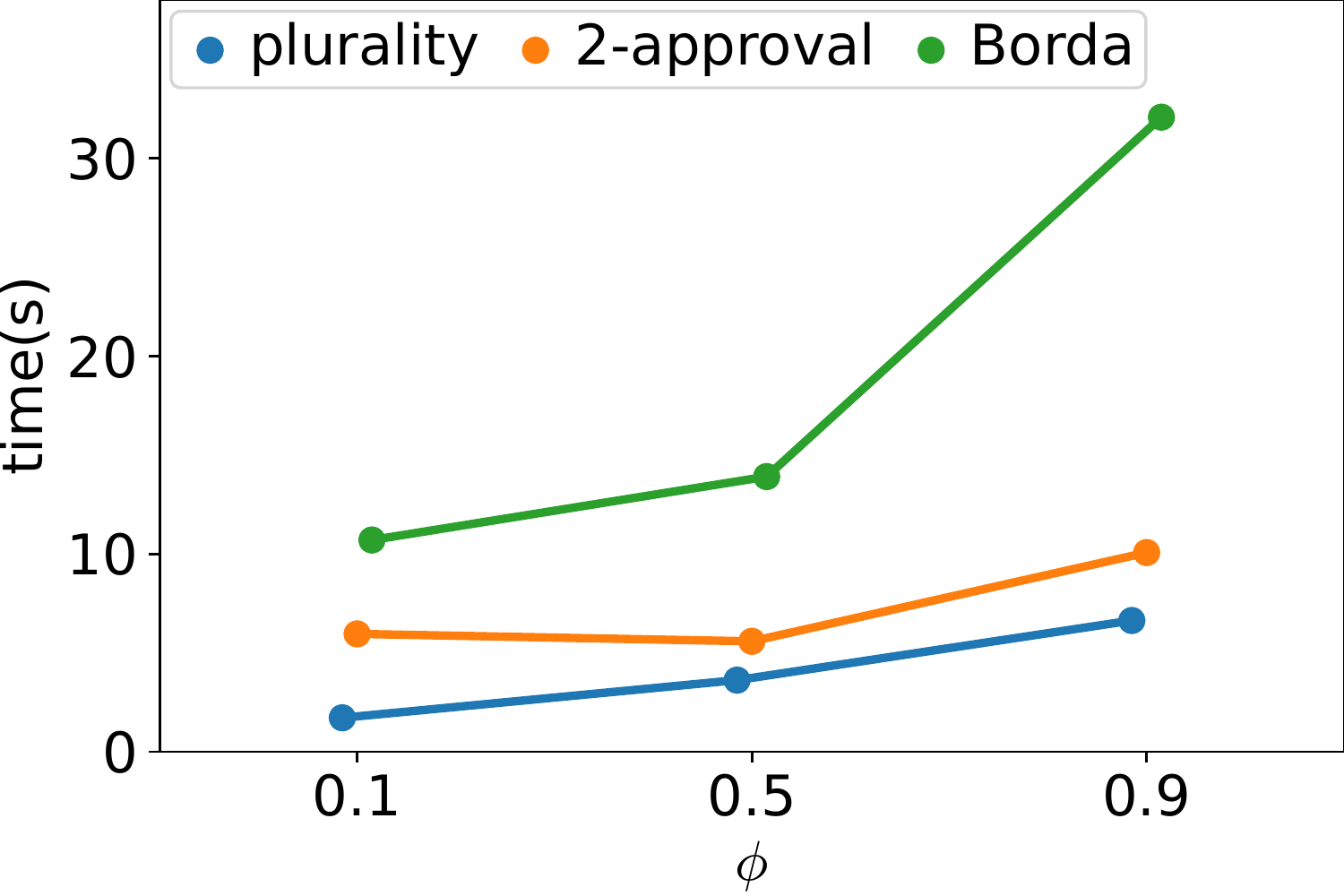}
	}~
	\caption{Average time over combined voting profiles. The $\phi$ has little impact over the running time when the Mallows models are combined with fully partitioned preferences or truncated rankings, while the running time increases with $\phi$ for voting profiles of the Mallows combined with partial orders, especially under the Borda rule. }
	\label{fig:combined}
\end{figure}

\begin{figure}[tb!]
	\centering
	\includegraphics[width=0.3\linewidth]{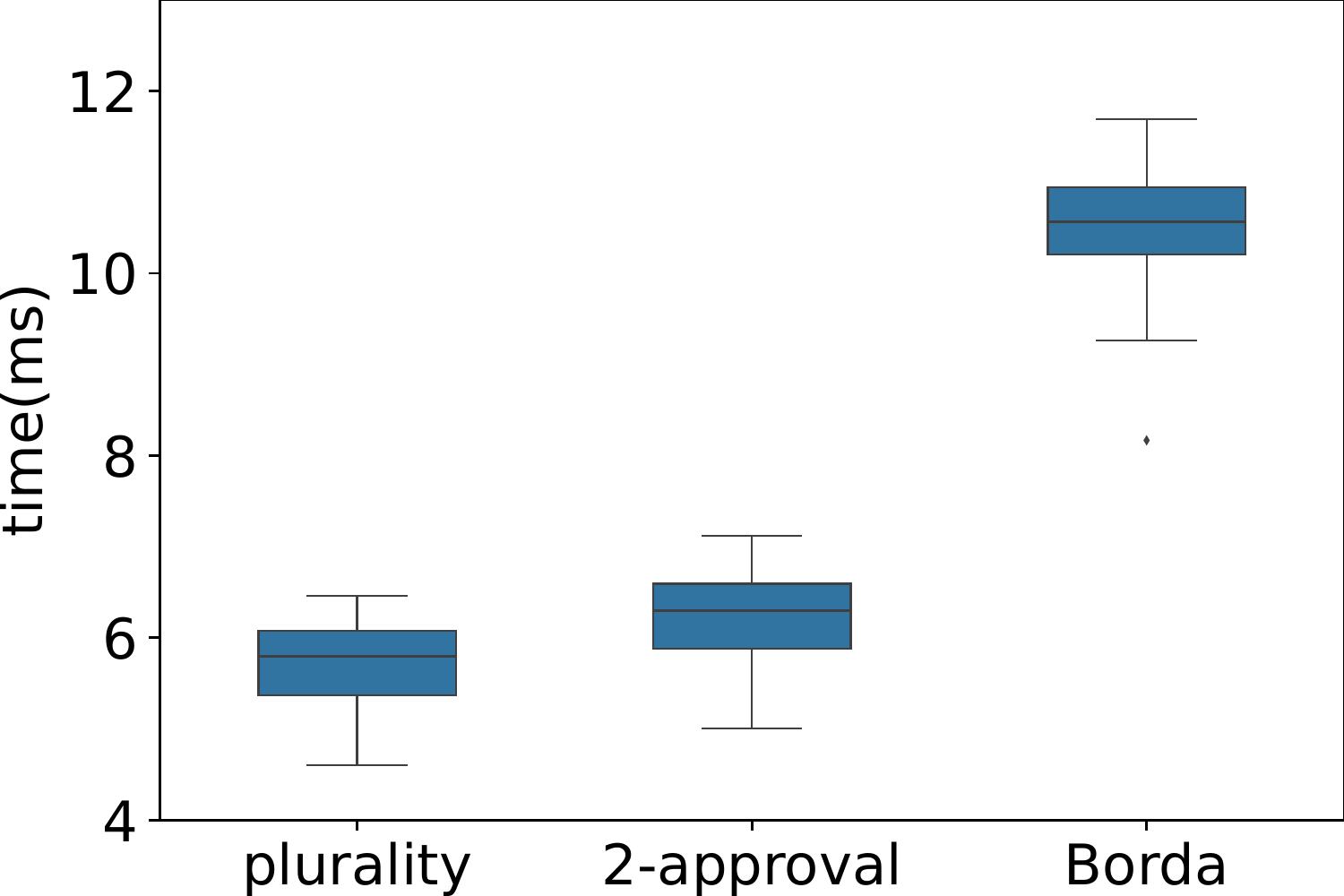}
	\caption{Running time on CrowdRank. Computing \mew under Borda is slower than under plurality or 2-approval.}
	\label{fig:crowdrank}
\end{figure}

\subsection{Real datasets}

Figure~\ref{fig:crowdrank} compares the running time of computing \mew under 3 voting rules for the 50 HITs in CrowdRank.
Recall that each HIT has its own set of 20 movies and 100 partial chains over these movies.
The experiment result shows that the voting rule has a significant impact on the running time.
It took much longer to compute \mew under the Borda rule, compared to plurality and 2-approval.

\begin{table}[b!]
	\caption{Computing \mew for real datasets}
	\begin{tabular}{@{}cccccc@{}}
		\toprule
		\multirow{2}{*}{Dataset} & \multicolumn{1}{c}{\multirow{2}{*}{\#cand}} & \multicolumn{1}{c}{\multirow{2}{*}{\#voters}} &                       \multicolumn{3}{c}{Running time (in seconds)}                        \\
		\cmidrule(l){4-6}     &          \multicolumn{1}{c}{}          &          \multicolumn{1}{c}{}          & \multicolumn{1}{c}{plurality} & \multicolumn{1}{c}{2-approval} & \multicolumn{1}{c}{Borda} \\ \midrule
		MovieLens         &                  200                   &                  6040                  &           \rev{137}           &           \rev{140}            &         \rev{344}         \\ \midrule
		Travel          &                   24                   &                  5456                  &          \rev{0.57}           &           \rev{0.49}           &        \rev{0.35}         \\ \bottomrule
	\end{tabular}
	\label{tab:movielens_n_travel}
\end{table}

Table~\ref{tab:movielens_n_travel} presents the running time of \mew over MovieLens and Travel, with number of items
$m$ (\ie movies and attraction categories, respectively), and  number of voters $n$. 
MovieLens is a partially partitioned voting profile of 200 movies.
It took about 38 minutes to compute the \mew under both plurality and 2-approval rules, and the winner is \e{American Beauty} in both cases.
In comparison, \e{Star Wars: Episode IV - A New Hope} is the winner under Borda, computed in around 100 minutes. 
Travel is a fully partitioned voting profile with 24 attraction categories. It took less than 1 second to compute the \mew under all three voting rules, which yields \e{museums} as the winner.
\section{Comparing \mew and \mpw}
\label{sec:mew_vs_mpw}

\rev{We now conduct a thorough comparison between \mew and \mpw to better demonstrate their differences.}
\mew and \mpw can be interpreted as different aggregation approaches across possible worlds.
Recall that given a general voting profile $\probaVP$, $\Omega(\probaVP) = \set{\completeVP_1, \ldots, \completeVP_\numPW}$ is the set of its possible worlds of $\probaVP$, and each $\completeVP_i$ is associated with a probability $p_i = \Pr(\completeVP_i \mid \probaVP)$ and $\sum_{i=1}^{\numPW} p_i = 1$.
Now let's see how the performance of a candidate $c$ is aggregated across possible worlds. Let $\mathds{1}()$ be the indicator function.

\begin{itemize}
  \itemsep -0.1em
  \item \mew: $\mathds{E}(\score(c, \probaVP)) = \sum_{i=1}^{\numPW} \score(c, \completeVP_i) \cdot p_i$
  \item \mpw: $\Pr(c \textsf{ wins} \mid \probaVP) = \sum_{i=1}^{\numPW} \mathds{1}(c \textsf{ wins} \mid \completeVP_i) \cdot p_i$
\end{itemize}

\mew estimates the average performance of a candidate, while \mpw estimates the probability that she wins.
As a result, \mpw ignores the possible worlds in which the candidate cannot win, \rev{putting certain candidates at a disadvantage}.

\begin{figure}[b!]
	\centering
	\subfloat[\revv{10 voters}, varying \#candidates]{
		\label{fig:mpw:candidates}
		\includegraphics[width=0.3\linewidth]{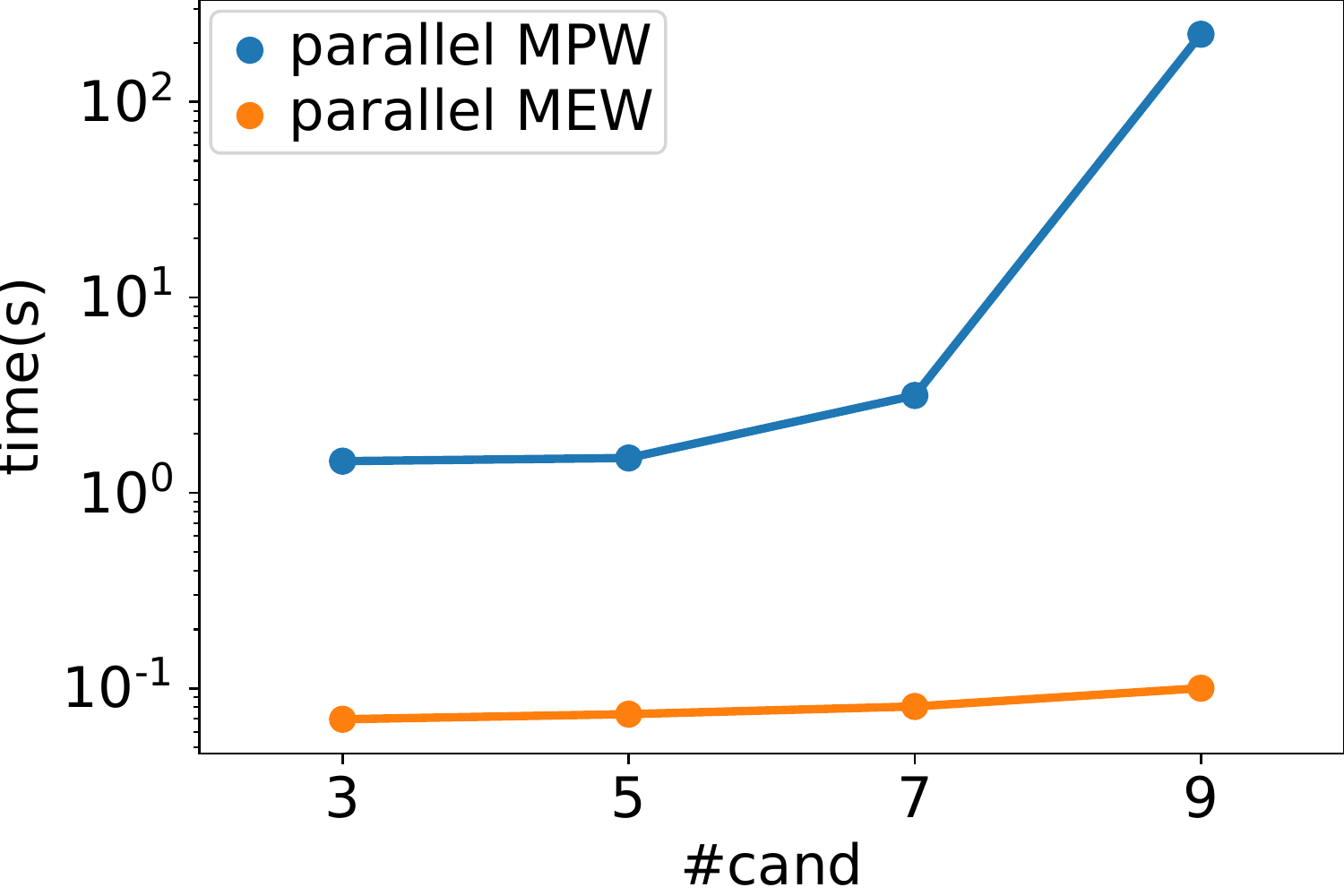}
	}\hspace{5em}
	\subfloat[\revv{9 candidates}, varying \#voters]{
		\label{fig:mpw:voters}
		\includegraphics[width=0.3\linewidth]{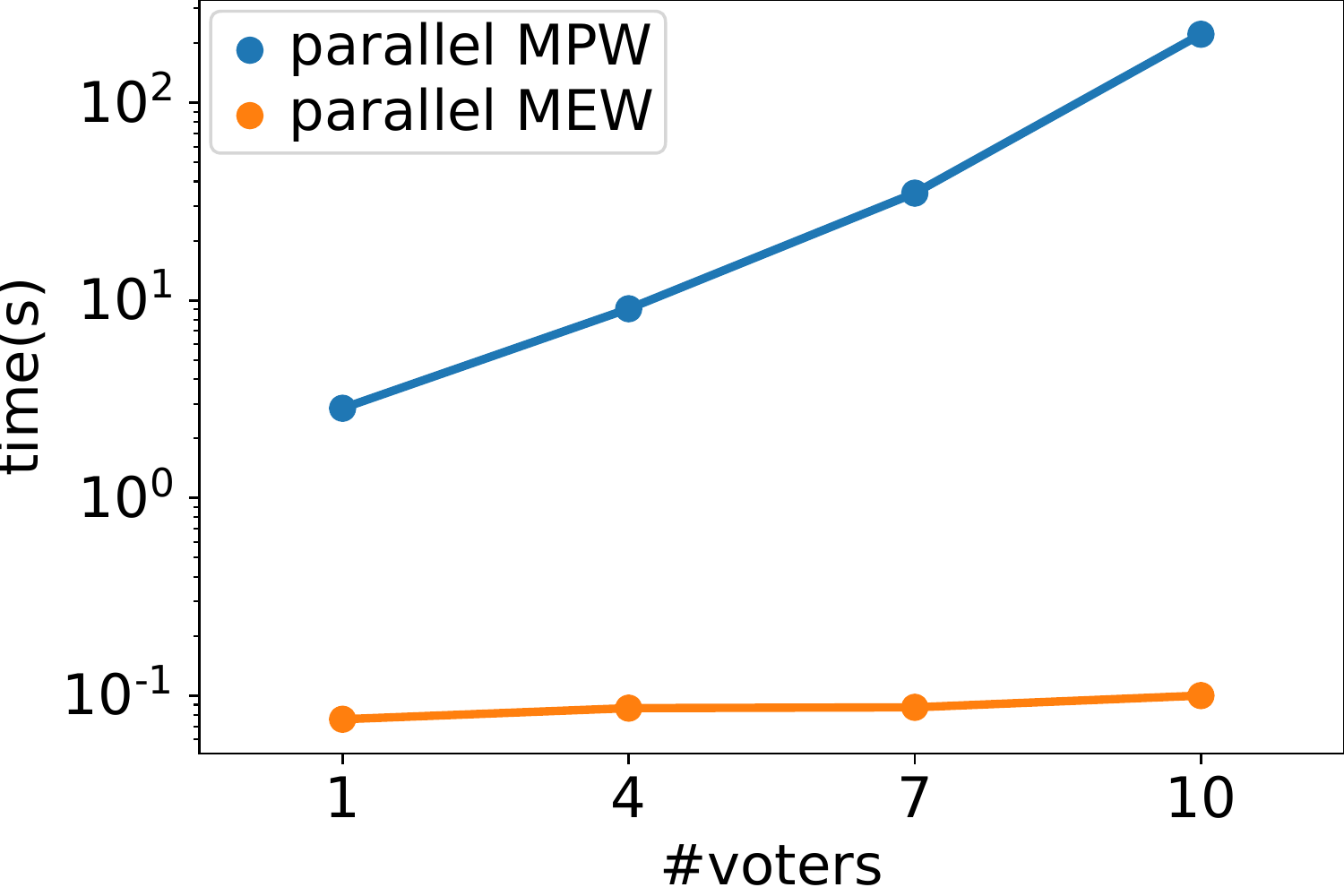}
	}
	\caption{\revv{Average time of parallel MPW and MEW, using 48 worker processes,  over partial voting profiles, fixing $\phi = 0.5$ and $p_{max}=0.1$, using the plurality rule. \mew scales much better than \mpw,  with both \#candidates and \# voters.}}
	\label{fig:mpw}
\end{figure}

\begin{example}
  \rev{Let $\probaVP = \set{\model_0}$ be a single-voter voting profile over 4 candidates $\set{a, b, c, d}$. 
  Assume $\model_0$ is a ranking model where $\Pr(\angs{b,a,c,d}) = \Pr(\angs{c,a,b,d}) = \Pr(\angs{d,a,b,c}) = 1/3$.
  Under Borda rule, \mew favors candidate $a$ who, despite losing in every possible world, enjoys the highest expected score of 2, while the MPWs are $\set{b, c, d}$, each winning in one possible world.}
\end{example}

\rev{\mew also quantifies the performance of candidates more granularly than \mpw, as illustrated in the next example.}

\begin{example}
  \rev{Let $\partialVP = \set{\partialOrder_0}$ be a single-voter voting profile over 4 candidates $\set{a, b, c, d}$. 
    Assume $\partialOrder_0 = \set{a \succ b, b \succ c, b \succ d}$.
    Under the Borda rule, \mew and \mpw agree that candidate $a$ is the winner, but they disagree on the performance of candidate $b$. \mpw cannot differentiate between $b$, $c$, and $d$, since none of them wins in any possible world, but \mew believes that $b$ outperforms $c$ and $d$, since $b$ has a higher expected score than $c$ and $d$.}
\end{example}

\rev{\mew also has a computational advantage over \mpw, making it practical for real-world applications: Although both are intractable in the general case, \mew enjoys linear complexity in the number of voters, while the running time of \mpw grows exponentially. We implemented the \mpw solver based on the \e{VotingResult} algorithm by Hazon \etal \cite{DBLP:journals/ai/HazonAKW12}, with the modification that the solver iterates over the completions of each voter's partial order, while the original algorithm assumes that each voter gives an explicit probability distribution over the rankings.} \revv{We compared performance of the parallel versions of \mpw and \mew under plurality, fixing the number of voters to 10 and varying the number of candidates from 3 to 9 (Figures~\ref{fig:mpw:candidates}), then fixing the number of candidates to 9 and varying the number of voters from 1 to 10 (Figure~\ref{fig:mpw:voters}). Observe that \mew scales much better than \mpw.}
\revv{The corresponding results under the Borda rule are available in Appendix~\ref{sec:appendix:experiments}.}

\section{Concluding Remarks}
\label{sec:conc}
 
In this paper we modeled uncertainty in voter preferences with the help of a framework that distinguishes between uncertainty in preference generation and uncertainty in preference observation, unifying incomplete and probabilistic voting profiles.  
We then proposed the \mEw (\mew) semantics for  positional scoring rules and established the theoretical hardness of this problem.  We identified tractable cases with the help of the uncertainty framework for voting profiles, and developed solvers.

Much exciting future work remains.  For example, the hardness of \esc is proved over only plurality, veto, and $k$-approval, which calls for investigation of other positional scoring rules such as Borda. When  \mew is intractable, it may be necessary to develop approximate solvers. \mew can also be extended to other score-based rules, such as Simpson and Copeland.
Another direction is to consider voter preferences represented by additional ranking models~\cite{marden1995analyzing} such as the Plackett-Luce (PL)~\cite{luce1959individual,plackett1975analysis} and  Thurstone-Mosteller (TM)~\cite{Thurstone1927-THUALO-2,RePEc:spr:psycho:v:16:y:1951:i:1:p:3-9}. For example, others~\cite{DBLP:conf/aaai/NoothigattuGADR18,DBLP:conf/uai/ZhaoLWKMSX18} have studied a preference aggregation method over PL models that is closely related to the \mew over PL models, and we plan to investigate this connection further in the future.

\bibliographystyle{ACM-Reference-Format}
\bibliography{references}

\newpage 
\appendix

\section{Additional proofs}
\label{sec:appendix:proofs}

\begin{reptheorem}{\ref{theorem:least_expected_regret_winner}}
    \theoremLeastExpectedRegretWinner
\end{reptheorem}

\begin{proof}
    Let $\probaVP$ denote a general voting profile with $\Omega(\probaVP)=\set{\enum[\numPW]{\VP}}$.
    The expected regret of a candidate $w$ can be rewritten as follows.
    
    \begin{align*}
        & \mathds{E}(\txt{Regret}(w, \probaVP)) \\
        =& \sum_{i=1}^z \txt{Regret}(c, \completeVP_i) \cdot \Pr(\completeVP_i \mid \probaVP) \\
        =& \sum_{i=1}^z \bigg( \max_{c \in C} \score(c, \completeVP_i) - \score(w, \completeVP_i) \bigg) \cdot \Pr(\completeVP_i \mid \probaVP) \\
        =& \sum_{i=1}^z \max_{c \in C} \score(c, \completeVP_i) \cdot \Pr(\completeVP_i \mid \probaVP)  - \mathds{E}(\score(w, \probaVP))\\
    \end{align*}
    
    The first term $\sum_{i=1}^z \max_{c \in C} \score(c, \completeVP_i) \cdot \Pr(\completeVP_i \mid \probaVP)$ is a constant value, when $\probaVP$ and the voting rule are fixed.
    Thus, $\mathds{E}(\txt{Regret}(w, \probaVP))$ is minimized by maximizing $\mathds{E}(\score(w, \probaVP))$, the expected score of the candidate $w$.
\end{proof}

\begin{reptheorem}{\ref{theorem:gigantic_election_winner}}
    \theoremGiganticElectionWinner
\end{reptheorem}

\begin{proof}
    Let $\probaVP$ denote a general voting profile with $\Omega(\probaVP)=\set{\enum[\numPW]{\VP}}$, and $\completeVP_{meta} = (\completeVP_1, \ldots, \completeVP_\numPW)$ denote the large meta profile where rankings in $\completeVP_i$ are weighted by $\Pr(\completeVP_i \mid \probaVP)$.
    According to the definition of the Meta-Election Winner, $\score(w, \completeVP_{meta}) = \max_{c \in C} \score(c, \completeVP_{meta})$.
    As a result, for any candidate $c$,
    \[
    \mathds{E}(\score(c, \probaVP)) = \sum_{\completeVP \in \Omega(\probaVP)} \score(c, \completeVP) \cdot \Pr(\completeVP \mid \probaVP) = \score(c, \completeVP_{meta})
    \]
    Her expected score in $\probaVP$ is precisely her score in $\completeVP_{meta}$.
    The two winner definitions are optimizing the same metric.
\end{proof}

\begin{reptheorem}{\ref{theorem:FCP_shaPcomplete_over_partialOrders}}
    \theoremFCPhardnessOverPartialOrders
\end{reptheorem}

\begin{proof}
    First, we prove its membership in \#P.
    The FCP is the counting version of the following decision problem: given a partial order $\partialOrder$, an item $c$, and an integer $j$, determine whether $\partialOrder$ has a linear extension $\ranking \in \Omega(\partialOrder)$ where $c$ is ranked at $j$.
    This decision problem is obviously in NP, meaning that the FCP is in \#P.
    
    Then, we prove that the FCP is \#P-hard by reduction.
    Recall that counting $|\Omega(\partialOrder)|$, the number of linear extensions of a partial order $\partialOrder$, is \#P-complete~\cite{DBLP:conf/stoc/BrightwellW91}.
    This problem can be reduced to the FCP by $|\Omega(\partialOrder)| = \sum_{j=1}^{m}{N(c@j \mid \partialOrder)}$.
    
    In conclusion, the FCP is \#P-complete.
\end{proof}

\begin{replemma}{\ref{lemma:REPt_shaPcomplete_over_partialOrders}}
    \lemmaREPtHardnessOverPartialOrders
\end{replemma}

\begin{proof}
    First, we prove that the REP-t is in FP$^{\#P}$.
    Recall that $\Omega(\partialOrder)$ is the linear extensions of a partial order $\partialOrder$, and $N(c@1 {\mid} \partialOrder)$ is the number of linear extensions in $\Omega(\partialOrder)$ where candidate $c$ is at rank 1.
    Then $\Pr(c@1 {\mid} \partialOrder) = N(c@1 {\mid} \partialOrder) / |\Omega(\partialOrder)|$.
    Consider that counting $N(c@1 \mid \partialOrder)$ is in \#P (Theorem~\ref{theorem:FCP_shaPcomplete_over_partialOrders}) and counting $|\Omega(\partialOrder)|$ is \#P-complete~\cite{DBLP:conf/stoc/BrightwellW91}, so $\Pr(c@1 {\mid} \partialOrder)$ is in FP$^{\#P}$.
    
    In the rest of this proof, we prove that the REP-t is \#P-hard by reduction from the \#P-complete problem of counting $|\Omega(\partialOrder)|$.
    
    Let $c^*$ denote an item that has no parent in $\partialOrder$.
    Let $\partialOrder_{-c^*}$ denote the partial order of $\partialOrder$ with item $c^*$ removed.
    If we are interested in the probability that $c^*$ is placed at rank 1, we can write $\Pr(c^*@1 {\mid} \partialOrder) = N(c^*@1 \mid \partialOrder) / |\Omega(\partialOrder)|$.
    The item $c^*$ has been fixed at rank 1, so any placement of the rest items will definitely satisfy any relative order involving $c^*$.
    That is to say, the placement of the rest items just needs to satisfy $\partialOrder_{-c^*}$, which leads to $N(c^*@1 \mid \partialOrder) = |\Omega(\partialOrder_{-c^*})|$.
    
    For example, let $\partialOrder' = \set{c_1 \succ c_4, c_2 \succ c_4, c_3 \succ c_4}$.   Then $N(c_1@1 \mid \partialOrder') = |\Omega(\partialOrder'_{-c_1})| = |\Omega(\set{c_2 \succ c_4, c_3 \succ c_4})|$.
    Then, we re-write $\Pr(c^*@1 {\mid} \partialOrder) = N(c^*@1 {\mid} \partialOrder) / |\Omega(\partialOrder)| = |\Omega(\partialOrder_{-c^*})| / |\Omega(\partialOrder)|$.
    The oracle for $\Pr(c^*@1 {\mid} \partialOrder)$ manages to reduce the size of the counting problem from $|\Omega(\partialOrder)|$ to $|\Omega(\partialOrder_{-c^*})|$.
    This oracle should be as hard as counting $|\Omega(\partialOrder)|$.
    Thus calculating $\Pr(c^*@1 {\mid} \partialOrder)$ is FP$^{\#P}$-hard.
    
    In conclusion, the REP-t is FP$^{\#P}$-complete.
\end{proof}

\begin{replemma}{\ref{lemma:REPb_shaPcomplete_over_partialOrders}}
    \lemmaREPbHardnessOverPartialOrders
\end{replemma}

\begin{proof}
    This proof adopts the same approach as the proof of Lemma~\ref{lemma:REPt_shaPcomplete_over_partialOrders}.
    
    Let $m$ be the number of items in the ranking model $\model$.
    For the membership proof that the REP-b is in FP$^{\#P}$, let $N(c@m {\mid} \partialOrder)$ denote the number of linear extensions in $\Omega(\partialOrder)$ where candidate $c$ is at the bottom rank $m$.
    Then $\Pr(c@m {\mid} \partialOrder) = N(c@m {\mid} \partialOrder) / |\Omega(\partialOrder)|$.
    Consider that counting $N(c@m {\mid} \partialOrder)$ is in \#P (Theorem~\ref{theorem:FCP_shaPcomplete_over_partialOrders}) and counting $|\Omega(\partialOrder)|$ is \#P-complete~\cite{DBLP:conf/stoc/BrightwellW91}, so $\Pr(c@m \mid \partialOrder)$ is in FP$^{\#P}$.
    
    In the proof of Lemma~\ref{lemma:REPt_shaPcomplete_over_partialOrders}, item $c^*$ is an item with no parent in the partial order $\partialOrder$.
    In the current proof, item $c^*$ is set to be an item with no child in $\partialOrder$.
    The $\partialOrder_{-c^*}$ still denotes the partial order of $\partialOrder$ but with item $c^*$ removed.
    Then the probability that item $c^*$ at the bottom rank $m$ is $\Pr(c^*@m \mid \partialOrder) = N(c^*@m \mid \partialOrder) / |\Omega(\partialOrder)| = |\Omega(\partialOrder_{-c^*})| / |\Omega(\partialOrder)|$.
    The oracle for $\Pr(c^*@m \mid \partialOrder)$ manages to reduce the size of the counting problem again from $|\Omega(\partialOrder)|$ to $|\Omega(\partialOrder_{-c^*})|$.
    Thus, this oracle is \#P-hard, and calculating $\Pr(c^*@m \mid \partialOrder)$ is FP$^{\#P}$-hard.
    
    In conclusion, the REP-b is FP$^{\#P}$-complete.
\end{proof}

\begin{reptheorem}{\ref{theorem:REP_shaPcomplete_over_partialOrders}}
    \theoremRepHardnessOverPartialOrders
\end{reptheorem}

\begin{proof}
    First, we prove that the REP is in FP$^{\#P}$.
    Recall that $\Omega(\partialOrder)$ is the linear extensions of a partial order $\partialOrder$, and $N(c@j {\mid} \partialOrder)$ is the number of linear extensions in $\Omega(\partialOrder)$ where candidate $c$ is at rank $j$.
    Then $\Pr(c@j {\mid} \partialOrder) = N(c@j {\mid} \partialOrder) / |\Omega(\partialOrder)|$.
    Consider that counting $N(c@j {\mid} \partialOrder)$ is \#P-complete (Theorem~\ref{theorem:FCP_shaPcomplete_over_partialOrders}) and counting $|\Omega(\partialOrder)|$ is \#P-complete~\cite{DBLP:conf/stoc/BrightwellW91} as well.
    So $\Pr(c@j {\mid} \partialOrder)$ is in FP$^{\#P}$.
    
    Lemma~\ref{lemma:REPt_shaPcomplete_over_partialOrders} demonstrates that REP-t, a special case of REP, is FP$^{\#P}$-hard.
    Thus REP is \#P-hard as well.
    
    In conclusion, REP is FP$^{\#P}$-complete.
\end{proof}

\begin{reptheorem}{\ref{theorem:reduction}}
    \theoremReductionEscToRep
\end{reptheorem}

\begin{proof}
    Recall that the \mew $w$ maximizes the expected score, \ie
    \[
    \score(w, \probaVP) = \max_{c \in C}\mathds{E}(\score(c, \probaVP))
    \]
    The voting profile $\probaVP$ contains $n$ ranking distributions $\set{\enum[n]{\model}}$, so
    \[
    \mathds{E}(\score(c, \probaVP)) = \sum_{i=1}^{n} {\mathds{E}(\score(c, \model_i))}
    \]
    where $\mathds{E}(\score(c, \model_i))$ is the expected score of $c$ from voter $v_i$.
    \[
    \mathds{E}(\score(c, \model_i)) = \sum_{j=1}^{m} Pr(c@j \mid \model_i) \cdot \vr_m(j)
    \]
    where $c@j$ denotes candidate $c$ at rank $j$, and $\vr_m(j)$ is the score of rank $j$.
    
    Let $\mathds{T}$ denote the complexity of calculating $Pr(c@j {\mid} \model_i)$.
    The original \mew problem can be solved by calculating $Pr(c@j {\mid} \model_i)$ for all $m$ candidates, $m$ ranks and $n$ voters, which leads to the complexity of $O(n\cdot m^2 \cdot \mathds{T})$.
\end{proof}

\begin{reptheorem}{\ref{theorem:REP_MEW_equivalent_under_k_approval}}
    \theoremRepMewEquivalentUnderKapproval
\end{reptheorem}

\begin{proof}
    The \esc problem has been reduced to the REP (Theorem~\ref{theorem:reduction}).
    This proof will focus on the other direction, \ie reducing the REP to the \esc problem.
    
    Let $\Pr(c@j \mid \model)$ denote the probability of placing candidate $c$ at rank $j$ over a ranking distribution $\model$.
    Let $\VP^{\model}$ denote a single-voter profile consisting of only this ranking distribution $\model$.
    
    When $k=1$, the REP can be reduced to solving the \esc problem under plurality or $1$-approval rule.
    \[
    \Pr(c@1 \mid \model) = \mathds{E}(\score(c \mid \VP^{\model}, 1\text{-approval}))
    \]
    
    When $k=m$, the REP can be reduced to solving the \esc problem under veto or $(m {-} 1)$-approval rule.
    \[
    \Pr(c@m \mid \model) = 1 - \mathds{E}(\score(c \mid \VP^{\model}, (m {-} 1)\text{-approval}))
    \]
    
    When $2 \leq k \leq m$, the REP can be reduced to solving the \esc problem twice under $k$-approval and $(k-1)$-approval rules.
    \begin{align*}
        \Pr(c@k \mid \model) 
        &= \mathds{E}(\score(c \mid \VP^{\model}, k\text{-approval})) \\
        &- \mathds{E}(\score(c \mid \VP^{\model}, (k-1)\text{-approval}))
    \end{align*}
\end{proof}

\begin{reptheorem}{\ref{theorem:shaPcomplete_for_expected_score_given_partialOrder_and_plurality}}
    \theoremHardnessForEscGivenPartialOrderAndPlurality
\end{reptheorem}

\begin{proof}
    Firstly, we prove the membership of the \esc problem as an FP$^{\#P}$ problem.
    Consider that the REP is FP$^{\#P}$-complete over partial orders (Theorem~\ref{theorem:REP_shaPcomplete_over_partialOrders}), and the \esc problem can be reduced to the REP (Theorem~\ref{theorem:reduction})
    So the \esc problem is in FP$^{\#P}$ for partial voting profiles.
    
    Secondly, we prove that the \esc problem is FP$^{\#P}$-hard, even for plurality rule, by  reduction from the REP-t that is FP$^{\#P}$-hard (Lemma~\ref{lemma:REPt_shaPcomplete_over_partialOrders}).
    
    Let $\partialOrder$ denote the partial order of the REP-t problem.
    Recall that the REP-t problem aims to calculate $\Pr(c@1 \mid \partialOrder)$ for a given item $c$.
    Let $\VP^{\partialOrder}$ denote a voting profile consisting of just this partial order $\partialOrder$.
    The answer to the REP-t problem is the same as the answer to the corresponding \esc problem, \ie $\Pr(c@1 \mid \partialOrder) = \mathds{E}(\score(c \mid \VP^{\partialOrder}, \text{plurality}))$.
    So the \esc problem is FP$^{\#P}$-hard, even for plurality voting rule.
    
    In conclusion, the \esc problem is FP$^{\#P}$-complete, under plurality rule.
\end{proof}

\begin{reptheorem}{\ref{theorem:shaPcomplete_for_expected_score_given_partialOrder_and_veto}}
    \theoremHardnessForEscGivenPartialOrderAndVeto
\end{reptheorem}

\begin{proof}
    This proof adopts the same approach as the proof of Theorem~\ref{theorem:shaPcomplete_for_expected_score_given_partialOrder_and_plurality}.
    
    Firstly, the membership proof that the ESC is in FP$^{\#P}$ is based on the conclusions that the REP is FP$^{\#P}$-complete over partial orders (Theorem~\ref{theorem:REP_shaPcomplete_over_partialOrders}), and that the ESC can be reduced to the REP (Theorem~\ref{theorem:reduction})
    So the ESC is in FP$^{\#P}$ for partial voting profiles.
    
    Secondly, we prove that the ESC is FP$^{\#P}$-hard, under veto voting rule, by reduction from the REP-b that is FP$^{\#P}$-hard (Lemma~\ref{lemma:REPb_shaPcomplete_over_partialOrders}).
    
    Let $\partialOrder$ denote the partial order of the REP-b problem.
    Recall that the REP-b problem aims to calculate $\Pr(c@m \mid \partialOrder)$ for a given item $c$.
    Let $\VP^{\partialOrder}$ denote a voting profile consisting of just this partial order $\partialOrder$.
    The answer to the ESC indirectly solves the REP-b, \ie $\Pr(c@m \mid \partialOrder) = 1 - \mathds{E}(\score(c \mid \VP^{\partialOrder}, \text{veto}))$.
    So the ESC problem is FP$^{\#P}$-hard under veto rule.
    
    In conclusion, the ESC is FP$^{\#P}$-complete, under veto rule.
\end{proof}

\begin{reptheorem}{\ref{theorem:shaPcomplete_for_expected_score_given_partialOrder_and_k_approval}}
    \theoremHardnessForExpectedScoreGivenPartialOrderAndKApproval
\end{reptheorem}

\begin{proof}
    Firstly, the proof that the Expected Score Computation (ESC) is in FP$^{\#P}$ is the same as the proof of Theorem~\ref{theorem:shaPcomplete_for_expected_score_given_partialOrder_and_plurality}.
    Now we prove that the ESC problem is FP$^{\#P}$-hard, under $k$-approval rule $\vr_m$, by reduction from the REP-t problem that is FP$^{\#P}$-hard (Lemma~\ref{lemma:REPt_shaPcomplete_over_partialOrders}).
    
    Let $\partialOrder$ denote the partial order of the REP-t problem.
    Recall that the REP-t problem aims to calculate $\Pr(c@1 \mid \partialOrder)$ for a given item $c$.
    Let $\partialOrder_+$ denote a new partial order by inserting $(k - 1)$ ordered items $d_1 \succ \ldots \succ d_{k-1}$ into $\partialOrder$ such that item $d_{k-1}$ is preferred to every item in $\partialOrder$.
    Such placement of items $\set{\enum[k-1]{d}}$ is to guarantee that all linear extensions of $\partialOrder_+$ start with $d_1 \succ \ldots \succ d_{k-1}$ and these linear extensions will be precisely the linear extensions of $\partialOrder$ after removing $\set{\enum[k-1]{d}}$.
    
    Let $\VP^{\partialOrder_+}$ denote a voting profile consisting of just this partial order $\partialOrder_+$.
    The answer to the ESC problem for item $c$ is $\mathds{E}(\score(c \mid \VP^{\partialOrder_+}, k\text{-approval}))$.
    Since there is only one partial order $\partialOrder_+$ in the voting profile, $\mathds{E}(\score(c \mid \VP^{\partialOrder_+}, k\text{-approval})) = \sum_{j=1}^{k} \Pr(c@j \mid \partialOrder_+)$.
    Recall that any linear extension of $\partialOrder_+$ always starts with $d_1 \succ \ldots \succ d_{k-1}$, so $\forall 1 \leq j \leq (k-1), \Pr(c@j \mid \partialOrder_+) = 0$, which leads to $\mathds{E}(\score(c \mid \VP^{\partialOrder_+}, k\text{-approval})) = \Pr(c@k \mid \partialOrder_+)$.
    Since $\partialOrder_+$ is constructed by inserting $(k-1)$ items before items in $\partialOrder$, $\Pr(c@k \mid \partialOrder_+) = \Pr(c@1 \mid \partialOrder)$.
    So $\mathds{E}(\score(c \mid \VP^{\partialOrder_+}, k\text{-approval})) = \Pr(c@1 \mid \partialOrder)$.
    The answer to the REP-t problem has been reduced to the ESC problem.
    So the ESC problem is FP$^{\#P}$-hard, under $k$-approval rule.
    
    In conclusion, the ESC problem is FP$^{\#P}$-complete, under the $k$-approval rule.
\end{proof}

\begin{reptheorem}{\ref{theorem:tractability_of_fullparVP}}
  \theoremTractabilityOfFP
\end{reptheorem}

\begin{proof}
    Any $\fullpar \in \fullparVP$ defines a set of consecutive ranks in the linear extensions of $\fullpar$ for each of its partitions of candidates.
    Any candidate is equally likely to be positioned at these ranks.
    So the REP can be solved in $O(1)$ for any candidate.
    Thus, the \mew problem can be solved in $O(nm^2)$ by calculating the expected scores of all candidates.
\end{proof}

\begin{reptheorem}{\ref{theorem:tractability_of_pchainVP}}
  \theoremTractabilityOfPC
\end{reptheorem}

\begin{proof}
  For any $\pchain \in \pchainVP$ and any candidate $c$, the $\Pr(c {\rightarrow} j \mid \pchain)$ is proportional to the degree of freedom to place the rest of the candidates, after fixing $c$ at rank $j$.
  \begin{itemize}
     \item If $c \not\in \pchain$, this is a trivial case where $c$ is equally likely to be placed at any rank, thus $\forall 1 \leq j \leq m, \Pr(c {\rightarrow} j \mid \pchain) = 1/m$.
    \item If $c \in \pchain$, let $K_l = |\set{c' \mid c' \succ_{\pchain} c}|$ be the number of items preferred to $c$ by $\pchain$ and $K_r = |\set{c' \mid c \succ_{\pchain} c'}|$ be the number of items less preferred to $c$ by $\pchain$, then $\Pr(c {\rightarrow} j \mid \pchain) \propto \binom{j-1}{K_l} \cdot \binom{m - j}{K_r}$ where .
  \end{itemize}

  It takes $O(nm^2)$ to obtain the expected scores of all candidates and to determine whether $w$ is a \mew.
\end{proof}

\begin{reptheorem}{\ref{theorem:tractability_of_parparVP}}
  \theoremTractabilityOfPP
\end{reptheorem}

\begin{proof}
  For any $\parpar \in \parparVP$ and any candidate $c$, the $\Pr(c {\rightarrow} j \mid \parpar)$ is proportional to the degree of freedom to place the rest of the candidates, after fixing $c$ at rank $j$.

 \begin{itemize}
     \item If $c \not\in \parpar$, this is a trivial case where $c$ is equally likely to be placed at any rank, thus $\forall 1 \leq j \leq m, \Pr(c {\rightarrow} j \mid \pchain) = 1/m$.
     \item If $c \in \parpar$, let $K_l = |\set{c' \mid c' \succ_{\parpar} c}|$ be the number of items preferred to $c$ by $\parpar$, $K_r = |\set{c' \mid c \succ_{\parpar} c'}|$ be the number of items less preferred to $c$ by $\parpar$, and $K_c$ be the number of items in the partition of $c$, then $\Pr(c {\rightarrow} j \mid \pchain) \propto \sum_{x=0}^{K_c - 1} \binom{j-1}{K_l + x} \cdot \binom{m - j}{K_r + K_c - 1 - x}$ where $x$ is the number of items from the same partition as $c$ and placed to the left of $c$.
   \end{itemize}

  It takes $O(nm^2)$ to obtain the expected scores of all candidates and to determine whether $w$ is a \mew.
\end{proof}

\begin{reptheorem}{\ref{theorem:tractability_of_rimVP}}
  \theoremTractabilityOfRIM
\end{reptheorem}

\begin{proof}
    Given any $\RIM \in \rimVP$ and any candidate $c$, the $Pr(c {\rightarrow} j \mid \RIM)$ for $j=1,\ldots,m$ can be calculated by Algorithm~\ref{alg:rim_rank_estimation} in $O(m^3)$.
    Algorithm~\ref{alg:rim_rank_estimation} is a variant of RIMDP~\cite{DBLP:conf/aaai/KenigIPKS18}.
    RIMDP calculates the marginal probability of a partial order over RIM via Dynamic Programming (DP).
    Algorithm~\ref{alg:rim_rank_estimation} is simplified RIMDP in the sense that Algorithm~\ref{alg:rim_rank_estimation} only tracks a particular item $c$, while RIMDP tracks multiple items to calculate the insertion ranges of items that satisfy the partial order.
    Note that Algorithm~\ref{alg:rim_rank_estimation} calculates all $m$ different values of $j$ simultaneously.
    So it takes $O(nm \cdot m^3) = O(nm^4)$ to obtain the expected scores of $m$ candidates over $n$ RIMs to determine \mew.
\end{proof}

\begin{reptheorem} {\ref{theorem:tractability_of_rimTrunVP}}
  \theoremTractabilityOfRimTrun
\end{reptheorem}

\begin{proof}
    Given any $(\RIM, \ranking^{(t, b)}) \in \rimTrunVP$, candidate $c$, and rank $j$, if $c$ is in the top or bottom part of $\ranking^{(t, b)}$, its rank has been fixed, which is a trivial case;
    If $c$ is in the middle part of $\ranking^{(t, b)}$, we just need to slightly modify Algorithm~\ref{alg:rim_rank_estimation} to calculate $\Pr(c {\rightarrow} j \mid \RIM, \ranking^{(t, b)})$.
    Line~\ref{alg:rim_rank_estimation:j} in Algorithm~\ref{alg:rim_rank_estimation} enumerates values for $j$ from 1 to $i$.
    The constraints made by $\ranking^{(t, b)}$ limits this insertion range of item $\bsigma(i)$.
    If $\bsigma(i)$ is in the top or bottom part of $\ranking^{(t, b)}$, its insertion position has been fixed by $\ranking^{(t, b)}$ and the inserted items of the top and bottom parts of $\ranking^{(t, b)}$ should be recorded as well by the state $\delta'$;
    If $\bsigma(i)$ is in the middle part of $\ranking^{(t, b)}$, $\bsigma(i)$ can be inserted into any position between the inserted top and bottom items.

    Theoretically, the algorithm needs to track as many as $(t+b+1)$ items.
    But $(t+b)$ items are fixed, which makes $c$ the only item leading to multiple DP states.
    The complexity of calculating $\Pr(c {\rightarrow} j \mid \RIM, \ranking^{(t, b)})$ for all $j$ values is $O(m^3)$.
    It takes $O(nm^4)$ to calculate the expected scores of all candidates across all voters to determine the \mew.
\end{proof}

\begin{reptheorem}{\ref{theorem:tractability_of_mallowsPartitionVP}}
  \theoremTractabilityOfMallowsFP
\end{reptheorem}

\begin{proof}
    Given any $(\mallows(\bsigma, \phi), \fullpar) \in \mallowsPartitionVP$, candidate $c$, and rank $j$, consider calculating $Pr(c {\rightarrow} j \mid \bsigma, \phi, \fullpar)$.
    Let $C_P$ denote the set of candidates in the same partition with $c$ in $\fullpar$.
    The relative orders between $c$ and items out of $C_P$ are already determined by $\fullpar$.
    That is to say, for a non-trivial $j$ value, $Pr(c {\rightarrow} j \mid \bsigma, \phi, \fullpar)$ is proportional to the exponential of the number of disagreed pairs within $C_P$.
    So we can construct a new Mallows model $\mallows'(\bsigma', \phi)$ over $C_P$.
    It has the same $\phi$ as $\mallows$ and its reference ranking $\bsigma'$ is shorter than but consistent with $\bsigma$.
    The $Pr(c {\rightarrow} j \mid \mallows', \fullpar)$ for all non-trivial $j$ values can be calculated in $O(|C_P|^3) < O(m^3)$ by Algorithm~\ref{alg:rim_rank_estimation}.

    The \mew problem can be solved in $O(nm^4)$ by calculating the expected scores of all candidates across all voters to determine whether $w$ is a \mew.
\end{proof}

\section{Tractability over RSM profiles}
\label{sec:appendix:rsm_profile}

RSM~\cite{DBLP:journals/tdasci/ChakrabortyDKKR21} denoted by $\mathsf{RSM}(\bsigma, \Pi, p)$ is another generalization of the Mallows.
It is parameterized by a reference ranking $\bsigma$, a probability function $\Pi$ where $\Pi(i, j)$ is the probability of the $j^{th}$ item selected at step $i$, and a probability function $p:\set{1,...,m-1} \rightarrow [0,1]$ where $p(i)$ is the probability that the $i^{th}$ selected item preferred to the remaining items.
In contrast to the RIM that randomizes the item insertion position, the RSM randomized the item insertion order.
In this paper, we use RSM as a ranking model, \ie $p \equiv 1$ such that it only outputs rankings.
This ranking version is named rRSM and denoted by $\RSM(\bsigma, \Pi)$.

\begin{example}
    $\RSM(\bsigma, \Pi)$ with $\bsigma=\angs{a, b, c}$ generates $\ranking {=} \angs{c, a, b}$ as follows.
    Initialize $\ranking_0 {=} \angs{}$. 
    When $i=1$, $\ranking_1 {=} \angs{c}$ by selecting $c$ with probability $\Pi(1,3)$, making the remaining $\bsigma = \angs{a, b}$.
    When $i=2$, $\ranking_2 {=} \angs{c, a}$ by selecting $a$ with probability $\Pi(2,1)$, making the remaining $\bsigma = \angs{b}$.
    When $i=3$, $\ranking {=} \angs{c, a, b}$ by selecting $b$ with probability $\Pi(3,1)$. 
    Overall, $\Pr(\ranking \mid \bsigma, \Pi) {=} \Pi(1,3) \cdot \Pi(2,1) \cdot \Pi(3,1)$.
\end{example}

\begin{theorem} \label{theorem:tractability_of_rsmVP}
    Given a positional scoring rule $\vr_m$, an RSM voting profile $\rsmVP = (\enum[n]{\RSM})$, and candidate $w$, determining $w \in \mew(\vr_m, \rsmVP)$ is in $O(nm^4)$.
\end{theorem}

\begin{proof}
    Given any $\RSM \in \rsmVP$, candidate $c$, and rank $j$, the $Pr(c@j \mid \RSM)$ is computed by Algorithm~\ref{alg:rsm_rank_estimation} in a fashion that is similar to Algorithm~\ref{alg:rim_rank_estimation}.
    This is also a Dynamic Programming (DP) approach.
    The states are in the form of $\angs{\alpha, \beta}$, where $\alpha$ is the number of items before $c$, and $\beta$ is that after $c$ in the remaining $\bsigma$.
    For state $\angs{\alpha, \beta}$, there are $(\alpha + 1 + \beta)$ items in the remaining $\bsigma$.
    Algorithm~\ref{alg:rsm_rank_estimation} only runs up to $i=(k-1)$ (in line~\ref{alg:rsm_rank_estimation:step}), since item $c$ must be selected at step $k$ and the rest steps do not change the rank of $c$ anymore.
    Each step $i$ generates at most $(i + 1)$ states, corresponding to $[0, \ldots, i]$ items are selected from items before $c$ in the original $\bsigma$.
    The complexity of Algorithm~\ref{alg:rsm_rank_estimation} is bounded by $O(m^2)$.
    It takes $O(nm^4)$ to obtain the expected scores of all candidates and to determine the \mew.
    
\end{proof}

\begin{algorithm}[tb!]
    \caption{REP solver for rRSM}
    \label{alg:rsm_rank_estimation}
    \textbf{Input}: Item $c$, rank $k$, $\RSM(\bsigma, \Pi)$ 
    \textbf{Output}: $\Pr(c@k \mid \bsigma, \Pi)$
    \begin{algorithmic}[1] 
        \STATE $\alpha_0 \defeq |\set{\sigma_i | \sigma_i \succ_{\bsigma} c}|$, $\beta_0 \defeq |\set{\sigma_i \mid c \succ_{\bsigma} \sigma_i}|$
        \STATE $\mathcal{P}_0 \defeq \set{\angs{\alpha_0, \beta_0}}$ and $q_0(\angs{\alpha_0, \beta_0}) \defeq 1$
        \FOR {$i=1, \ldots, (k - 1)$} \label{alg:rsm_rank_estimation:step}
        \STATE $\mathcal{P}_i \defeq \set{}$
        \FOR {$\angs{\alpha, \beta} \in \mathcal{P}_{i-1}$}
        \IF {$\alpha > 0$}
        \STATE Generate a new state $\angs{\alpha', \beta'} = \angs{\alpha - 1, \beta}$.
        \IF {$\angs{\alpha', \beta'} \notin \mathcal{P}_i$}
        \STATE $\mathcal{P}_i.add(\angs{\alpha', \beta'})$
        \STATE $q_i(\angs{\alpha', \beta'}) \defeq 0$
        \ENDIF
        \STATE $q_i(\angs{\alpha', \beta'}) \pluseq q_{i-1}(\angs{\alpha, \beta}) \cdot \sum_{j=1}^{\alpha} {\Pi(i, j)}$
        \ENDIF
        \IF {$\beta > 0$}
        \STATE Generate a new state $\angs{\alpha', \beta'} = \angs{\alpha, \beta - 1}$.
        \IF {$\angs{\alpha', \beta'} \notin \mathcal{P}_i$}
        \STATE $\mathcal{P}_i.add(\angs{\alpha', \beta'})$
        \STATE $q_i(\angs{\alpha', \beta'}) \defeq 0$
        \ENDIF
        \STATE $q_i(\angs{\alpha', \beta'}) \pluseq q_{i-1}(\angs{\alpha, \beta}) \cdot \sum_{j=\alpha + 2}^{\alpha + 1 + \beta} {\Pi(i, j)}$
        \ENDIF
        \ENDFOR
        \ENDFOR
        \RETURN  $\sum_{\angs{\alpha, \beta} \in \mathcal{P}_{k-1}} {q_{k - 1}(\angs{\alpha, \beta}) \cdot \Pi(k, \alpha + 1)}$
    \end{algorithmic}
\end{algorithm}

\begin{example}
Let $\RSM(\bsigma, \Pi)$ denote a RSM where $\bsigma=\angs{\sigma_1, \sigma_2, \sigma_3, \sigma_4}$, and $\Pi = [[0.1, 0.3, 0.4, 0.2],$ $[0.2, 0.5, 0.3], [0.3, 0.7],[1]]$.
Assume we are interested in $\Pr(\sigma_2 @ 3 \mid \bsigma, \Pi)$, the probability of item $\sigma_2$ placed at rank $3$.
\begin{itemize}
    \itemsep -0.1em
    \item Before running RSM, there is $\alpha_0 = 1$ item before $\sigma_2$ and $\beta_0 = 2$ items after $\sigma_2$ in $\bsigma$. So the initial state is $\angs{\alpha_0, \beta_0} = \angs{1, 2}$, and $q_0(\angs{1, 2}) = 1$.
    \item At step $i = 1$, the selected item can be either from $\set{\sigma_1}$ or $\set{\sigma_3, \sigma_4}$. So two new states are generated here.
    \begin{itemize}
        \item The $\sigma_1$ is selected with probability $\Pi(1, 1) = 0.1$, which generates a new state $\angs{0, 2}$, and $q_1(\angs{0,2}) = q_0(\angs{1,2}) \cdot \Pi(1, 1) = 0.1$.
        \item An item $\sigma \in \set{\sigma_3, \sigma_4}$ is selected with probability $\Pi(1, 3) + \Pi(1, 4) = 0.6$, which generates a new state $\angs{1, 1}$, and $q_1(\angs{1, 1}) = q_0(\angs{1,2}) \cdot 0.6 = 0.6$.
    \end{itemize}
    So $\mathcal{P}_1 {=} \set{\angs{0, 2}, \angs{1, 1}}$, $q_1 {=} \set{\angs{0, 2} {\mapsto} 0.1, \angs{1, 1} {\mapsto} 0.6}$.
    \item At step $i = 2$, iterate states in $\mathcal{P}_1$.
    \begin{itemize}
        \item For state $\angs{0, 2}$, the selected item must be from the last two items in the remaining reference ranking. A new state $\angs{0, 1}$ is generated with probability $\Pi(2, 2) + \Pi(2, 3) = 0.8$.
        \item For state $\angs{1, 1}$, the selected item is either the first or last item in remaining reference ranking. A new state $\angs{0, 1}$ is generated with probability $\Pi(2, 1) = 0.1$, and another state $\angs{1, 0}$ is generated with probability $\Pi(2, 3) = 0.3$.
    \end{itemize}
    So $\mathcal{P}_2 = \set{\angs{0, 1}, \angs{1, 0}}$ and
    \begin{itemize}
        \item[$\square$] $q_2(\angs{0, 1}) = q_1(\angs{0, 2}) \cdot 0.8 + q_1(\angs{1, 1}) \cdot 0.1 = 0.1 \cdot 0.8 + 0.6 \cdot 0.1 = 0.14$
        \item[$\square$] $q_2(\angs{1, 0}) = q_1(\angs{1, 1}) \cdot 0.3 = 0.6 \cdot 0.3 = 0.18$
    \end{itemize}
    \item At step $i = 3$, item $\sigma_2$ must be selected to meet the requirement. For each state $\angs{\alpha, \beta} \in \mathcal{P}_2$, the rank of $\sigma_2$ is $(\alpha + 1)$ in the corresponding remaining ranking. So $\Pr(\sigma_2 @ 3 {\mid} \bsigma, \Pi) = q_2(\angs{0, 1}) \cdot \Pi(3, 1) + q_2(\angs{1, 0}) \cdot \Pi(3, 2) = 0.14 \cdot 0.3 + 0.18 \cdot 0.7 = 0.168$.
\end{itemize}
\end{example}

\section{Additional experiments}
\label{sec:appendix:experiments}

\revv{Figure~\ref{fig:mpw} in Section~\ref{sec:mew_vs_mpw} has demonstrated that \mew is much more scalable than \mpw under the plurality rule. Figure~\ref{fig:mpw_borda} presents results of a similar experiment under the Borda rule, with up to $6$ candidates and up to $15$ voters. We first fixed the number of voters to 5 and varied the number of candidates from 3 to 6, then fixed the number of candidates to 5 and varied the number of voters from 1 to 15. In this experiment, \mew is still much more scalable than \mpw.}

\begin{figure}[tb!]
  \centering
  \subfloat[\revv{5 voters}, varying \#candidates]{
    \label{fig:mpw_borda:candidates}
    \includegraphics[width=0.3\linewidth]{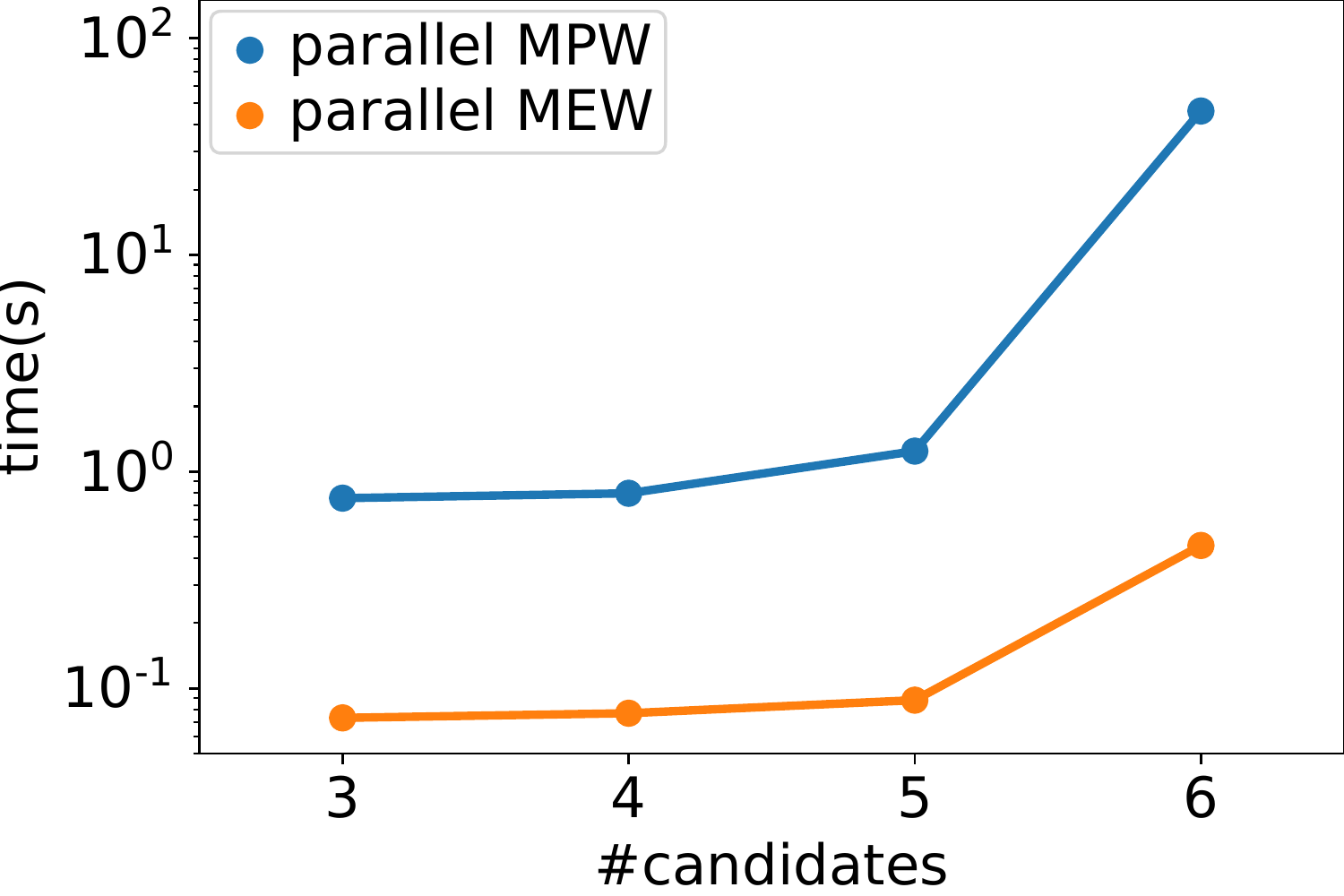}
  }\hspace{5em}
  \subfloat[\revv{5 candidates}, varying \#voters]{
    \label{fig:mpw_borda:voters}
    \includegraphics[width=0.3\linewidth]{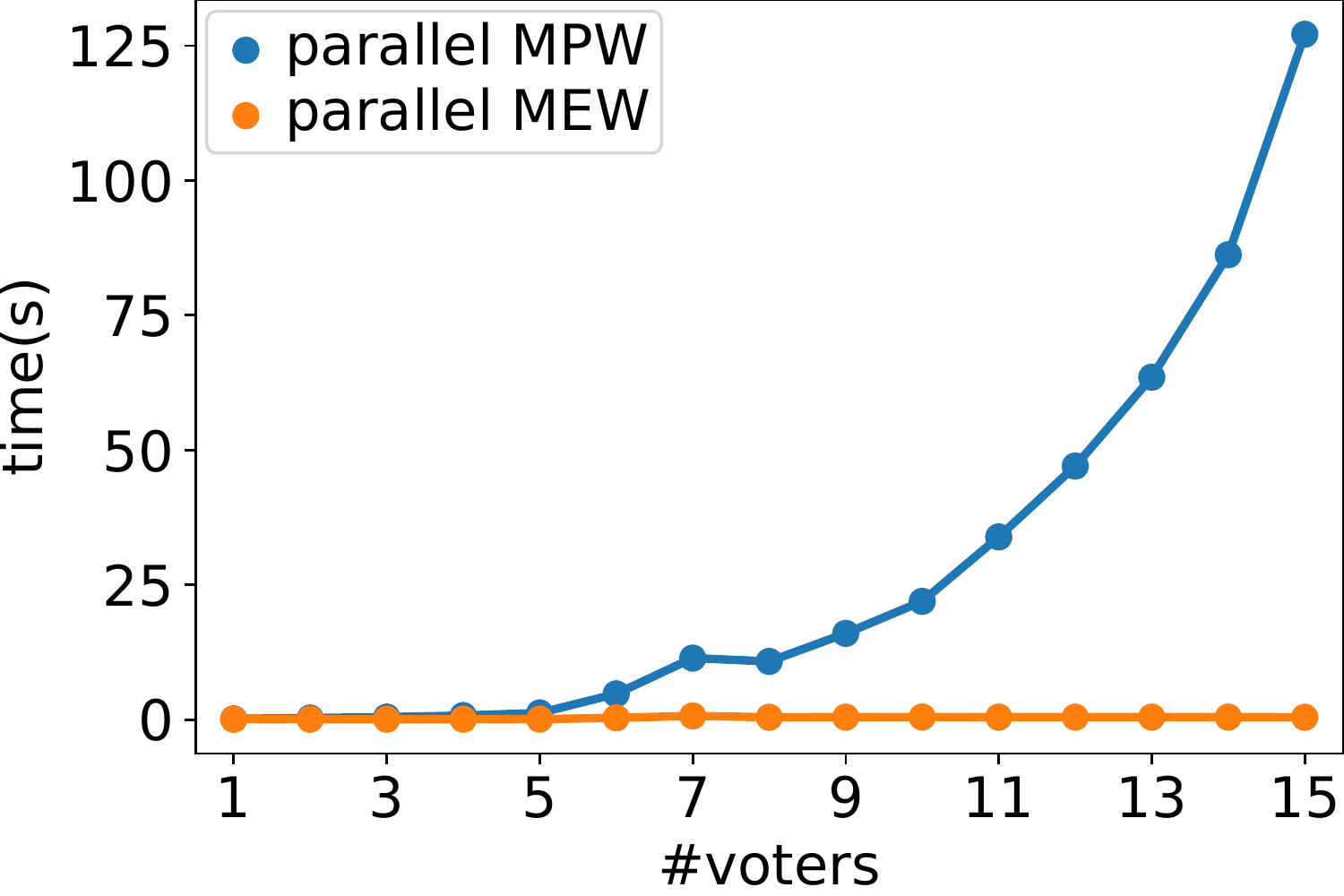}
  }
  \caption{\revv{Average time of parallel MPW and MEW, using 48 worker processes, under Borda, over partial voting profiles, fixing $\phi = 0.5$ and $p_{max}=0.1$. \mew scales much better than \mpw,  with both \#candidates and \#voters.}}
  \label{fig:mpw_borda}
\end{figure}

\received{April 2022}
\received[revised]{July 2022}
\received[accepted]{August 2022}
  
\end{document}
\endinput